\documentclass{aa}

\usepackage{graphicx}
\usepackage{txfonts}

\usepackage{subcaption}
\usepackage{lscape}
\usepackage{placeins}
                                
\usepackage{natbib}
\usepackage{graphicx}
\usepackage{txfonts}
\usepackage{color}
\usepackage{booktabs}
\usepackage[dvipsnames,table]{xcolor}
\usepackage[breaklinks,colorlinks=true,citecolor=blue,linkcolor=magenta,urlcolor=blue]{hyperref}
\usepackage{numdef}
\usepackage{caption}
\usepackage{rotating}%
\usepackage[flushleft]{threeparttable}
\usepackage{tikz}

\newcommand{\termconfig}[6][]{%
  \ensuremath{\smash{\mathrm{{#2\,#3}}\;^{#4}\mathrm{#5}_{#6}\ifx#1\empty\else^{\mathrm{#1}}\fi}}%
}

\newcommand{\msun}{${M}_{\odot}$}

\newcommand{\teff}{$T_\mathrm{eff}$}
\newcommand{\logg}{$\log g$}
\newcommand{\logy}{$\log n(\mathrm{He})/n(\mathrm{H})$}
\newcommand{\kms}{km\,s$^{-1}$}

\newcommand{\Z}{[Fe/H]}

\newcommand{\lsiv}{\object{LS\,IV\-$-$14$\degr$\-116}}
\newcommand{\feige}{\object{Feige\,46}}
\newcommand{\EC}{EC\,22536\-$-$5304}
\newcommand{\HD}{HD\,127493}
\num\def\BD+25{BD+25$^{\circ}\,$4655}
\newcommand{\RE}{\object{RE\,0503$-$289}}

\newcommand{\tlusty}{\textsc{Tlusty}}
\newcommand{\synspec}{\textsc{Synspec}}
\newcommand{\synthe}{\textsc{Synthe}}
\num\def\A9{\textsc{Atlas}{\footnotesize9}}
\num\def\A12{\textsc{Atlas}{\footnotesize12}}

\newcommand{\detboth}{\tikz[baseline=0.ex,x=1.2ex,y=1.2ex,line width=0.12ex,inner sep=0.1ex]{\draw (0.5,0.5) circle (0.5); \draw (0.5,0) -- (0.5,1); \draw (0,0.5) -- (1,0.5);}}
\newcommand{\detLS}{\tikz[baseline=0.ex,x=1.2ex,y=1.2ex,line width=0.125ex,inner sep=0.1ex]{\draw (0.5,0.5) circle (0.5);}}
\newcommand{\detEC}{\tikz[baseline=0.ex,x=1.2ex,y=1.2ex,line width=0.125ex,inner sep=0.1ex]{\draw (0.5,0) -- (0.5,1); \draw (0,0.5) -- (1,0.5);}}
\newcommand{\nodet}{\tikz[baseline=0.ex,x=1.2ex,y=1.2ex,line width=0.12ex,inner sep=0.1ex]{\draw (0,0.5) -- (1,0.5);}}

\begin{document}

\title{Evidence for neutron capture in heavy-metal hot subdwarfs}
\subtitle{Far-UV spectroscopy of \EC\ and \lsiv}

\author{
M. Dorsch\inst{1}
\and C. S. Jeffery\inst{2}
\and J. Deprince\inst{3,4}
\and D. J. Dougan\inst{5}
\and S. Beauraind\inst{3}
\and H. Dupuis\inst{3}
\and T. Battich\inst{6}
\and P. Quinet\inst{3}
\and U. Heber\inst{7}
\and L. J. A. Scott\inst{8}
\and S. Geier\inst{1}
}

\institute{
Institut für Physik und Astronomie, Universität Potsdam, Haus 28, Karl-Liebknecht-Str.\ 24/25, 14476 Potsdam, Germany\\
\email{dorsch@uni-potsdam.de}
\and Armagh Observatory and Planetarium, College Hill, Armagh BT61 9DG, United Kingdom
\and Atomic Physics and Astrophysics, Université de Mons (UMONS), Mons, Belgium
\and Astronomy and Astrophysics Institute, Université Libre de Bruxelles (ULB), Brussels, Belgium
\and Astrophysics Research Centre, Queen's University Belfast, Belfast, BT7 1NN, Northern Ireland, United Kingdom
\and Instituto de Astrofísica de La Plata, CONICET, Avenida Centenario (Paseo del Bosque) S/N, B1900FWA La Plata, Argentina
\and Dr.\ Remeis-Sternwarte \& ECAP, Friedrich-Alexander University Erlangen-Nürnberg, Sternwartstr.\ 7, 96049 Bamberg, Germany
\and School of Mathematics, Statistics and Physics, Newcastle University, Newcastle upon Tyne NE1 7RU, United Kingdom
}

\date{Received January XX, 20XX}

  \abstract
  {
  Most hot subdwarfs (sdO/B) are low-mass core-helium-burning stars formed through binary interaction. A subgroup of intermediate He-rich sdOBs shows extreme heavy-metal ($Z>30$) enrichments exceeding $10^4$ times solar, particularly in Zr or Pb.
  }
  {
  We analyse the first ultraviolet spectra of the ``heavy metal'' subdwarfs LS\,IV$-$14$\degr$116 (Zr-rich) and EC\,22536$-$5304 (Pb-rich) to determine their abundance patterns and test nucleosynthesis models. The targets probe contrasting evolutionary channels: the close binary EC\,22536$-$5304 with an sdF companion and the apparently single LS\,IV$-$14$\degr$116, likely formed by a white-dwarf merger.
  }
  {
  Both stars show exceptionally rich heavy-element spectra dominated by ions in stages \textsc{iii-vi}, many transitions being absent from standard line lists. We compiled literature energy levels, wavelengths, and oscillator strengths and implemented them in the spectral-synthesis code \textsc{Synspec}. Where data were unavailable we computed oscillator strengths for \ion{As}{iii}, \ion{Se}{iii}, \ion{Hf}{iv}, and \ion{Tl}{iv}. New photoionisation cross-sections for \ion{Pb}{iii-vi} enabled the first non-LTE models of multiply ionised Pb.
  }
  {
  In LS\,IV$-$14$\degr$116 we detect 16 light and 24 heavy metals (Ga-Bi); Br, Nb, Mo, Pd, In, Sb, Te, and Xe are measured in an sdO/B star for the first time. EC\,22536$-$5304 is even more enriched, with 13 light and 26 heavy metals detected, including first detections of La, Ce, Pr, Nd, Er, Yb, Lu, Hf, Ta, W, Os, Pt, Hg, Tl, and Bi, mainly in stages \textsc{iv-v}. LS\,IV$-$14$\degr$116 peaks at $\sim$4.3 dex for Sr-Sn, declining to 3.1 dex at Pb and 2.3 dex at Bi, whereas EC\,22536$-$5304 shows 1-3 dex enhancements from Ga-Sn, rising to $\sim$4 dex for La-Tl and reaching 6.2 dex for Pb and 5.4 dex for Bi. Both stars are Fe-poor ($-0.8$ and $-2.5$ dex).
  }
  {
  The photospheres of both stars are strongly enriched in heavy elements relative to their initial compositions. The abundance patterns cannot be explained by atomic diffusion alone and retain a clear signature of nucleosynthesis. The distribution in EC\,22536$-$5304 closely matches predictions of $i$-process nucleosynthesis during its formation, providing strong evidence for $i$-process self-enrichment in hot subdwarfs. The contrasting abundance patterns imply different evolutionary histories and $i$-process conditions: EC\,22536$-$5304 likely formed via Roche-lobe overflow onto its sdF-type companion, whereas the single star LS\,IV$-$14$\degr$116 most likely originated from the merger of two low-mass white dwarfs. These results suggest that heavy metals in other He-rich sdOB and sdO stars are also self-synthesised and that $i$-process nucleosynthesis may be widespread among He-rich hot subdwarfs.
  }


   \maketitle

\section{Introduction}

Most subdwarf B (sdB) stars and subdwarf O (sdO) stars have masses close to 0.5\,\msun\ and are found on the extended horizontal branch (EHB) or near the helium main sequence \citep[HeMS; see reviews by][]{Heber2009,Heber2016,Heber2024}. The absence of extended hydrogen-rich envelopes in these stars inhibits their evolution towards the asymptotic giant branch (AGB) and is widely attributed to stripping via binary interaction. 
Observationally, between 50 and 70\,\% of sdB stars are found in binary systems \citep{Maxted2001, Stark2003}, implying the existence of multiple post-red giant branch (RGB) evolutionary channels \citep{Han2002}: (i) common-envelope ejection in a red giant-white dwarf system, (ii) common-envelope ejection in a red giant-M-dwarf system, (iii) stable mass transfer from a red giant to an F-, G-, or K-type companion, and (iv) the merger of two helium white dwarfs.

The majority of hot subdwarfs (90\%) exhibit helium-poor surface compositions \citep{Dawson2024}. 
If sdB stars originate from stripped red-giant cores, any residual envelope is expected to be helium-enriched through first dredge-up or flash-driven convection at helium ignition \citep{MillerBertolami2008}. The commonly invoked explanation for the observed helium depletion is that, after contraction onto the HeMS/EHB at $T_{\rm eff}\lesssim 40\,000$\,K, gravitational settling becomes efficient enough to rapidly transform an initially helium-rich surface into a helium-poor one \citep{Michaud2011, Hu2011}. 
Also the abundances of light metals (C, N, O, Ne) should affected by past helium fusion and the hydrogen CNO-cycle \citep{Battich2018}, as well as atomic diffusion \citep{Byrne2018_CE}. 

A minority of hot subdwarfs ($\approx$10\,\%) show extremely helium-rich surfaces ($80 - 100\%$ helium by number). 
These extreme He-sdOs lie at higher temperatures \citep{Stroeer2007} and they have been identified with the products of double helium WD mergers \citep{Saio2000, Zhang2012, Yu2021}, and show surface CNO abundances affected by both hydrogen and helium fusion. 
A similar number of hot subdwarfs are intermediate in helium (iHe-sdOBs) with surfaces having $10 - 80\%$ helium \citep{Dawson2026}. 
These may either be in a process of transition from red-giant core to EHB \citep{MillerBertolami2008} or simply have insufficient hydrogen to quench the helium as the product of some other history, such as helium WD mergers. 

At least twelve iHe-sdOBs have extraordinary surface compositions, showing zirconium, lead and other heavy elements overabundant by factors of 10\,000 or more when compared to the Sun, thus earning the nickname ``heavy-metal stars''. 
These heavy-metal stars appear to divide into two major groups; the warmer ones are extremely lead-rich \citep[e.g.][]{Naslim2013, Naslim2020, Jeffery2017, Nemeth2021} whilst the coolest are zirconium-rich \citep{Naslim2011, Latour2019b, Ostensen2020}. 
The latter are best represented by the zirconium star \lsiv, with optical lines due to Ge, Ga, Se, Kr, Y, Sr, Zr, and Sn that indicate overabundances between $10^3$ and $10^5$ times solar for these elements \citep[][]{Naslim2011,Dorsch2020}. 

The lead-rich \EC\ \citep{Jeffery2019} is of particular interest because it is the first iHe-sdOB for which the primordial metallicity can be inferred via its F-type companion \citep[\Z\,=\,$-$2,][]{Dorsch2021}.
This low metallicity indicates that it is part of the Galactic halo, like \lsiv\ \citep{Randall2015}. 
\EC\ is the first heavy-metal iHe-sdOB found in a binary, and the most lead-rich one. 
Given its 457\,d orbital period \citep{Dorsch2021}, it is very likely that the system formed through stable Roche lobe overflow from a red giant progenitor to the sdF-type companion. This presents a unique opportunity to compare observed abundances with well-constrained evolutionary models.

A first set of detailed nuclear synthesis calculations for post-RLOF systems was performed by \cite{Battich2023}. 
This exploratory analysis showed that neutron capture processes can take place during the late helium flashes in the cores of such hot subdwarf progenitors. 
Neutron production is initiated by proton ingestion into the He-burning layer. Protons are captured by \(^{12}\mathrm{C}\) via \(^{12}\mathrm{C}(p,\gamma)^{13}\mathrm{N}\), which beta decays to \(^{13}\mathrm{C}\). The resulting \(^{13}\mathrm{C}\) acts as a neutron source through \(^{13}\mathrm{C}(\alpha,n)^{16}\mathrm{O}\). 
The  neutron densities predicted by \cite{Battich2023} are in the range of $i$-process nucleosynthesis, allowing the production of heavy elements up to bismuth \citep[see][for a review]{Wiedeking2025_irev}.
These models demonstrated for the first time that not only diffusion, but also self-synthesis of heavy elements can contribute to the extreme enrichments observed in heavy metal subdwarfs. 

Encouraged by these results, we have conducted a detailed analysis of high-quality Hubble Space Telescope (HST) observations of \EC, the most lead-rich star known, as well as \lsiv, the prototype zirconium star.  
In parallel, \cite{Battich2025} carried out comprehensive nucleosynthesis simulations specifically for \EC, and compared their predictions to a preliminary version of the metal abundances measured as part of this work. 

\section{Observations}
\label{sec:observations}

HST/STIS E140M observations of both stars were obtained in Cycle 30 under programme ID 17072 \citep{Dorsch2022_HST}. The spectra cover 1143 -- 1730\,\AA\ at a resolving power of $R=45\,800$ and a median signal-to-noise ratio of about 20. The programme also included an E140H spectrum of the heavy-metal iHe-sdO \HD\ \citep{Dorsch2019}, spanning 1163.3 -- 1356.9\,\AA\ at even higher resolution ($R$ = 114\,000) and reaching signal-to-noise ratio of 85 at 1300\,\AA. 
It was used to refine line wavelengths and resolve blended features, aiding the interpretation of the \EC\ and \lsiv\ spectra.

Both \lsiv\ and \EC\ have been extensively observed in the optical and near-UV ranges as well. While the focus of the present analysis is on the HST spectra, we used high-resolution UVES spectra ($R$ $\approx$ 40\,000) of both stars to constrain metals that show spectral lines only in the optical or both ranges. These spectra are the same as used by \cite{Dorsch2020} for \lsiv\ (SNR = 200) and \cite{Dorsch2021} for \EC\ (SNR = 50). 

\section{Atomic data}
\label{sec:atomic_data}

The far-UV spectra of both stars contain numerous lines from multiply ionised heavy metals. Because standard line lists omit many of these ions and are incomplete for others, we compiled line positions and oscillator strengths from various sources and significantly expanded the line list used by \citet{Dorsch2019} in their analysis of the heavy-metal sdOBs  HZ\,44 and HD\,127493. This compilation began with R.\,L.\,Kurucz's line list\footnote{\url{http://kurucz.harvard.edu/linelists/gfnew/gfall08oct17.dat}} \citep{Kurucz2018} and was expanded using data from the NIST database\footnote{\url{https://physics.nist.gov/PhysRefData/ASD/lines_form.html}} \citep{NIST_ASD}, the Atomic Line List v3.00b5\footnote{\url{https://linelist.pa.uky.edu/newpage/}} \citep{vanHoof2018}, the DREAM database\footnote{\url{https://agif.umons.ac.be/databases/dream.html}} \citep{Quinet2020_DREAM}, 
the TOSS database \citep{vo:TOSS_legacy}, 
and additional individual papers. 
More details are given for each heavy metal in Sect.\ \ref{sec:abundance_analysis}. 
While we have aimed to compile the best data available for each ion in stages \textsc{iii-v}, and to be as complete as possible in the $Z$ = 31 to 83 range, some sources may have been missed given the large number of heavy ions present in both stars. 

\subsection{New atomic data for \ion{As}{iii}, \ion{Se}{iii}, \ion{Hf}{iv} and \ion{Tl}{iv}}
\label{sect:umons}

Oscillator strengths were not available for many transitions originating from heavy ions detected in the spectra of \lsiv\ and \EC. As further discussed in Sect.\ \ref{sec:abundance_analysis}, this includes \ion{As}{iii}, \ion{Se}{iii}, \ion{Hf}{IV} and \ion{Tl}{IV} lines. For these ions, new sets of oscillator strengths and line positions were computed at the University of Mons. To do so, we used the pseudo-relativistic Hartree-Fock (HFR) method \citep{Cowan1981} that was modified to take core-polarization (CPOL) effects into account through the introduction of a polarization potential \citep[HFR+CPOL, see][]{Quinet1999,Quinet2002}. For each ion, radial parameters were refined via a semi-empirical least-squares fit to available experimental energy levels.

In this paper, we provide the HFR atomic data for the lines of interest in \ion{As}{iii}, \ion{Se}{iii}, \ion{Hf}{iv} and \ion{Tl}{iv} for the observed spectra, only summarising the important details of our atomic computations. More detail about our calculations and fits, in addition to atomic data for other lines in these ions, as well as a detailed discussion and comparison with the results obtained with an independent method (fully relativistic multiconfiguration Dirac-Hartree-Fock) will be presented in a future paper (Deprince et al., in prep.). 

The HFR multiconfiguration models considered for these four ions are given in Table \ref{tab:atomic_conf}, accompanied by the CPOL parameters used for each ion, namely the core dipole polarizability, $\alpha_d$, and the cut-off radius of the ionic core, $r_c$. The models were built by considering all configurations for which some energy levels are experimentally known, in addition to configuration interactions motivated based on previous works concerning similar ions (e.g., same isoelectronic sequence).
The list of the \ion{As}{iii}, \ion{Se}{iii}, \ion{Hf}{iv} and \ion{Tl}{iv} transitions computed with HFR and observed in this work (see Sect.\ \ref{sec:abundance_analysis}) is given in Table \ref{tab:lines_HFR} with the calculated oscillator strengths and observed wavelengths. 

\begin{table}[]
    \centering
    \caption{List of the lines computed by HFR which are observed in the \lsiv\ (\ion{As}{iii}, \ion{Se}{iii}) and \EC\ (\ion{Hf}{iv}, \ion{Tl}{iv}) spectra in this work. Observed wavelengths are given in \AA.}
    \label{tab:lines_HFR}
\resizebox{\columnwidth}{!}{
    \begin{tabular}{ccccr}
    \toprule
    \toprule
    Ion   & Wavelength & Lower level & Upper level & $\log gf$  \\
    \midrule
    \ion{As}{iii}    & 1172.150 & 4s$^2$4p $^2$P$_{1/2}$ & 4s4p$^2$ $^2$D$_{3/2}$ & $-1.158$\\
                     & 1209.280 & 4s$^2$4p $^2$P$_{3/2}$ & 4s4p$^2$ $^2$D$_{5/2}$ & $-1.005$\\
                     & 1274.310 & 4s4p$^2$ $^2$D$_{5/2}$ & 4p$^3$ $^2$P$_{3/2}$   & $-0.152$\\
                     & 3922.498 & 4s$^2$5s $^2$S$_{1/2}$ & 4s$^2$5p $^2$P$_{3/2}$ & $ 0.184$\\
                     & 4037.035 & 4s$^2$5s $^2$S$_{1/2}$ & 4s$^2$5p $^2$P$_{1/2}$ & $-0.129$\\
                     \midrule
    \ion{Se}{III}    &          3387.233 & \termconfig[o]{4s^2}{4p5s}{1}{P}{1} & \termconfig[]{4s^2}{4p5p}{1}{D}{2} & $0.257$ \\
&3413.931 & \termconfig[o]{4s^2}{4p4d}{3}{F}{4} & \termconfig[]{4s^2}{4p5p}{3}{D}{3} & $0.200$ \\
&3428.416 & \termconfig[o]{4s^2}{4p4d}{3}{F}{2} & \termconfig[]{4s^2}{4p5p}{3}{D}{1} & $-0.378$ \\
&3457.816 & \termconfig[o]{4s^2}{4p5s}{3}{P}{2} & \termconfig[]{4s^2}{4p5p}{3}{S}{1}  & $-0.022$ \\
&3543.637 & \termconfig[o]{4s^2}{4p4d}{3}{F}{3} & \termconfig[]{4s^2}{4p5p}{3}{D}{2} & $0.008$ \\
&3570.201 & \termconfig[o]{4s^2}{4p5s}{3}{P}{1} & \termconfig[]{4s^2}{4p5p}{3}{P}{0} & $-0.420$ \\
&3637.525 & \termconfig[o]{4s^2}{4p5s}{3}{P}{2} & \termconfig[]{4s^2}{4p5p}{3}{P}{2} & $0.269$ \\
&3711.683 & \termconfig[o]{4s^2}{4p5s}{3}{P}{0} & \termconfig[]{4s^2}{4p5p}{3}{D}{1} & $-0.083$ \\
&3738.728 & \termconfig[o]{4s^2}{4p5s}{3}{P}{1} & \termconfig[]{4s^2}{4p5p}{3}{D}{2} & $0.239$ \\
&3742.929 & \termconfig[o]{4s^2}{4p4d}{3}{F}{2} & \termconfig[]{4s^2}{4p5p}{1}{P}{1}  & $-0.596$ \\
&3800.941 & \termconfig[o]{4s^2}{4p5s}{3}{P}{2} & \termconfig[]{4s^2}{4p5p}{3}{D}{3} & $0.409$ \\
&3849.640 & \termconfig[o]{4s^2}{4p5s}{3}{P}{2} & \termconfig[]{4s^2}{4p5p}{3}{P}{1} & $-0.678$ \\
&4046.730 & \termconfig[o]{4s^2}{4p5s}{1}{P}{1} & \termconfig[]{4s^2}{4p5p}{3}{P}{1} & $-0.324$ \\
&4083.173 & \termconfig[o]{4s^2}{4p5s}{3}{P}{0} & \termconfig[]{4s^2}{4p5p}{1}{P}{1} & $-0.798$ \\
&4169.070 & \termconfig[o]{4s^2}{4p5s}{3}{P}{1} & \termconfig[]{4s^2}{4p5p}{1}{P}{1} & $-0.168$ \\
&5232.730 & \termconfig[o]{4s^2}{4p5s}{1}{P}{1} & \termconfig[]{4s^2}{4p5p}{1}{P}{1} & $-0.589$ \\
&5898.110 & \termconfig[o]{4s^2}{4p4d}{3}{P}{2} & \termconfig[]{4s^2}{4p5p}{3}{P}{1} & $-0.568$ \\
&6023.580 & \termconfig[o]{4s^2}{4p4d}{3}{D}{2} & \termconfig[]{4s^2}{4p5p}{3}{S}{1} & $-0.632$ \\
&6303.800 & \termconfig[o]{4s^2}{4p4d}{3}{D}{3} & \termconfig[]{4s^2}{4p5p}{3}{P}{2} & $-0.252$ \\                 
\midrule
    \ion{Hf}{iv}     & 1305.211 & 4f$^{14}$5d $^2$D$_{3/2}$ & 4f$^{14}$6p $^2$P$_{3/2}$     & $-1.062$ \\
                     & 1357.339 & 4f$^{14}$6p $^2$P$_{1/2}$ & 4f$^{14}$6d $^2$D$_{3/2}$     & $ 0.353$ \\
                     & 1390.381 & 4f$^{14}$5d $^2$D$_{5/2}$ & 4f$^{13}$5d$^2$ $^2$D$_{3/2}$ & $-0.134$ \\
                     & 1491.669 & 4f$^{14}$5d $^2$D$_{3/2}$ & 4f$^{14}$6p $^2$P$_{1/2}$     & $-0.420$ \\
                     & 1528.777 & 4f$^{14}$6p $^2$P$_{3/2}$ & 4f$^{14}$6d $^2$D$_{5/2}$     & $ 0.556$ \\
                     & 1717.181 & 4f$^{14}$6s $^2$S$_{1/2}$ & 4f$^{14}$6p $^2$P$_{3/2}$     & $ 0.091$ \\
    \midrule
    \ion{Tl}{iv}     & 1272.872 & 5d$^9$6s $^1$D$_2$ & 5d$^9$6p $^1$P$_1$ & $-0.970$\\
                     & 1272.946 & 5d$^9$6s $^1$D$_2$ & 5d$^9$6p $^1$P$_1$ & $-0.669$\\
                     & 1304.482 & 5d$^9$6s $^3$D$_1$ & 5d$^9$6p $^1$D$_2$ & $-1.277$\\
                     & 1304.517 & 5d$^9$6s $^3$D$_1$ & 5d$^9$6p $^1$D$_2$ & $-1.101$\\
                     & 1337.081 & 5d$^9$6s $^3$D$_3$ & 5d$^9$6p $^3$D$_3$ & $-0.682$\\
                     & 1337.136 & 5d$^9$6s $^3$D$_3$ & 5d$^9$6p $^3$D$_3$ & $-0.557$\\
                     & 1358.496 & 5d$^9$6s $^1$D$_2$ & 5d$^9$6p $^1$D$_2$ & $-1.205$\\
                     & 1358.578 & 5d$^9$6s $^1$D$_2$ & 5d$^9$6p $^1$D$_2$ & $-1.029$\\
                     & 1374.619 & 5d$^9$6s $^3$D$_1$ & 5d$^9$6p $^3$F$_2$ & $-0.379$\\
                     & 1377.703 & 5d$^9$6s $^3$D$_3$ & 5d$^9$6p $^3$P$_2$ & $-0.463$\\
                     & 1377.793 & 5d$^9$6s $^3$D$_3$ & 5d$^9$6p $^3$P$_2$ & $-0.287$\\
                     & 1404.613 & 5d$^9$6s $^3$D$_2$ & 5d$^9$6p $^3$D$_3$ & $-0.472$\\
                     & 1404.668 & 5d$^9$6s $^3$D$_2$ & 5d$^9$6p $^3$D$_3$ & $-0.347$\\
                     & 1412.922 & 5d$^9$6s $^1$D$_2$ & 5d$^9$6p $^3$P$_1$ & $-1.208$\\
                     & 1413.008 & 5d$^9$6s $^1$D$_2$ & 5d$^9$6p $^3$P$_1$ & $-0.907$\\
                     & 1434.738 & 5d$^9$6s $^1$D$_2$ & 5d$^9$6p $^3$F$_2$ & $-1.066$\\
                     & 1434.811 & 5d$^9$6s $^1$D$_2$ & 5d$^9$6p $^3$F$_2$ & $-0.890$\\
                     & 1449.517 & 5d$^9$6s $^3$D$_2$ & 5d$^9$6p $^3$P$_2$ & $-1.248$\\
    \bottomrule
    \end{tabular}
}
\end{table}

\subsection{Atomic data for \ion{Pb}{iii-vi}}

Energy levels, line positions, and oscillator strengths are sufficient to calculate line strengths under the local thermal equilibrium (LTE) approximation. 
However, heavy-metal hot subdwarfs can reach temperatures of up to 40\,000\,K, where non-LTE effects are significant. Determining level populations under non-LTE conditions requires solving the statistical equilibrium equations, which in turn depend on detailed opacity data, particularly photoionisation cross sections.

As lead is by far the most abundant heavy metal in the atmosphere of \EC, photoionisation cross sections were computed for \ion{Pb}{iii-vi}, as presented by \cite{Dougan2025_Pb}. 
In brief, level structures were calculated using the General Relativistic Atomic Structure Package \citep[\textsc{grasp}$^0$;][]{Dyall1989}, and the photoionisation cross sections were obtained with the $R$-matrix method \citep[see][for a review]{Burke2011}. 
The cross-sections were smoothed following \cite{Bautista1998} for the use with \tlusty. 
These data enabled us to construct the first non-LTE models for lead in heavy-metal subdwarfs.

\section{Metal abundance analysis}
\label{sec:abundance_analysis}

Our models of \lsiv\ and \EC\ were computed using \tlusty\ 205 non-LTE atmospheres and \synspec\ 51 synthetic spectra \citep[for a detailed description, see][]{Hubeny2017a}. 
\synspec\ was updated to treat ionised heavy metals in LTE using partition functions, following \cite{Dorsch2019} and \cite{Chayer2006}; the  updated version is available online\footnote{\url{https://github.com/mattidorsch/synspec51_fork}}. In the present analysis, Pb is the only heavy metal treated in full non-LTE. 
We used the same Pb and lighter-metal model atoms for both stars, as listed in Table \ref{tab:modelatoms}. 

The previous abundance studies of \EC\ were limited to optical spectroscopy. 
Since the surface gravity and helium abundance were already well determined from the optical spectrum, we kept the values of \cite{Dorsch2021}.
Like its sdF-type companion, \EC\ has a low iron abundance, as discussed in more detail in Sect.\ \ref{sect:iron}. 
We then performed a simultaneous $\chi^2$ fit to the optical UVES and far-UV HST spectra for \teff\ and all metal abundances, including heavy metals, as well as the microturbulence.
The sdF spectrum was modelled as in \cite{Dorsch2021} using a grid of \A12/\synthe\ models \citep{Kurucz1993,Kurucz1996}; it does not affect the UV spectrum.

For the hot subdwarf component, synthetic spectra were computed for each metal separately using \synspec, based on a single \tlusty\ model atmosphere for each \teff, \logg\ combination. Each spectrum was divided by a pure H+He spectrum from the same atmosphere, yielding normalized absorption profiles for the individual metals. The full synthetic spectrum was obtained by multiplying these metal profiles, interpolated to the desired abundances, with the H+He spectrum. This procedure is equivalent to adding metal opacities in the limit of weak lines, and remains a good approximation provided that saturated lines from different metals do not overlap significantly. The method has been successfully applied to sdB and B-type stars in previous studies \citep[e.g.][]{Irrgang2014,Dorsch2021,Schaffenroth2021}. %
Based on these fits, we adopt $T_\mathrm{eff} = 37750 \pm 500$\,K, $\log g = 5.80 \pm 0.10$, $\log n(\mathrm{He}) / n(\mathrm{H}) = -0.08 \pm 0.05$, and a microturbulence of $\xi = 0.7 \pm 0.5$\,\kms\ for \EC. 

We applied the same method for \lsiv, this time starting from the model of \cite{Dorsch2020}. 
The new STIS-E140M spectrum and the co-added UVES spectrum from \cite{Dorsch2020} were fit simultaneously.
Free parameters were \teff, 52 metal abundances, and the microturbulence. 
To limit computation time, the helium abundance was kept fixed to the value used by \cite{Dorsch2020} (\logy\ = $-$0.6). 
Multi-mode pulsations with periods comparable to the HST exposure time broaden the spectral lines of \lsiv\ through phase averaging. The dominant mode spans about 11\,\kms\ in radial velocity \citep{Jeffery2015}, which we model as a macroturbulent broadening of $\zeta$ = 11\,\kms. 
The UV spectrum of \lsiv\ is more crowded than that of \EC\ due to its higher iron-group abundances, which can be an issue with the multiplication method. 
Saturated intrinsic blends, spectral lines absent from the model, and lines known to be poorly reproduced were carefully removed before the final fit. 
Based on these fits, we adopt $T_\mathrm{eff} = 35230 \pm 500$\,K, $\log g = 5.80 \pm 0.10$, $\log n(\mathrm{He}) / n(\mathrm{H}) = -0.6 \pm 0.05$, and a microturbulence of $\xi = 0.8 \pm 1.0$\,\kms\ for \lsiv. 

The best-fit abundances of \lsiv\ and \EC\ are summarised in Table~\ref{tab:abu}, while Table~\ref{tab:ion_detection} provides an overview of the detected heavy ions and the availability of atomic data. %
The final abundances for both \lsiv\ and \EC\ are based on a final visual inspection of a fully self-consistent \tlusty/\synspec\ model, constructed without the multiplication method, which led to minor modifications to Table \ref{tab:abu}. The quoted uncertainties reflect our assessment of the robustness of each abundance measurement, based on the number of available lines, the degree of blending, and the overall fit quality. While formal statistical uncertainties are considered, these represent only a minor component of the total error budget. 
The following paragraphs describe the abundance determination and atomic data used for each metal in more detail, focusing on trans-iron elements. 
Metals not mentioned for a given star had no detectable spectral lines. 

\subsection{Light metals}

The abundances of most light metals up to sulphur -- specifically C, N, O, Mg, and Si for \EC, and additionally Ne and Ar for \lsiv\ -- are well constrained from optical and near-UV spectra. Since their spectral lines in the HST data are consistent with those in the optical range, our best-fit abundances closely match the results of \citet{Dorsch2020,Dorsch2021}. 

Several light metals cannot be measured from the optical spectra alone. 
Neon in \lsiv\ is well constrained by multiple \ion{Ne}{ii} lines in the near-UV, whereas in \EC\ only an upper limit can be derived from \ion{Ne}{ii} 4409.3\,\AA\ owing to the lack of near-UV data. Aluminium is clearly detected in \EC\ through the strong, isolated \ion{Al}{iii} lines at 1384.13, 1605.77, and 1611.87\,\AA. The non-detection of \ion{Al}{iii} 1611.87\,\AA\ in \lsiv\ suggests a sub-solar abundance, but we refrain from stating an upper limit based on a single line. 
The optical phosphorus abundance in \EC\ was based on two weak lines, whereas the stronger STIS \ion{P}{iii-iv} transitions (e.\,g.\ \ion{P}{iii} 1334.813, 1344.317\,\AA\ and \ion{P}{iv} 1377.275, 1467.427, 1484.509, 1487.789, 1489.087\,\AA) yield a more reliable and slightly higher abundance. In \lsiv, only the \ion{P}{iii} lines are detectable. 
The strongest sulphur lines in \EC\ are \ion{S}{iii} 1194.06, 1201.73\,\AA, \ion{S}{iv} 1286.07, 1623.962, 1629.17\,\AA, and \ion{S}{v} 1501.76\,\AA. In \lsiv, the \ion{S}{iii} and \ion{S}{v} lines are present but blended with stronger lines, and we only state an upper limit. Argon is not detected in \EC; in \lsiv\ it is detected only in the near-UV, yielding the same abundance as found by \citet{Dorsch2020}. Calcium is not detected in either star, nor are F, Na, Cl, or K.

\subsection{Iron-peak elements}
\label{sect:iron}

\EC\ has an extremely low iron abundance of approximately $-2.8$\,dex relative to the solar value by number, substantially lower than that of its sdF-type companion ($-1.95$\,dex; \citealt{Dorsch2021}). Consequently, iron lines are very weak in \EC, with only a few \ion{Fe}{iv} transitions detected at 1568.28, 1609.10, 1616.68, 1631.08, and 1724.06\,\AA. The strongest isolated nickel lines are \ion{Ni}{iv} lines at 1398.19, 1411.45, 1421.22, 1430.19, 1520.62, 1525.31, and 1534.71\,\AA. Although \lsiv\ also has a subsolar iron abundance ($-0.8$\,dex), its spectrum exhibits many strong iron and nickel lines. 

Titanium \textsc{iv} lines at 1451.74 and 1467.34\,\AA\ are clearly detected in both stars; \lsiv\ also shows \ion{Ti}{iii} 1286.37, 1293.23, 1294.7, 1455.20\,\AA, and \ion{Ti}{iv} 1183.63, 1195.20\,\AA. 
\ion{Ti}{iii} are systematically stronger than both UV and optical \ion{Ti}{iv} lines; we rely on \ion{Ti}{iv} for consistency between the two stars.  

No scandium and vanadium lines are detected in either star. 
The strongest predicted \ion{V}{iv} transitions, at 1355.13 and 1403.61\,\AA, are absent, and \ion{V}{iv} 1426.65\,\AA\ is blended with \ion{C}{iii} 1426.74\,\AA. In \lsiv, the upper limit is based on \ion{V}{iv} 1321.92\,\AA. 

Chromium lines are weak but clearly present in \EC; the abundance is based on \ion{Cr}{iv} 1306.05, 1319.70, 1325.03, 1339.19, 1355.33, 1361.63\,\AA. 
In \lsiv, about 15 \ion{Cr}{iv} lines are detected; the most isolated is \ion{Cr}{iv} 1332.44\,\AA. 

Manganese is not detected in \EC, with a sub-solar upper limit constrained by the non-detection of \ion{Mn}{iv} 1664.76 and 1666.99\,\AA. 
\ion{Mn}{iv} 1257.30, 1664.72, 1664.76, 1666.99\,\AA\ are detected in \lsiv, next to weaker lines. 

Cobalt \textsc{iii-iv} lines are strong in \lsiv, but not clearly detected in \EC; the strongest predicted line, \ion{Co}{iv} 1521.64\,\AA, coincides with a weak unidentified feature.

No copper lines are detected in \EC, limited by wavelength accuracy, although this remains consistent with their expected weakness. The strongest predicted transitions, \ion{Cu}{v} 1159.76\,\AA\ and \ion{Cu}{iii} 1593.75, 1642.20, and 1722.37\,\AA\ are detected in \lsiv\ but not clearly in \EC. 

In contrast, zinc lines are strong in \EC\ and very strong in \lsiv, allowing precise abundance determinations. Small wavelength corrections were applied to several \ion{Zn}{iii-iv} lines based on the \lsiv\ spectrum.

\begin{figure*}
\centering
\includegraphics[width=0.97\textwidth]{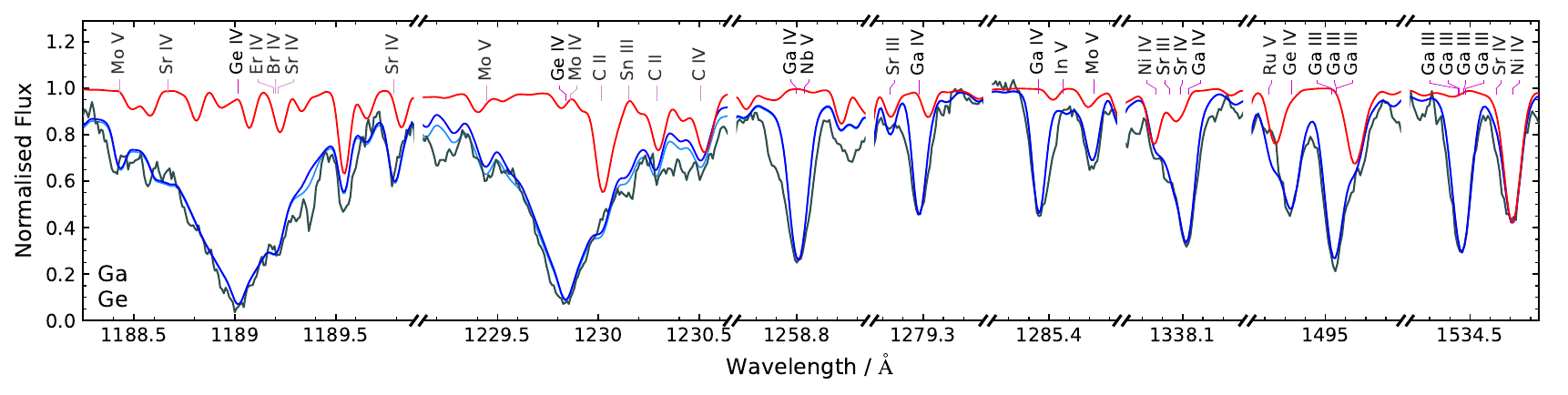}\\[-17.3pt]
\includegraphics[width=0.97\textwidth]{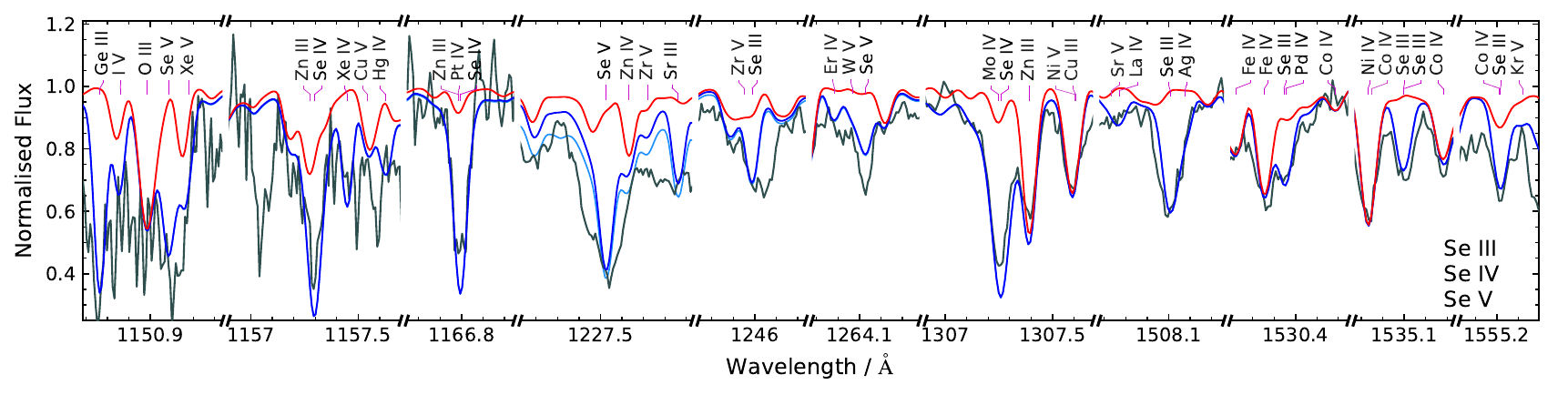}\\[-17.3pt]
\includegraphics[width=0.97\textwidth]{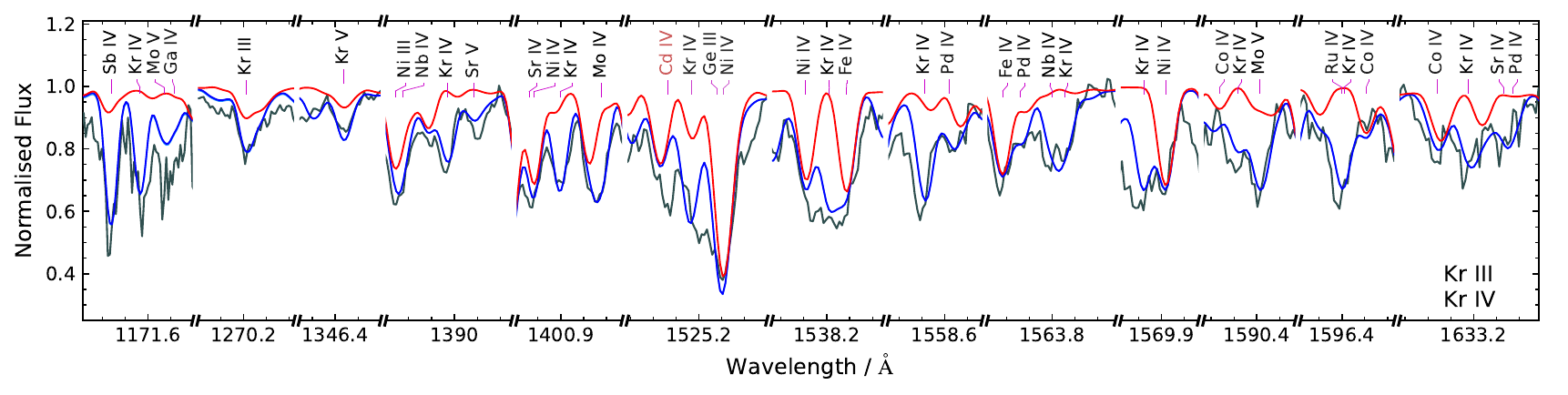}\\[-17.3pt]
\includegraphics[width=0.97\textwidth]{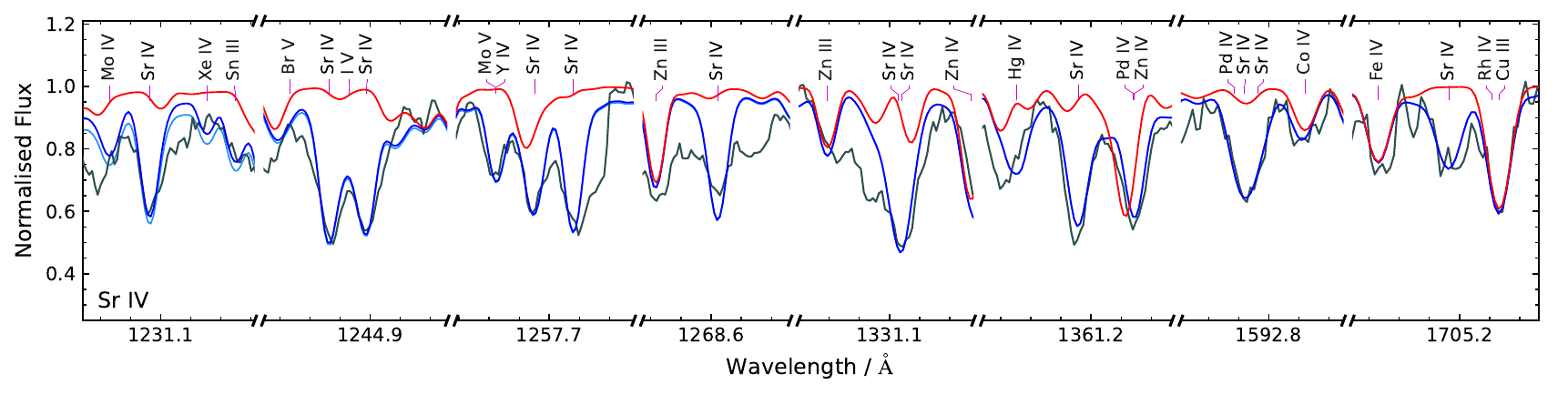}\\[-17.3pt]
\includegraphics[width=0.97\textwidth]{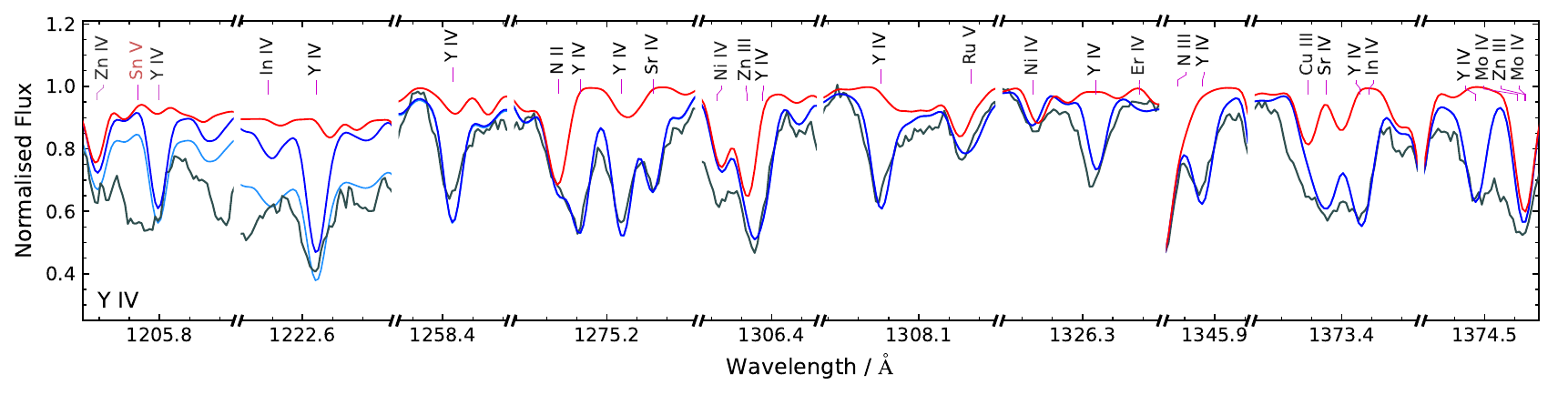}\\[-17.3pt]
\includegraphics[width=0.99\textwidth]{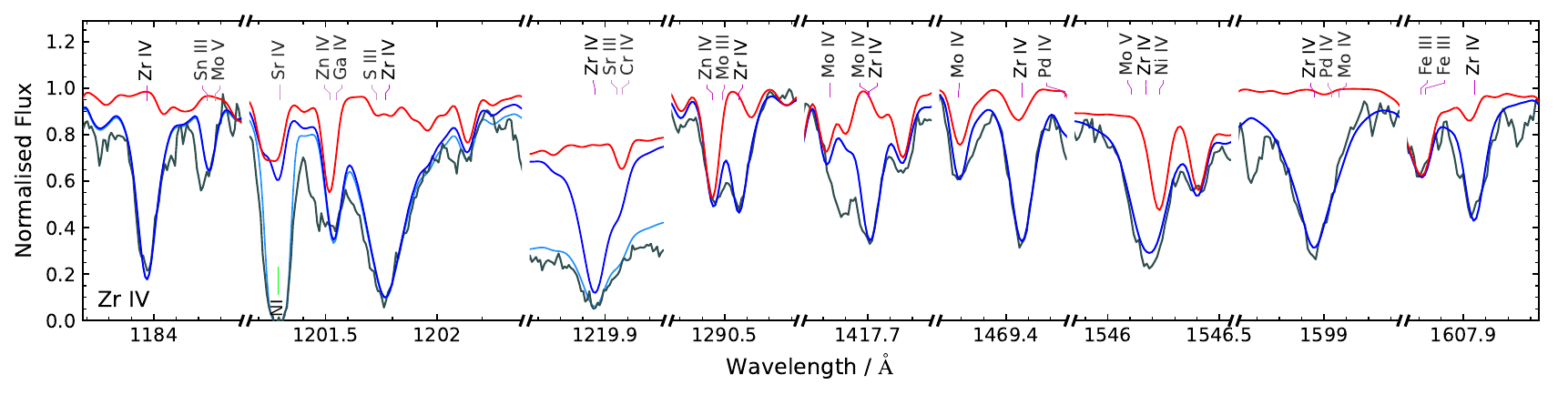}%
\vspace{-7pt}
\caption{The strongest Ga, Ge, Se, Kr, Sr, Y, and Zr lines in the STIS/E140M spectrum of \lsiv\ (grey), compared to our best model with (blue) and without (red) heavy metals ($Z>30$). The light blue model further includes interstellar absorption. 
}
\label{fig:detail_lsiv_1}
\end{figure*}

\subsection{Lighter trans-iron metals}

\begin{figure}
  \centering
  \includegraphics[width=0.99\columnwidth]{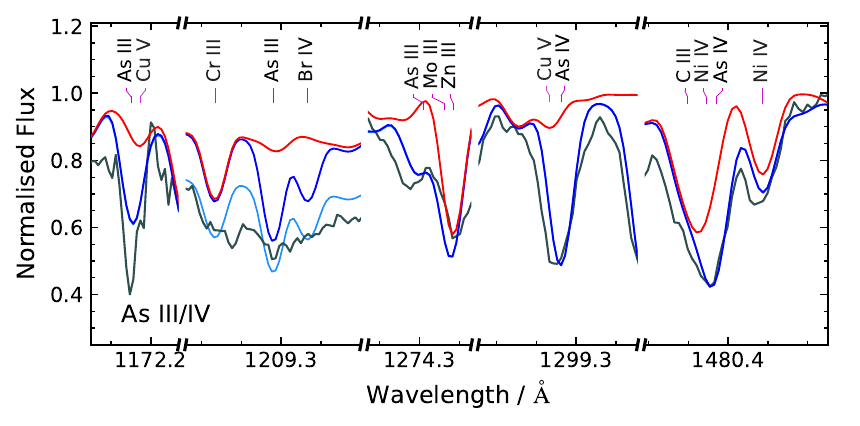}\\[-17.5pt]
  \includegraphics[width=0.99\columnwidth]{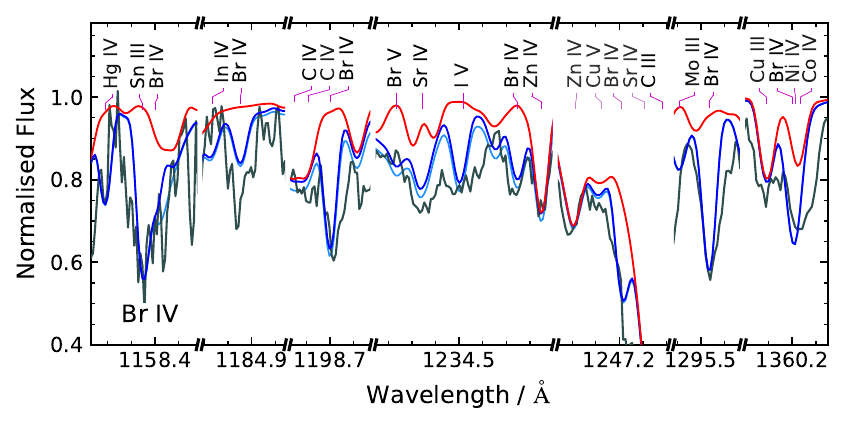}
  \vspace{-5pt}
  \caption{\textit{Top:} the strongest \ion{As}{iii-iv} lines in \lsiv, with available oscillator strengths (see Fig.\ \ref{fig:asiv_seiv_rbiv} for \ion{As}{iv} lines without). 
  \textit{Bottom}: similar for \ion{Br}{iv}; note also \ion{Hg}{iv} 1158.191\,\AA. 
  See Fig.\ \ref{fig:detail_lsiv_1} for line definitions. 
  }
  \label{fig:detail_lsiv_As}
\end{figure}

\begin{figure*}
\centering
\includegraphics[width=0.99\textwidth]{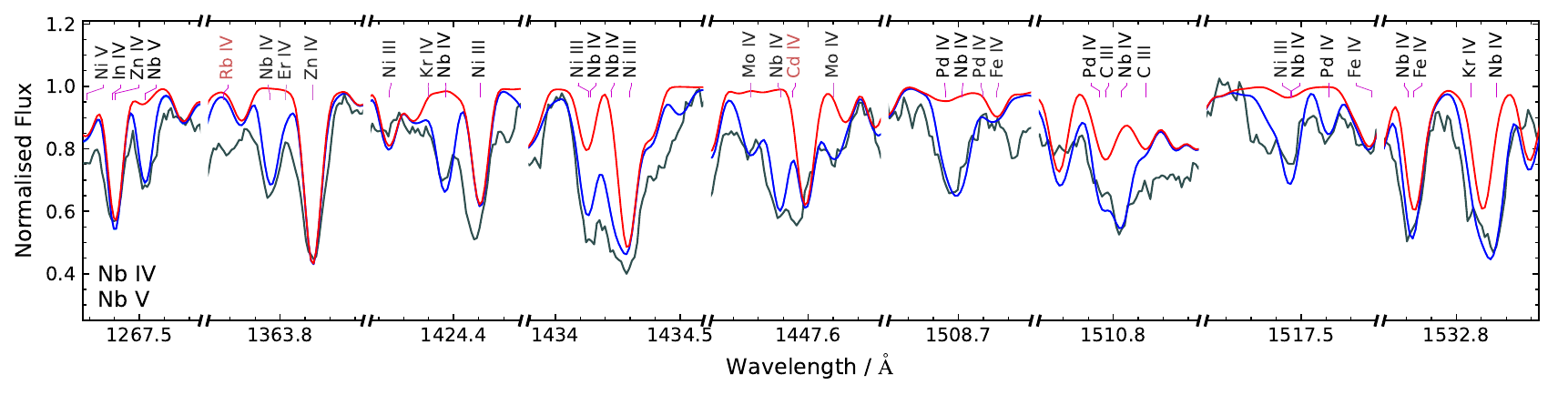}\\[-18.3pt]
\includegraphics[width=0.99\textwidth]{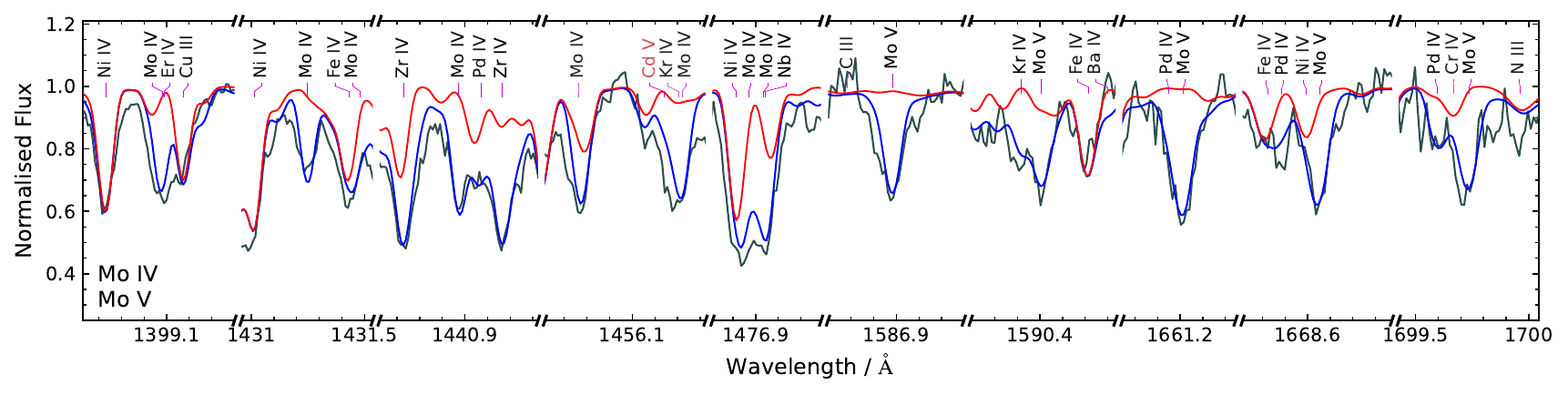}\\[-18.3pt]
\includegraphics[width=0.99\textwidth]{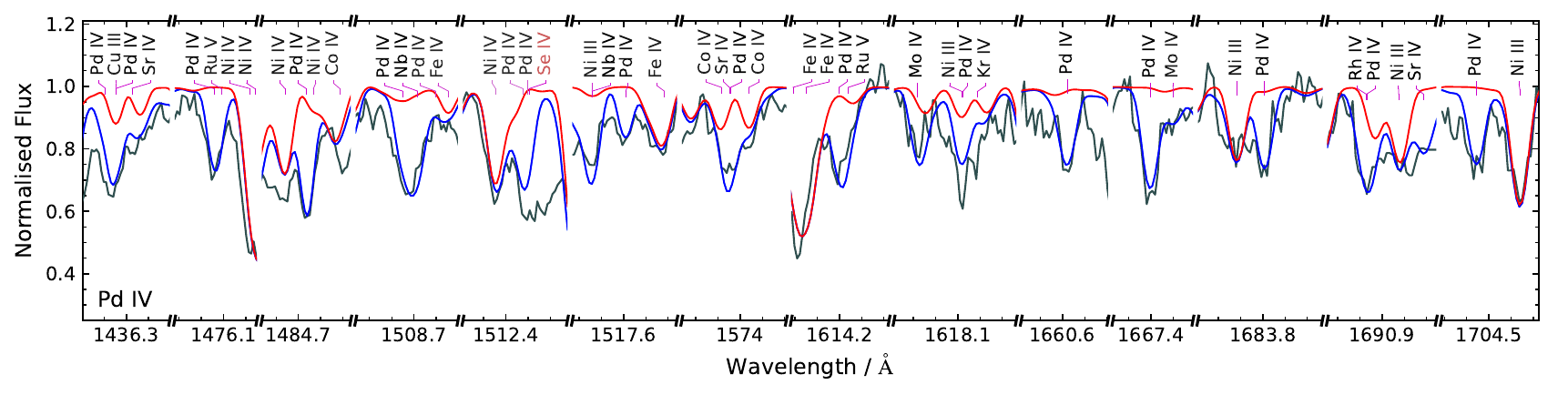}\\[-18.3pt]
\includegraphics[width=0.99\textwidth]{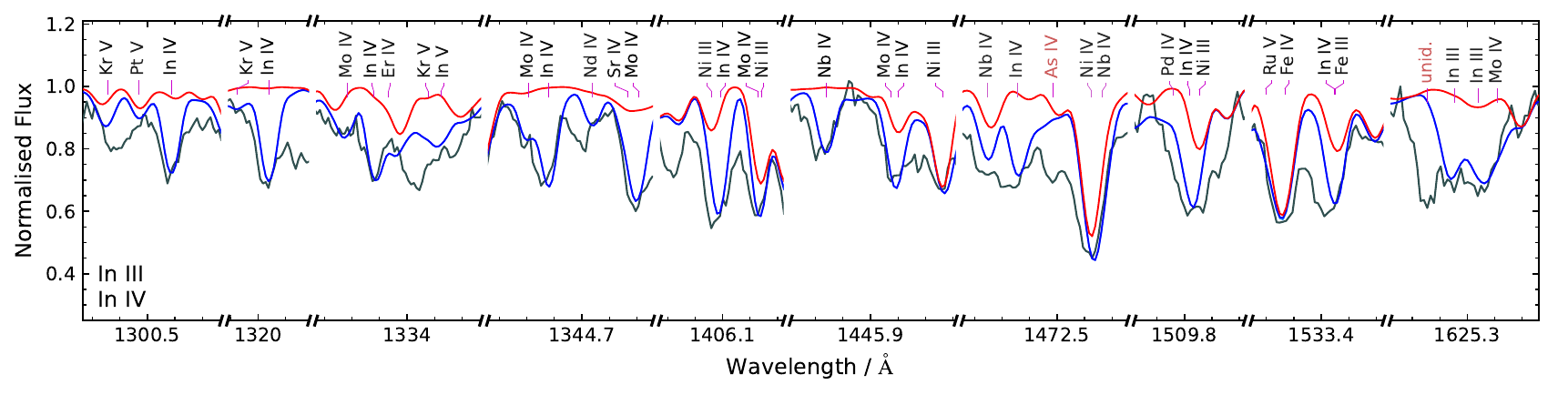}\\[-18.3pt]
\includegraphics[width=0.99\textwidth]{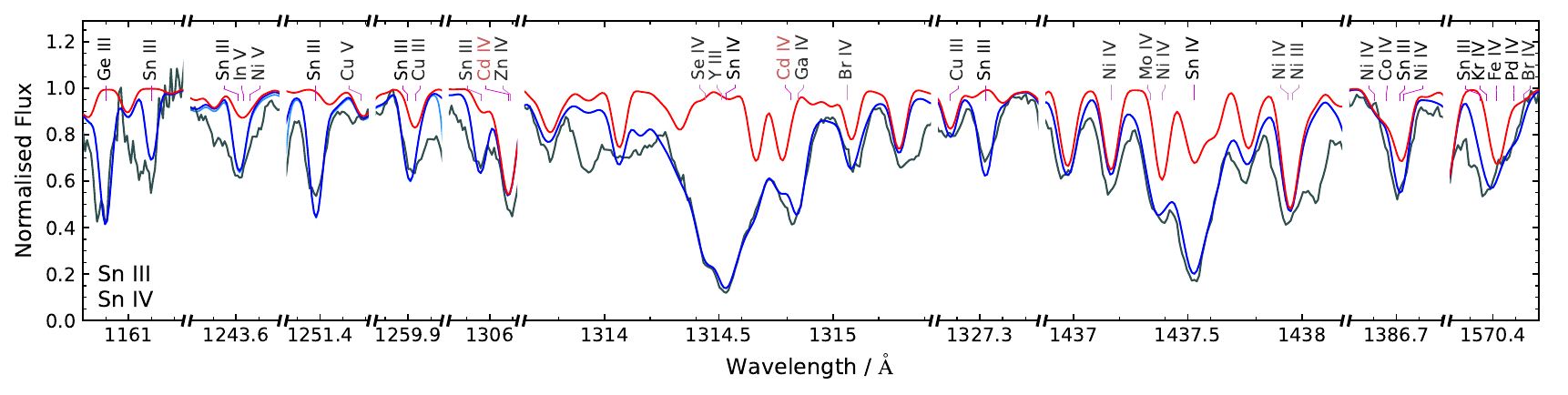}\\[-18.3pt]
\includegraphics[width=0.99\textwidth]{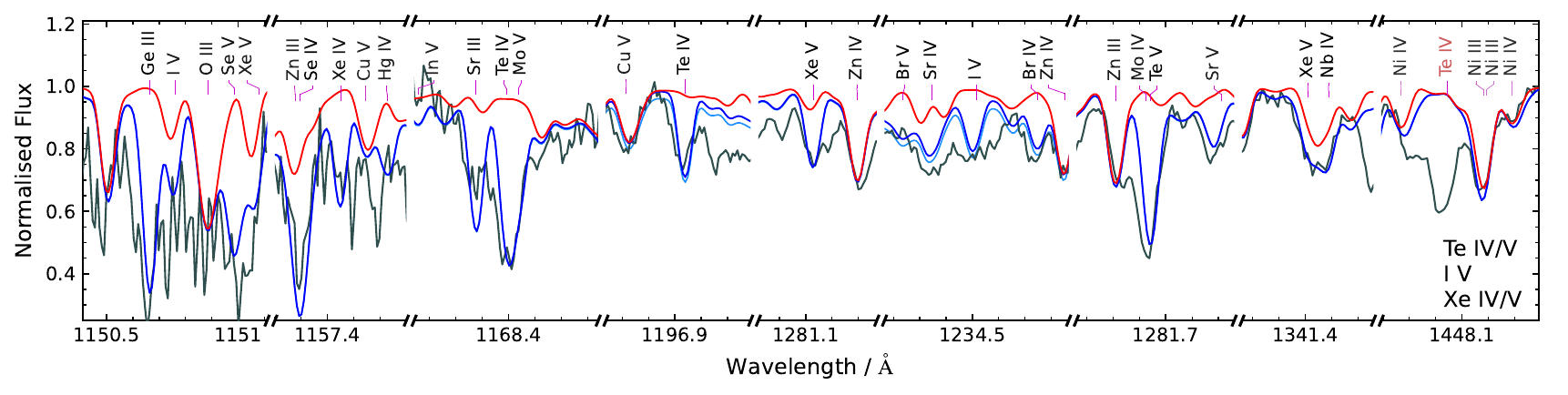}\\%
\vspace{-7pt}
\caption{Nb ($Z$\,=\,41), Mo (42), Pd (46), Cd (48), In (49),  Sn (50), Te (52), I (53), and Xe (54) lines in \lsiv, like Fig.\ \ref{fig:detail_lsiv_1}.
}
\label{fig:detail_lsiv_2}
\end{figure*}

\begin{figure}
  \centering
  \includegraphics[width=0.99\columnwidth]{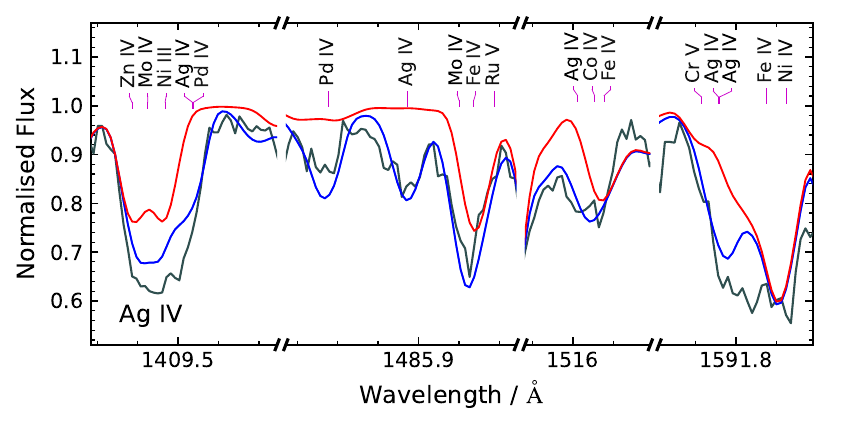}
  \vspace{-5pt}
  \caption{Possible detection of \ion{Ag}{iv} in \lsiv, like Fig.\ \ref{fig:detail_lsiv_1}. }
  \label{fig:detail_lsiv_ag}
\end{figure}

\begin{figure}
  \centering
\includegraphics[width=0.99\columnwidth]{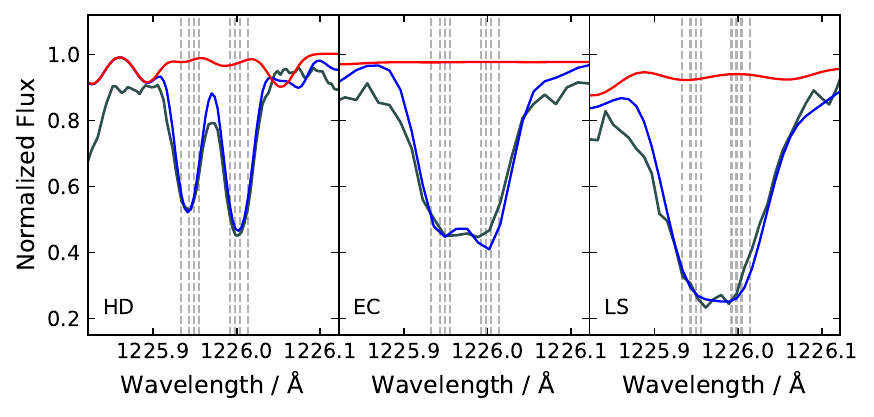}\\[-16pt]
\includegraphics[width=0.99\columnwidth]{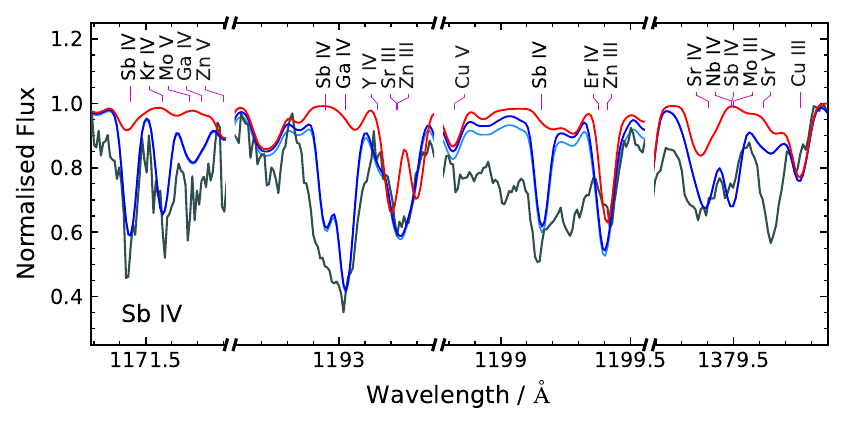}
  \vspace{-5pt}
  \caption{\textit{Top:}
  HFS splitting of \ion{Sb}{v} 1226\,\AA\ in \HD, \EC, and \lsiv. Split component wavelengths are indicated. The HFS splits are clearly resolved in \HD\ because it was observed at $R = 114\,000$. 
  \textit{Bottom:} the strongest \ion{Sb}{iv} lines in \lsiv, like Fig.\ \ref{fig:detail_lsiv_1}. 
  }
  \label{fig:detail_sb_hfs}
\end{figure}

Comparatively light trans-iron elements up to the lanthanides ($Z=57$) are more abundant in \lsiv\ than in \EC. 
In the following paragraphs, the detection of each metal is thus discussed mainly for the example of \lsiv\ and the strongest lines are shown in Figures \ref{fig:detail_lsiv_1} to \ref{fig:detail_sb_hfs}. 

\vspace{-10pt}\paragraph{Gallium. }

Gallium and germanium abundances in \lsiv\ have already been estimated from optical spectra \citep{Dorsch2020}, with both ions also detected in UV spectra of HZ\,44 and \HD\ by \cite{Dorsch2019}. 

For gallium, the UV features are consistent with the optical determinations. 
The resonance lines at 1495.045 and 1534.462\,\AA\ are the strongest \ion{Ga}{iii} features in \lsiv. We adopt fine-structure oscillator strengths from \cite{Nielsen2005} and include hyperfine structure (HFS) using magnetic dipole constants from \citet[][]{Dutta2013_GaIII_HFS}, derived with the relativistic coupled-cluster method (see also Appendix\ \ref{sect:HFS}). 
Although the hyperfine components are not resolved, including HFS is important as it significantly increases the equivalent widths of both lines. 

Strong \ion{Ga}{iv} lines are present at 1258.80, 1295.88, 1303.54, and 1338.13\,\AA, for which we use the atomic data of \citet{Rauch2015_Ga}. 
Both the \ion{Ga}{iii} resonance lines and the \ion{Ga}{iv} transitions are detected in \EC\ as well. No \ion{Ga}{v} features are identified in either star; the strongest predicted line at 1150.22\,\AA\ is blended with unidentified absorption. 

\vspace{-10pt}\paragraph{Germanium.}

For \ion{Ge}{iv} we use data from \cite{Dutta2011_GeIV}, supplemented with data from strengths from the ALL database. 
The \ion{Ge}{iv} resonance doublet at 1189.017 and 1229.838\,\AA\ is very strong in \lsiv\ (Fig.~\ref{fig:detail_lsiv_1}) and remains clearly detectable in \EC.
Besides these, the strongest \ion{Ge}{iv} lines occur at 1494.83 and 1500.55\,\AA. 
In \lsiv, the resonance lines are so strong that their profiles are strongly affected by Stark broadening, which we model using the Stark widths predicted by \citet{Elabidi2023_Stark_GaIV_GeIV}. 
Experimental wavelengths from \citet{Ryabtsev1993_GeIII}, together with preliminary \ion{Ge}{iii} oscillator strengths (Claudio Mendoza, priv.\ comm.), enabled the identification of \ion{Ge}{iii} lines at 1150.67, 1159.18, 1159.71, 1160.90, 1173.91, and 1183.49\,\AA, as well as the 1600\,\AA\ intercombination transition, the latter detected only in \lsiv.
Using the data of \citet{Rauch2012_Ge}, we also identify the weak \ion{Ge}{v} line at 1163.39\,\AA\ in \lsiv, while \ion{Ge}{v} 1165.26\,\AA\ is blended.

The  optical $3d^{10} 5s$ -- $3d^{10} 5p$ doublet \ion{Ge}{iv} 3554.257, 3676.735\,\AA\ is significantly too weak in our model at the best-fit abundances obtained from UV \ion{Ge}{iii-v} lines. It is unclear whether this discrepancy arises from non-LTE effects or other modelling limitations.  %

\vspace{-10pt}\paragraph{Arsenic. }

The strongest \ion{As}{iii} features in \lsiv\ occur at 1172.13, 1196.94, and 1209.28\,\AA, with wavelengths from \cite{Churilov1996_AsIII_AsIV}. 
Among these, only \ion{As}{iii} 1172.15\,\AA\ is unblended; 1196.94\,\AA\ is blended with \ion{Te}{iv} 1196.90\,\AA, and 1209.28, 1213.99\,\AA\ lie within the wings of the interstellar Ly\,$\alpha$ line (Fig.~\ref{fig:detail_lsiv_As}). 
The optical \ion{As}{iii} 3922.47, 4037.04\,\AA\ were previously identified in \lsiv\ by \cite{Dorsch2020}.
Because no oscillator strengths were available for these transitions, we computed new \ion{As}{iii} $f$-values (Sect.~\ref{sect:umons}), enabling a consistent treatment of both the UV and optical lines, all of which are well reproduced by our model.

The strong and isolated \ion{As}{iv} 1299.262\,\AA\ intercombination line is clearly detected in \lsiv, with an oscillator strength measured by \citet{Curtis1992_AsIV}. The \ion{As}{iv}\,1480.37\,\AA\ line is also detected; its wavelength was measured by \citet{Churilov1996_AsIII_AsIV}, and its lifetime by \citet{Pinnington1981_AsIII_IV_V}. 
Other lines in \lsiv\ coincide with \ion{As}{iv} positions measured by \citet{Churilov1996_AsIII_AsIV}, e.\,g.\ at 
1347.48, 1351.32, 1353.92, 1355.05, and 1472.47\,\AA\ (see Fig.\,\ref{fig:asiv_seiv_rbiv}).  
However, oscillator strengths for these transitions are not available. 
All optical \ion{As}{iv} lines of \cite{Churilov1996_AsIII_AsIV}  are too weak to be detected in the UVES spectrum of \lsiv. 
None of the \ion{As}{v} lines measured by \cite{Joshi1986_AsV} are detectable in the HST spectrum of \lsiv.

In \EC, the strongest arsenic lines, \ion{As}{iv} 1299.26, 1480.37\,\AA, are blended with the much stronger \ion{Ce}{v} 1299.30\,\AA\, and \ion{C}{iii} 1478.33\,\AA, respectively, preventing an abundance measurement. 

\vspace{-10pt}\paragraph{Selenium.}

\ion{Se}{iii}, \ion{Se}{iv}, and \ion{Se}{v} show strong lines in \lsiv\ (Fig.\ \ref{fig:detail_lsiv_1}). Several strong \ion{Se}{iii} lines were already identified in the UVES spectrum of \lsiv\ by \cite{Dorsch2020}, but no oscillator strengths were available at the time. 
New \ion{Se}{iii} atomic data, computed as summarised in Sect.~\ref{sect:umons}, now enable us to model 17 \ion{Se}{iii} transitions in the UVES spectrum. The energy levels were optimised using the experimental measurements of \citet{Tauheed2012}, although small wavelength adjustments were still required.
Updated wavelengths are provided in Table~\ref{tab:lines_HFR}, and the strongest \ion{Se}{iii} lines in the UVES spectrum are shown in Fig.~\ref{fig:seiii_uves}. 
Only \ion{Se}{iii} 1245.988\,\AA\ (\termconfig[o]{4s^2}{4p^2}{1}{D}{2} - \termconfig[]{4s}{4p^3}{2}{D}{3}) is securely identified in the UV range, and the selenium abundance in \lsiv\ is primarily derived from the UVES spectrum. 
In addition, the theoretical study of \cite{Kitoviene2024_SeIIIBrIVKrV} predicts strong \ion{Se}{iii} lines in the UV, involving upper 4$s^2$\,4$p$\,4$f$ $^1\mathrm{F}_3$, $^3\mathrm{F}_3$, and $^1\mathrm{G}_3$ levels, whose energies remain poorly constrained owing to the lack of experimental data; these lines are likely present in \lsiv. 

Based on the energy levels from \cite{Kelly1987}, as provided by NIST, several strong \ion{Se}{iv} lines are also identified in \lsiv, listed in Table \ref{tab:seiv} (see Fig.\,\ref{fig:asiv_seiv_rbiv}). 
Several of these transitions were originally identified experimentally by \cite{Rao1931_SeIV_SeV} and \cite{Gautam1972_SeIV}. 
Oscillator strengths are available for only three of these lines from beam-foil lifetime measurements by \cite{Bahr1982_Se}.
However, the derived line strengths are too strong compared to those of \ion{Se}{iii} and \ion{Se}{v}. 

Oscillator strengths for \ion{Se}{v} were computed by \cite{Rauch2017_SeSrTeI}, who also identified four \ion{Se}{v} lines in the spectrum of \RE. 
The strongest, at 1227.527\,\AA, is clearly isolated in \EC\ and consistent with \ion{Se}{iv} 1307.24\,\AA. 
In \lsiv, the \ion{Se}{v}  line is blended with at least two unidentified lines. 
Additional \ion{Se}{v} lines are detected in \lsiv\ at 1150.986 and 1264.129\,\AA, with equivalent widths between 10 and 20\,m\AA.

\vspace{-10pt}\paragraph{Bromine. }

Given the strong enrichment in neighbouring metals, we expect \lsiv\ to be enriched in bromine as well. 
Data for for \ion{Br}{iv-v} was provided by \citet{Riyaz2014_BrIV, Riyaz2014_BrV}. 
We identify \ion{Br}{iv} lines at 1158.4, 1184.852, 1198.702, 1234.746, 1247.209, 1295.536 and 1360.200\,\AA, which is the first detection of \ion{Br}{iv} in a star.
The strongest line is \ion{Br}{iv} 1295.536\,\AA\ at about 20\,m\AA\ (see Fig.\ \ref{fig:detail_lsiv_As}).
These lines were used to determine the bromine abundance in \lsiv, but no bromine lines are detectable in \EC. 
\ion{Br}{v} lines are weak even in \lsiv. 
\ion{Br}{v} 1155.650, 1234.236, 1244.633, 1371.306\,\AA\ improve the fit quality, but at equivalent widths of only about 10 m\AA\, they are blended with stronger lines. 
None of the \ion{Br}{iii} lines measured experimentally by \cite{Joshi1986_BrIII} and \cite{Jabeen2015_BrIII} are detectable in \lsiv. 

\vspace{-10pt}\paragraph{Krypton.}

The krypton abundance in \lsiv\ was already determined by \cite{Dorsch2020} from \ion{Kr}{iii} lines in its UVES spectrum. 
Next to the optical lines, \ion{Kr}{iii} 1270.21\,\AA\ is also detected using data from \cite{Raineri1998}. 
Krypton is one of the metals identified in white dwarfs by \cite{Rauch2016b}, who also provide line positions and oscillator strengths for \ion{Kr}{iv-vii} lines in the UV range. 
The HST spectrum of \lsiv\ shows 15 \ion{Kr}{iv} lines at equivalent widths larger than 15\,m\AA, the strongest being at 1558.514, 1596.409\,\AA\ (Fig.\ \ref{fig:detail_lsiv_1}). 
The abundance originally derived from the optical/near-UV \ion{Kr}{iii} lines is consistent with the \ion{Kr}{iv} lines in the HST spectrum. 
Several \ion{Kr}{v} lines are also detected in \lsiv, e.g. at 1346.438\,\AA, but are typically weaker than 10\,m\AA. 
The upper limit for krypton in \EC\ is based on the non-detection of \ion{Kr}{iv} 1525.17, 1538.21\,\AA; \ion{Kr}{v} 1589.269\,\AA\ is close to the detection threshold. 

\vspace{-10pt}\paragraph{Rubidium.}

Although oscillator strengths for \ion{Rb}{iv} are currently unavailable, accurate line positions have been measured by \citet{Persson1985_RbIV}. 
Nearly all of the 20 strongest lines reported in that work correspond to otherwise unidentified features in the HST spectrum of \lsiv, with \ion{Rb}{iv} 1383.47, 1400.58, 1407.09\,\AA\ being the most prominent examples (see Fig.\,\ref{fig:asiv_seiv_rbiv}). 
While the identification of individual lines remains uncertain, the large number of matches suggests a high rubidium enrichment, similar to that of neighbouring metals. 

Extensive \ion{Rb}{v} oscillator strengths were published by \citet{Radziute2022_RbV}, with experimental energies from \citet{Persson1984_RbV_energy} and \citet{O'Sullivan1989_RbV_energy}, but the predicted STIS-range transitions are not covered by these measurements, and the theoretical energies are insufficiently accurate for reliable identification. The level structure of \ion{Rb}{iii} similarly produces no strong transitions in the STIS/E140M range, but several optical transitions are observable using data by \cite{Zhang2014_RbIII}. The \ion{Rb}{iii} 3492.677\,\AA\ line measured experimentally \citep{Reader1972_RbIII} coincides with a strong feature in the UVES spectrum of \lsiv, and would be consistent with the weaker \ion{Rb}{iii} 3330.155, 3511.088, 3636.744\,\AA.  
Reliable \ion{Rb}{iv} oscillator strengths remain essential to determine the rubidium abundance and should be calculated to enable such measurements. 

\vspace{-10pt}\paragraph{Strontium.}

The strontium abundance of \lsiv\ was previously determined from optical \ion{Sr}{ii-iii} lines \citep{Naslim2011, Dorsch2020}. 
Its UV spectrum shows about 150 \ion{Sr}{iv} and 30 \ion{Sr}{iii} lines with equivalent widths of 15 to 22\,m\AA\ (see Fig.\ \ref{fig:detail_lsiv_1}), and many weaker features, similar to the HST/GHRS spectrum of its spectroscopic twin, \feige\ \citep{Latour2019b}.
Two \ion{Sr}{v} lines, at 1390.084 and 1396.308\,\AA, are identified with strengths near 10\,m\AA; the remaining lines are blended with stronger features. 
For the present analysis, \ion{Sr}{iv-v} atomic data were taken from \citet{Rauch2017_SeSrTeI}, while \ion{Sr}{iii} data were adopted from R.\ Kurucz's database\footnote{\url{http://kurucz.harvard.edu/atoms.html}}. 
The \ion{Sr}{iv} lines in the UV range are consistent with optical \ion{Sr}{iii} transitions; \ion{Sr}{ii} 4077.714, 4215.524\,\AA\ are expected to be affected by stronger non-LTE effects and appear too weak in our LTE models. 

Strontium lines are considerably weaker in \EC, yet several \ion{Sr}{iv} transitions (1244.137, 1244.763, 1244.888, 1257.780, 1331.129, and 1361.159\,\AA, see Fig.\ \ref{fig:detail_lsiv_4}) are still detected, with equivalent widths of 4 to 8\,m\AA. \ion{Sr}{iv} 1408.678\,\AA\ is blended with \ion{Ni}{iv} 1408.715\,\AA, while the strongest strontium line, \ion{Sr}{iv} 1347.901\,\AA, is blended with \ion{C}{iii} 1347.947\,\AA. 

\vspace{-10pt}\paragraph{Yttrium.}

The yttrium abundance in \lsiv\ was already determined by \cite{Naslim2011} and \cite{Dorsch2020} based on the optical \ion{Y}{iii} 4039.6, 4040.1\,\AA\ doublet. 
The strongest yttrium lines in the UV spectrum originate from \ion{Y}{iv}, with about 20 lines with strengths between 10 and 20\,m\AA, as predicted using atomic data from \cite{Loginov2001_YIV}. The strongest lines \ion{Y}{iv} lines, at 1205.796, 1222.645, 1257.523, 1258.433, 1275.112, 1275.248, 1306.369, 1307.974, 1326.344, 1345.859, 1373.461, and 1374.470\,\AA, are shown in Fig.\ \ref{fig:detail_lsiv_1}. 
All \ion{Y}{iii} lines predicted in the UV range using data from \cite{Fernandez2020} are too weak to be detected. 
The \ion{Y}{iv} lines observed in the UV range require a lower yttrium abundance than the \ion{Y}{iii} 4039.6, 4040.1\,\AA\ doublet; this may be caused by non-LTE effects or underestimated oscillator strengths for the \ion{Y}{iii} doublet. 
Experimental wavelengths and theoretical oscillator strengths for \ion{Y}{v} were provided by \cite{Reader2016_YV}. However, none of these lines are detected in either star. 

In \EC, \ion{Y}{iv} 1222.645, 1258.433, 1275.112, 1275.248\,\AA\ are detected at strengths close to 5\,m\AA\ (see Fig.\ \ref{fig:detail_lsiv_4}). 
Since these lines are also identified in \lsiv, confirming their positions, they could be used to determine an abundance despite their weakness.

\vspace{-10pt}\paragraph{Zirconium.}

\lsiv\ was first identified as a heavy-metal star from its strong optical \ion{Zr}{iv} lines \citep{Naslim2011}.
Its UV spectrum exhibits a further 15 strong \ion{Zr}{iv} lines with equivalent widths between 15 and 600\,m\AA\ (Fig.~\ref{fig:detail_lsiv_1}).

Here we use atomic data from \citet{Rauch2017_ZrXe}, based on experimental wavelengths from \cite{Acquista1980_ZrIV}. 
The strongest line is \ion{Zr}{iv} 1201.765\,\AA, followed by \ion{Zr}{iv} 1219.862\,\AA\ (in the Ly$_\alpha$ wing) and 1183.973\,\AA. 
Stark broadening is important for the 1201.765\,\AA\ line in \lsiv; however, the calculations of \citet{Majlinger2017_ZrIV_Stark} and \cite{Elabidi2021_ZrIV_Stark} do not include these transitions; we require a damping parameter of $\log \Gamma_\mathrm{Stark}$ $\approx$ $-6.2$ to match the observation. 
Additional strong lines are seen at 1546.171 and 1598.948\,\AA.
The optical \ion{Zr}{iv} lines are consistent with these UV lines. 

\lsiv\ further shows several \ion{Zr}{iii} lines, most notably at 1612.332\,\AA\ at a strength of 25\,m\AA. 
Here we use data from R.\ Kurucz's website, based on wavelengths from \cite{Reader1997_ZrIII}. 
\ion{Zr}{iii} 1356.95\,\AA\ is blended with unidentified lines, while \ion{Zr}{iii} 1593.53, 1631.28, 1638.30, and 1703.26\,\AA\ are weak. 
The only clearly identified \ion{Zr}{v} line using the \cite{Rauch2017_ZrXe} data is at 1303.933\,\AA\ with a strength of 10\,m\AA. It is slightly blended with the weaker \ion{Mo}{iv} 1303.893\,\AA.

In \EC, \ion{Zr}{iv} 1201.769, 1219.862, 1469.472, 1598.948, and 1607.948\,\AA\ are strong and isolated at strengths between 15 and 30\,m\AA\ (see Fig.\ \ref{fig:detail_lsiv_4}), allowing a reliable abundance determination.  
\ion{Zr}{iv} 1183.973\,\AA\ is also strong, but seems to be blended with an unidentified line. 

\vspace{-10pt}\paragraph{Niobium.}

The UV spectrum of \lsiv\ exhibits more than 50 \ion{Nb}{iv} lines, with predicted equivalent widths of 10 to 25\,m\AA, based on line positions and oscillator strengths from \citet{Tauheed2005_NbIV}. The strongest  lines are shown in Fig.\,\ref{fig:detail_lsiv_2}, including \ion{Nb}{iv} 1363.762, 1424.368, 1434.223, 1447.491, 1508.721, 1510.832, 1517.461, and 1532.981\,\AA. Data for \ion{Nb}{v} are available on R.\ Kurucz's website, with the strongest lines observed at 1212.15 and 1267.52\,\AA. This represents the first detection of \ion{Nb}{iv-v} lines in a stellar spectrum.

The strongest predicted niobium lines in \EC, \ion{Nb}{v} 1212.15, 1258.84\,\AA, are both blended, so we report only an upper limit, which is expected to be close to the true abundance. 

\vspace{-10pt}\paragraph{Molybdenum.}

About 200 \ion{Mo}{iv-v} lines with equivalent widths of 10 to 22\,m\AA\ are detected in \lsiv, representing the first detection of Mo in a hot subdwarf star. 
We use data from \cite{Rauch2016_Mo} for \ion{Mo}{iv-vi} and R.\ Kurucz's list for \ion{Mo}{iii}. The strongest isolated lines include \ion{Mo}{iv} 1643.095\,\AA\ and \ion{Mo}{v} 1586.883, 1661.215, 1668.652, and 1699.736\,\AA\ (see Fig.\ \ref{fig:detail_lsiv_2}). Although about ten \ion{Mo}{iii} lines are predicted to exceed 10\,m\AA, none are clearly identified, likely due to inaccurate wavelengths.

Molybdenum lines are weak in \EC, but \ion{Mo}{v} 1590.405, 1661.215, 1699.736\,\AA\ are clearly detected at strengths between 5 and 10\,m\AA. 

\vspace{-10pt}\paragraph{Technetium (undetected).} The radioactive Tc might be detectable in \lsiv\ if present in a similar quantity as Nb or Mo; however, the experimental wavelengths required for \ion{Tc}{iv-v} are not available, preventing a meaningful analysis. This situation  is analogous to that described by \cite{Werner2015_Tc} for hot WDs, where the absence of reliable line positions similarly precluded a detection.

\vspace{-10pt}\paragraph{Ruthenium, rhodium and palladium.}

Atomic data for ruthenium, rhodium and palladium are scarce; NIST lists next to no atomic data for any of these metals in states \textsc{iii} to \textsc{v}. 
However, \cite{Kurucz2018} provides extensive line lists up to \ion{Ru}{v}, \ion{Rh}{v}, and \ion{Pd}{v}, as computed with Cowan's Hartree-Fock code, which allows us to study these elements. 
In particular, their triply ionised states are predicted to produce strong lines in the 1450-1725\,\AA\ range of the \lsiv\ spectrum. 

The situation is best for \ion{Pd}{iv}, because its laboratory UV spectrum was analysed by \citet{Barakat1985_1985_PdIV_a, Barakat1985_1985_PdIV_b}, providing precise energy levels and line positions. 
Indeed, about 70 \ion{Pd}{iv} lines with strengths greater than 15\,m\AA\ are detected in the HST spectrum of \lsiv, mostly between 1470 and 1725\,\AA. 
The most prominent examples are located at 1476.057, 1484.736, 1573.958, 1660.625, 1667.397, 1682.904, 1683.808, 1699.596, and 1704.443 \AA\ (see Fig.\ \ref{fig:detail_lsiv_2}). 
This represents the first detection of any palladium lines in a hot star. According to \citet{Kurucz2018}, no \ion{Pd}{v} lines are expected within the E140M wavelength range. 

Because no experimental energy levels are available for \ion{Ru}{iv-v} and \ion{Rh}{iv-v}, their predicted line positions are uncertain by roughly 2-20\,\AA, corresponding to typical theoretical energy-level uncertainties of 100-1000\,cm$^{-1}$. Several unidentified features in the STIS spectrum of \lsiv\ are likely due to \ion{Ru}{iv} in particular, but secure identifications require more accurate wavelength measurements.
No Ru, Rh, or Pd lines are detected in \EC, although \ion{Pd}{iv} 1484.736\,\AA\ is close to the detection threshold. 

\vspace{-10pt}\paragraph{Silver.}

The spectrum of \ion{Ag}{iv} has been studied in detail by \cite{Ankita2020_AgIV}, who provide both experimental wavelengths and theoretical oscillator strengths. 
\ion{Ag}{iv} 1485.872\,\AA\ is consistent with a fairly isolated line in \lsiv\ (see Fig.\ \ref{fig:detail_lsiv_ag}); predicted lines at \ion{Ag}{iv} 1515.100, 1516.012, 1558.149\,\AA\ are consistent with the observation, but blended with stronger lines. 

Experimental wavelengths in the UV range for \ion{Ag}{v} are also available \citep{Kildiyarova1995_AgV}, but the lack of oscillator strengths prevents clear identifications. 
No silver lines are detected in \EC. 

\vspace{-10pt}\paragraph{Cadmium.}

The UV spectrum of \lsiv\ shows dozens of strong \ion{Cd}{iv} lines, identified using the experimental wavelengths of \citet{Joshi1980_CdIV}. Notable examples include \ion{Cd}{iv} 1167.277, 1294.078, 1337.299, 1418.092, 1418.882, 1482.940, and 1513.883\,\AA\ (Fig.\ \ref{fig:asiv_seiv_rbiv}), with equivalent widths up to $\sim$70\,m\AA. Many \ion{Cd}{iv} lines arising from the $^{2}D$ ground term exhibit the expected fine-structure splitting into $^{2}D_{5/2}$ and $^{2}D_{3/2}$, including the lines at 1492.965, 1501.517, and 1513.883\,\AA, several of which are resolved in our HST spectrum.

Several features may also be attributed to \ion{Cd}{v}, for example at 1419.966 and 1502.650\,\AA, using experimental wavelengths from \citet{vanKleef1982_CdV}. However, \ion{Cd}{iv} is expected to dominate in the atmosphere of \lsiv, and lines identified as blends between \ion{Cd}{iv} and \ion{Cd}{v} (e.g.\ 1337.30, 1492.97, 1513.88\,\AA) are likely primarily due to \ion{Cd}{iv}. Unfortunately, no oscillator strengths are available for either ion. %

\ion{Cd}{iv} lines are much weaker in \EC\ than in \lsiv, to the point that clear identification is not possible without oscillator strengths. \ion{Cd}{iv} 1337.30, 1482.94\,\AA\ match otherwise unidentified lines at strengths of 5 to 10\,m\AA. 

\vspace{-10pt}\paragraph{Indium.}

Detailed experimental studies of \ion{In}{iii-v} and corresponding Cowan-code oscillator strengths were presented by \cite{Varshney2017_InIII, Varshney2013_InIV, Varshney2016_InV}. Using these data, we identify 25 \ion{In}{iv} lines in the UV spectrum of \lsiv, each with an estimated equivalent width above 10\,m\AA. Prominent, isolated \ion{In}{iv} lines include 1300.588, 1320.040, 1333.879, 1344.579, 1406.092, and 1446.000\,\AA. The strong \ion{In}{iv} 1381.786\,\AA\ feature is likely blended with \ion{As}{iv} 1381.76\,\AA, a line identified by \cite{Rao1931_SeIV_SeV} but lacking an oscillator strength. The strongest indium features are shown in Fig.~\ref{fig:detail_lsiv_2}.

Five \ion{In}{v} lines were detected in the DO white dwarf \RE\ by \cite{Rauch2020_Cu_In}. In \lsiv, these lines are predicted to have strengths of order 10\,m\,\AA\ but are blended with stronger features and therefore remain undetected. 

In addition, the \ion{In}{iii} 1625.3\,\AA\ resonance line ($5s_{1/2}$ -- $5p_{3/2}$) is detected in \lsiv. 
Given that this transition involves the $5s$ ground state and that the dominant isotope $^{115}$In has nuclear spin $I$ = 9/2, significant hyperfine splitting is expected (see Appendix\ \ref{sect:HFS}). 
Magnetic dipole hyperfine constants computed by \citet{Roy2014_InIII_HFS} using a relativistic coupled-cluster approach yield $A(5s_{1/2}) = 21.4$\,GHz and $A(5p_{3/2}) = 0.7$\,GHz. 
Using these values, we identify the \ion{In}{iii} 1625.3\,\AA\ feature in our spectrum of \lsiv\ as a resolved hyperfine doublet; the HFS of the upper level is negligible. 

\ion{In}{iv-v} lines are weaker in \EC, and only an upper abundance limit can be derived, based on the non-detection of \ion{In}{iv} 1320.04, 1344.58, 1406.09, and 1509.82\,\AA.

\vspace{-10pt}\paragraph{Tin.}

The optical spectrum of \lsiv\ shows strong \ion{Sn}{iv} lines at 3862.1 and 4217.2\,\AA\ \citep{Dorsch2020}, but the tin lines in its UV spectrum are even stronger. 
Using atomic data from \cite{Haris2012_SnIII}, we identified 14 \ion{Sn}{iii} lines in \lsiv\ with equivalent widths between 10 and 30\,m\AA. The strongest of these is the \ion{Sn}{iii}\,1251.383\,\AA\ resonance line, the only \ion{Sn}{iii} line detected in both stars. 
For this line and \ion{Sn}{iii}\,1347.637\,\AA, the observed spectrum is better reproduced using the lower oscillator strengths from \cite{Colon2010_SnIII}, which we adopt here. 

The strongest tin lines in both stars are the \ion{Sn}{iv} resonance lines at 1314.530 and 1437.527\,\AA, for which we use atomic data by \citet{Kaur2020_SnIV} (see Figs.\ \ref{fig:detail_lsiv_2} and \ref{fig:detail_Pb}). In \lsiv\ they reach equivalent widths of about 720\,m\AA\ and 520\,m\AA, respectively, and are sufficiently saturated that their strengths are strongly influenced by Stark broadening, which we model using theoretical widths from \citet{deAndresGarcia2016_SnIV_Stark}. 
The inclusion of Stark broadening further leads to more consistent abundances obtained from \ion{Sn}{iii} and \ion{Sn}{iv} lines. 

A \ion{Sn}{v} line at 1160.755\,\AA\ was previously identified in \RE\ by \cite{Werner2012_KrXe}. This star also exhibits additional \ion{Sn}{v} lines at 1205.727, 1254.116, and 1355.627\,\AA, with measured wavelengths and identifications by \cite{Wu_1967_Sn}. 
These lines are also detected in \lsiv, but unfortunately, no oscillator strengths are available. 

Similar to the optical \ion{Ge}{iv} lines, \ion{Sn}{iv} 3861.207, 4216.192\,\AA\ are too weak in our model at the best-fit abundance obtained from UV \ion{Sn}{ii-iv} lines. 
This explains our lower tin abundance compared to \cite{Dorsch2020}. 

\vspace{-10pt}\paragraph{Antimony.}

The dominant ionisation stages of antimony in \lsiv\ and \EC\ are \ion{Sb}{iv} and \ion{Sb}{v}; none of the \ion{Sb}{iii} lines listed by \cite{Andersen1977_AsIII_SeIV_SbIII_TeIV} and \cite{Tauheed1992_TeIV_IV_intercom} are detected. 
We use the oscillator strengths for \ion{Sb}{iv} from \cite{Joensson2012_SbIV}; however, since their line positions are often  inaccurate by several \AA, we adopt the wavelengths measured by \cite{Rana2001_SbIV} using normal incidence spectroscopy. 
We clearly detect the strong \ion{Sb}{iv} 1171.436, 1192.934, 1199.150\,\AA\ lines in \lsiv\ (see Fig.\ \ref{fig:detail_lsiv_2}), which is the first detection of \ion{Sb}{iv} in any star. 

The strongest antimony line observed in both \lsiv\ and \EC\ is the \ion{Sb}{v} 1226\,\AA\ resonance transition (\termconfig[]{4d^{10}}{5s}{2}{S}{1/2} -- \termconfig[o]{4d^{10}}{5p}{2}{P}{1/2}), which was previously used by \citet{Werner2018_BrSb} to determine the antimony abundance in \RE. We adopt the experimental oscillator strength from \citet{Pinnington1985b}, as reported by \citet{morton00}. In our spectra of \lsiv\ and \EC, the feature appears unusually broad and is best reproduced as a split line; the doublet is clearly resolved in \HD\ at 1225.941 and 1226.002\,\AA\ (see Fig.\ \ref{fig:detail_sb_hfs}). The splitting is likely due to hyperfine structure, with additional broadening arising from the presence of two natural isotopes of antimony with different nuclear spins ($I_{121}=5/2$, $I_{123}=7/2$); we assume a solar isotopic ratio.  
We implemented HFS splits as described in Appendix\ \ref{sect:HFS}; a satisfactory fit to the observed profile is achieved with magnetic dipole hyperfine constants of $A \approx 35$\,GHz for the $^{121}$Sb ground state (27\,GHz for $^{123}$Sb) and $A \approx 9$\,GHz (7\,GHz for $^{123}$Sb) for the upper level; only the doublet splitting caused by the lower level is clearly resolved. While this is the first observational analysis of this hyperfine structure, the derived splitting is consistent with the Dirac-Fock prediction of \citet{Beck1991_Sb_Ahf} for the $^{121}$Sb ground state. 
The \ion{Sb}{iv} 1499.3\,\AA\ resonance line also likely exhibits hyperfine (triplet) splitting, but it is weak in \EC\ and blended with \ion{Ni}{iv} and \ion{Zn}{iii} in \lsiv. 

\vspace{-10pt}\paragraph{Tellurium.}

\cite{Marcinek1994_TeIV_IV} computed oscillator strengths for several \ion{Te}{iv} multiplet transitions, which are still the only data available for the HST and UVES spectral ranges. 
We converted their configuration interaction oscillator strengths to individual fine structure transitions using Wigner 6j symbols. %
Our line positions are based on experimental energy levels from \cite{Crooker1964_TeIII_to_VI}. %
This allowed us to clearly detect the \ion{Te}{iv}\,1168.391, 1196.938\,\AA\ \termconfig[\circ]{5s^2}{5p}{2}{P}{} -- \termconfig[\circ]{5s}{5p^2}{2}{D}{} resonance doublet lines in \lsiv, at strengths close to 20\,m\AA\ (see Fig.\ \ref{fig:detail_lsiv_2}). 
The lowest-lying transitions in \ion{Te}{iv} are the intercombination transitions \termconfig[\circ]{5s^2}{5p}{2}{P}{} -- \termconfig[\circ]{5s}{5p^2}{4}{P}{}, the positions of which were measured by \cite{Tauheed1992_TeIV_IV_intercom}. 
The strongest of these transitions is clearly observed \lsiv\ at 1448.043\,\AA, while 1355.554, 1439.532\,\AA\ are blended. 
Unfortunately, no oscillator strengths are available for these transitions. 
\ion{Te}{iv} 1168.391\,\AA\ is clearly detected in \EC, but \ion{Te}{iv} 1196.938\,\AA\ is not. 

We adopt atomic data for \ion{Te}{v} from \citet{Ekman2013_TeV} and experimental wavelengths from \citet{Tauheed2000_TeV}.
The \ion{Te}{v} 1281.634\,\AA\ resonance line is clearly detected in both \lsiv\ and \EC. %
Other strong transitions, such as \ion{Te}{v} 1236.259 and 1406.549\,\AA, are blended with stronger lines in both stars.
The \ion{Te}{vi} 1313.874\,\AA\ feature, identified in \RE\ by \citet{Rauch2017_SeSrTeI}, is not present in \lsiv\ owing to its lower \teff.

\vspace{-10pt}\paragraph{Iodine.}

Iodine was detected in \RE\ by \citet{Rauch2017_SeSrTeI}, but the \ion{I}{vi} lines used in that study are not excited in our stars. No oscillator strengths are available for \ion{I}{iv}, and all lines in the HST range listed by \citet{Tauheed1991_IIV} are intercombination transitions that are not detected in \lsiv.

The situation is more favourable for \ion{I}{v}. The lowest-lying transitions in this ion are the intercombination lines \termconfig[\circ]{5s^2}{5p}{2}{P}{} -- \termconfig[\circ]{5s}{5p^2}{4}{P}{}, which are covered by the HST range. Lifetimes for these transitions were measured by \citet{Ansbacher1991_IV}, and their wavelengths refined by \citet{Tauheed1992_TeIV_IV_intercom}. Although partially blended with unidentified features, \ion{I}{v} 1234.516\,\AA\ is detected in \lsiv, while \ion{I}{v} 1244.831\,\AA\ is blended with the stronger \ion{Sr}{iv} 1244.763 and 1244.888\,\AA\ lines. The inclusion of \ion{I}{v} 1150.760\,\AA\ also improves the spectral fit for \lsiv, though this feature is blended with unidentified lines (see Fig.\ \ref{fig:detail_lsiv_2}). Line positions between the 5$s^2$5$d$, 5$p^3$, 5$s$5$p$5$d$, and 5$s$5$p$6$s$ configurations of \ion{I}{v} were measured by \citet{Tauheed1998_IV}, but unambiguous identification of these transitions is not possible without oscillator strengths. 

No iodine lines are detected in \EC, but a rough upper limit could be estimated from \ion{I}{v} 1244.831\,\AA, which is also blended with \ion{Sr}{iv} in this star. 

\vspace{-10pt}\paragraph{Xenon.}

Singly ionised xenon was already detected in the chemically peculiar hot BHB star Feige\,86 \citep{Bonifacio1995_F86,Castelli1997_CP,Nemeth2017_F86_XeII}, where the enrichment is likely caused by diffusion \citep[e.g.][]{Michaud2008}. 

Strong \ion{Xe}{vi} and \ion{Xe}{vii} lines were observed in the hot white dwarf \RE\ by \citet{Rauch2017_ZrXe}, who also provided oscillator strengths for \ion{Xe}{iv} and \ion{Xe}{v}, which we use here. Xenon lines are weak in the HST spectrum of \lsiv, but several are identifiable, including \ion{Xe}{iv} 1157.450\,\AA\ and \ion{Xe}{v} 1151.078, 1225.097, and 1281.13\,\AA, at equivalent widths between 10 and 15\,m\AA\ (see Fig.\ \ref{fig:detail_lsiv_2}). 
We can thus determine a xenon abundance for \lsiv, although the weakness of the lines results in increased uncertainty.  
No xenon lines are detected in \EC, although \ion{Xe}{v} 1281.13\,\AA\ is isolated and close to the detection threshold. 

\vspace{-10pt}\paragraph{Caesium.}

Strong caesium \textsc{iv-vi} lines were recently detected in FUSE spectrum of the cool (\teff\ = 49500\,K) DO white dwarf HD\,149499B by \cite{Chayer2023_Cs}. 
However, none of the lines reported in that study are detectable in \lsiv\ or \EC; most predicted \ion{Cs}{iv-vi} transitions lie below 1150\,\AA, outside the wavelength range covered by HST. The only \ion{Cs}{iv} line with an experimental wavelength within the HST range is 1282.659\,\AA\ \citep{Reader1983_CsIV_BaV}. This line lies close to a sharp, otherwise unidentified feature at 1282.606\,\AA\ in \lsiv, but in the absence of additional supporting lines, this identification remains uncertain.

\vspace{-10pt}\paragraph{Barium.}

The strongest barium line in \lsiv, \ion{Ba}{iv} 1503.576\,\AA, is blended with two nickel lines, while the weak \ion{Ba}{iv} lines in \EC\ are blended with unidentified features. As a result, we provide only an upper limit for the barium abundance in both stars, based on the atomic data from \citet{Sharma2014_BaIV}.
We also searched for \ion{Ba}{v} lines as predicted by \cite{Rauch2014_Ba}, but none were detected. 

\subsection{Lanthanides}

\begin{figure}
  \centering
  \includegraphics[width=0.99\columnwidth]{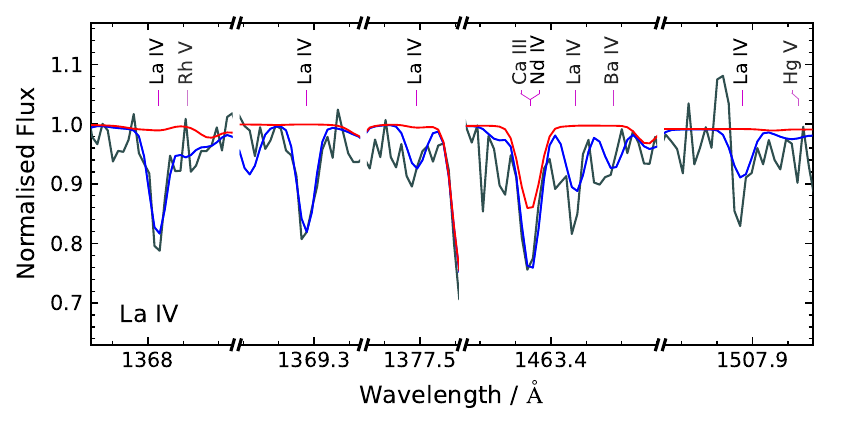}
  \vspace{-5pt}
  \caption{The strongest \ion{La}{iv} lines in \EC, like Fig.\ \ref{fig:detail_lsiv_1}. }
  \label{fig:detail_ec22_La}
\end{figure}

\begin{figure*}
\centering
\includegraphics[width=0.99\textwidth]{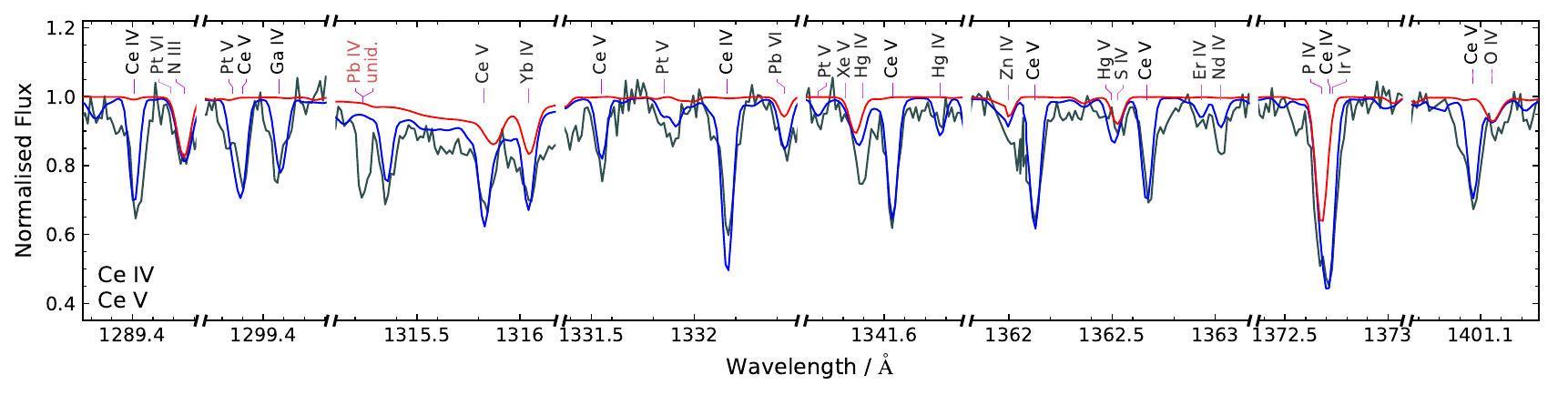}\\[-18.3pt]
\includegraphics[width=0.99\textwidth]{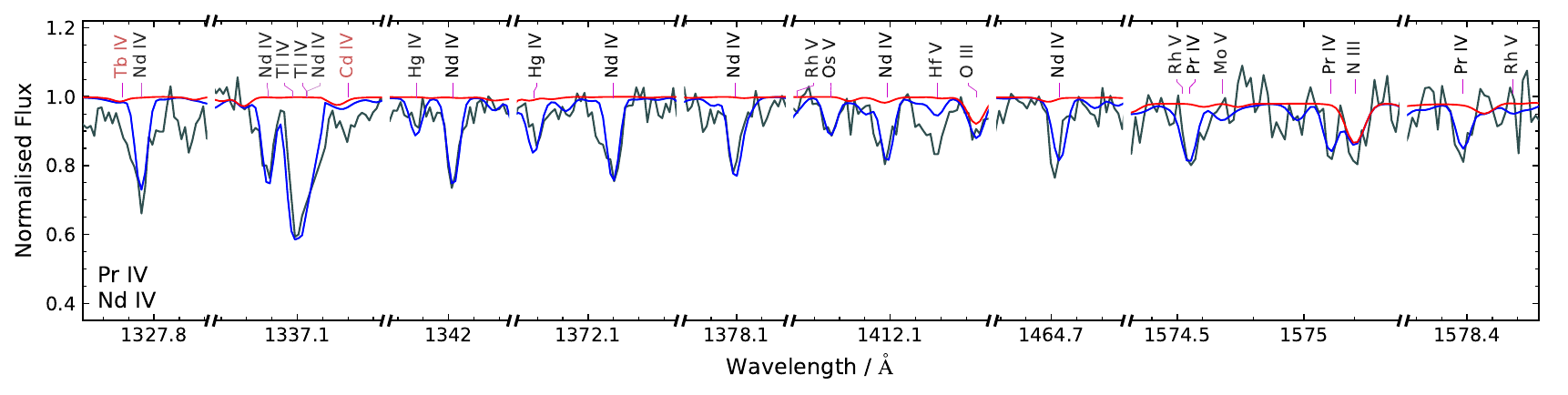}\\[-18.3pt]
\includegraphics[width=0.99\textwidth]{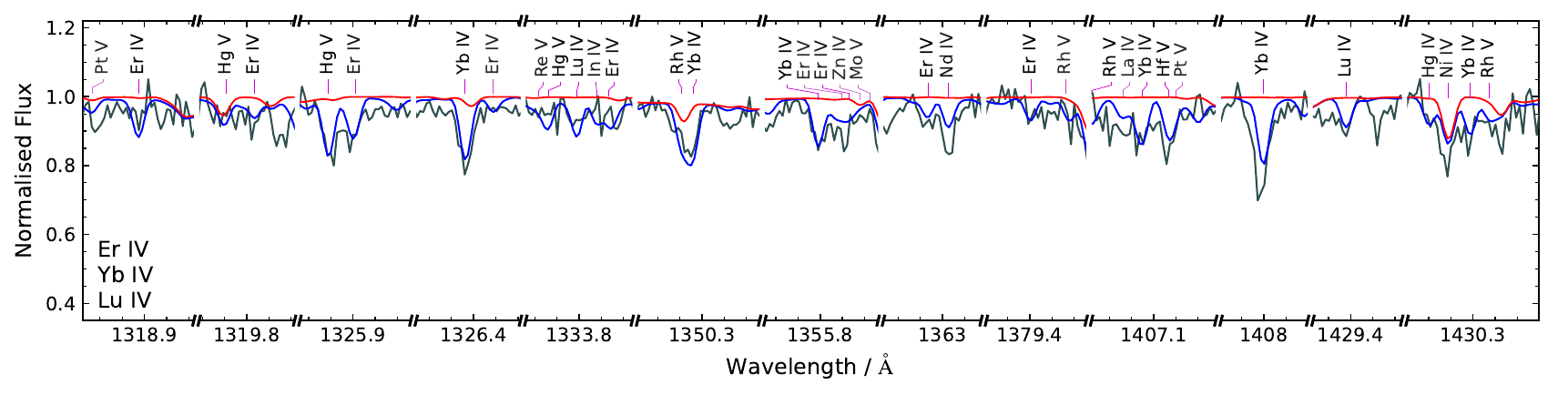}\\[-18.3pt]
\includegraphics[width=0.99\textwidth]{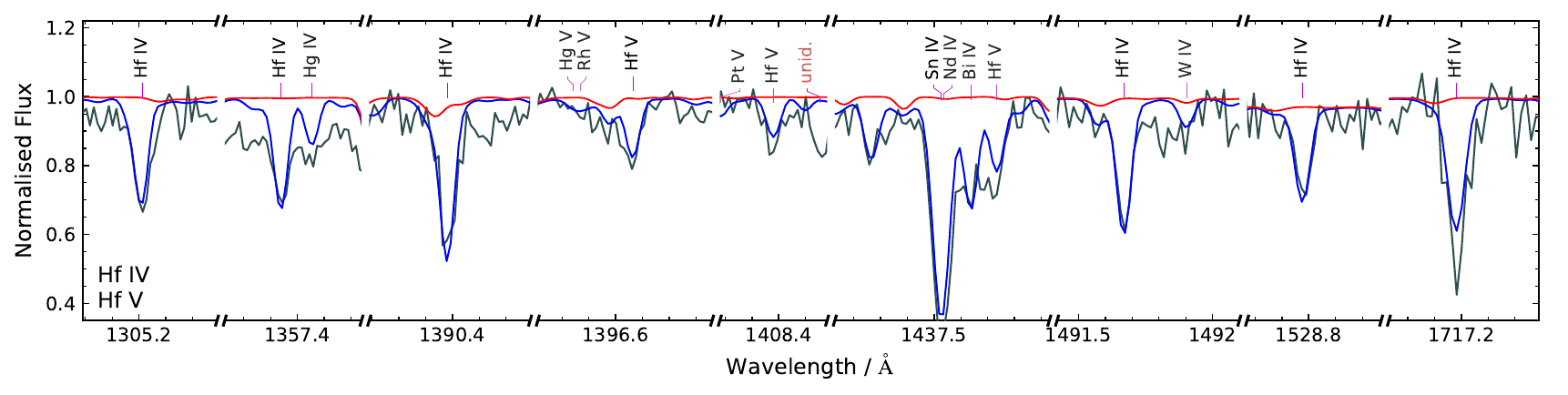}\\[-18.3pt]
\includegraphics[width=0.99\textwidth]{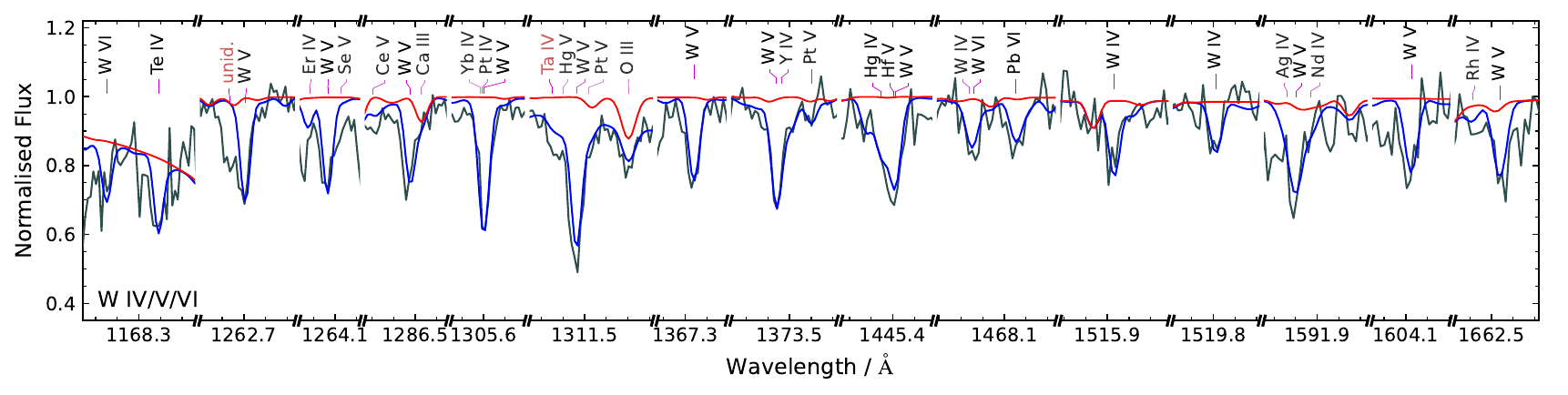}\\[-18.3pt]
\includegraphics[width=0.99\textwidth]{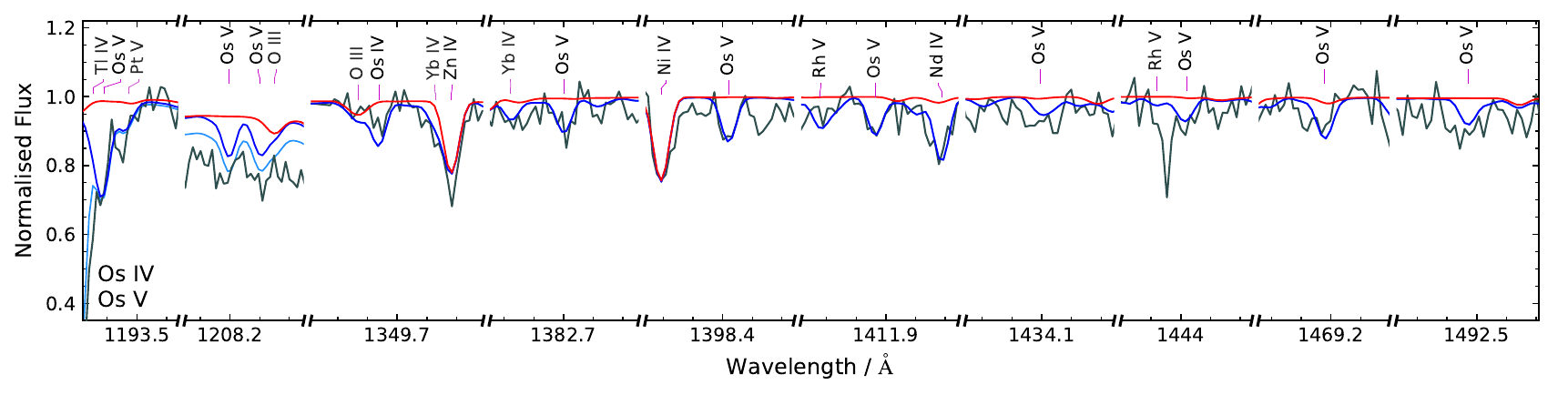}%
\vspace{-5pt}
\caption{Ce ($Z$\,=\,58), Pr (59), Nd (60), Er (68), Yb (70), Lu (71), Hf (72), W (74), and Os (76) lines in \EC, like Fig.\ \ref{fig:detail_lsiv_1}.
}
\label{fig:detail_EC_0}
\end{figure*}

Starting from praseodymium, triply ionised lanthanides have partially filled 4$f$ subshells, which give rise to numerous low-lying energy levels. In particular, for elements such as neodymium and erbium, the 4$f$ -- 5$p$ transition arrays fall within the STIS/E140M spectral range \citep{Cowan1973_Lanth}. We report the first detection of spectral lines from triply ionised lanthanides in any astronomical object. These lines are clearly identifiable in \EC\ and, to a lesser extent, in \lsiv. Lanthanides have not previously been observed in hot subdwarf stars. The strongest detected lines are shown in Figs.\ \ref{fig:detail_ec22_La} and \ref{fig:detail_EC_0}. 

\vspace{-10pt}\paragraph{Lanthanum.}

Theoretical oscillator strengths for \ion{La}{iv} were computed by \citet{Karacoban2020_LaIV,Karacoban2020_LaIVs}, yielding reasonably accurate wavelengths despite relying on the older energy levels from \citet{Martin1978} rather than the more precise data of \citet{Epstein1979_LaIV}. After applying small wavelength corrections, several \ion{La}{iv} lines were identified in \EC, the strongest at 1368.029 and 1369.278\,\AA\ with equivalent widths of about 15\,m\AA\ (Fig.\ \ref{fig:detail_ec22_La}). No \ion{La}{v} lines are expected, as its level structure lacks low-lying transitions within the STIS/E140M wavelength range \citep{Epstein1976_LaV}. 

\ion{La}{iv} lines are weaker in \lsiv\ and  are blended with stronger lines due the higher iron group abundances in \lsiv. 
As a result, we could only determine an upper limit, mostly based on the non-detection of \ion{La}{iv} 1368.029\,\AA. 

\vspace{-10pt}\paragraph{Cerium.}

Cerium lines are prominent in \EC, with approximately 20 \ion{Ce}{iv} and \ion{Ce}{v} lines detected at strengths between 10 and 25\,m\AA. For \ion{Ce}{iv}, we use atomic data from \cite{CarvajalGallego2021_CeIV} and adopt experimental wavelengths from \cite{Reader2009_CeIV}. The strongest \ion{Ce}{iv} lines are at 1289.41, 1332.16, and 1372.72\,\AA.
We use data from \cite{Wajid2021_CeV} for \ion{Ce}{v}. The most prominent \ion{Ce}{v} lines are found at 1315.826, 1341.640, 1362.125, 1362.668, and 1401.064\,\AA, all with strengths close to 20\,m\AA. 

In \lsiv, the \ion{Ce}{iv} 1332.16\,\AA\ line is strong and relatively unblended. Although no other isolated cerium features are present, this line matches the corresponding feature observed in \EC. We therefore use it to estimate the cerium abundance in \lsiv, which is consistent with the slightly blended \ion{Ce}{iv} 1372.72\,\AA. 

\vspace{-10pt}\paragraph{Praseodymium.}

We used data from \cite{EnzongaYoca2013_PrIV} to search for \ion{Pr}{iv}; indeed, the \ion{Pr}{iv} resonance triplet at 1574.827, 1578.384, 1578.384\,\AA\ is detected in \EC\ at EWs between 10 and 16\,m\AA. 
Atomic data for highly ionised Pr was provided by \cite{CarvajalGallego2023_PrV_NdV_PmV}, but the only two experimentally confirmed lines in the STIS/E140M range \citep[1234.070, 1342.775\,\AA;][]{Kaufman1967_PrV} originate from high-lying levels and are very weak. 
No praseodymium lines are detected in \lsiv. %

\vspace{-10pt}\paragraph{Neodymium.}

After cerium, the strongest lanthanide lines in the STIS spectrum of \EC\ originate from \ion{Nd}{iv}.
We use atomic data from \cite{EnzongaYoca2014_NdIV}, based on experimental wavelengths from \cite{Wyart2007_NdIV}. 
The strongest lines, \ion{Nd}{iv} 1327.750, 1336.985, 1342.019, 1344.737, 1372.199, and 1378.094\,\AA, are 4$f^3$ -- 4$f^2$5d transitions. 
As shown by \cite{Meftah2008_NdV}, the level structure of \ion{Nd}{v} does not produce strong transitions in the STIS wavelength range.
Their energy levels are thus only used to compute the \ion{Nd}{v} partition function. 

Almost all \ion{Nd}{iv} lines in \lsiv\ are blended with stronger features from other elements; the only relatively unblended line is \ion{Nd}{iv} 1344.737\,\AA. We therefore derive only an upper abundance limit, which is likely close to the true value.

\vspace{-10pt}\paragraph{Promethium to holmium (undetected).} 

The available atomic data for ions from promethium (Pm; $Z$\,=\,61) to holmium (Ho; $Z$\,=\,67) are insufficient for reliable spectral line identification. 
For promethium (Pm), samarium (Sm), and europium (Eu) in ionisation stage \textsc{iv}, theoretical energy levels have been calculated by \citet{Dzuba2003_NdPmSmEu_IV_tau}.
However, these calculations are not accurate enough for definitive line identification and none of the predicted \ion{Sm}{iv} levels exhibit LS- or intermediate-coupling allowed dipole transitions.
Despite these limitations, the calculations predict twelve 4$f$ $^5$I -- $5d$ ${^5}$K transitions for \ion{Pm}{iv} in the 1300 to 1450\,\AA\ range and three 4$f$ $^5\mathrm{L}_6$ -- 5$d$ ${^7}\mathrm{K}_{5,6,7}$ transitions for \ion{Eu}{iv} between 1550 and 1750\,\AA. 
This suggests that these lines could be detectable in \EC, given the expected strong enrichment\footnote{Besides Pm, which is radioactive and has only short-lived isotopes. }. 
It is the lack of experimental UV wavelength measurements that currently limits our ability to confirm these predictions\footnote{An exception might be the unpublished dissertation by \citet{Elias1972_Lanth_UV}, which is not accessible to us.}. 
Although accurate line positions are available for gadolinium (Gd) \textsc{iv} \citep{Kielkopf1970_GdIV} and terbium (Tb) \textsc{iv} \citep{Spector1976_TbIV}, their line strengths are likely near the detection threshold in \EC, and without oscillator strengths, they cannot be confidently identified. 
For triply and quadruply ionised dysprosium (Dy) and holmium (Ho), no accurate energy level data are available.  %

\vspace{-10pt}\paragraph{Erbium and thulium. }

\cite{Chikh2021_ErIV} provide both experimental wavelengths and theoretical oscillator strengths for \ion{Er}{iv}. We used their data to search for \ion{Er}{iv} in \EC. 
While these lines are weak, we detect \ion{Er}{iv} 1325.911\,\AA, which is only slightly affected by a blend with two weaker \ion{Nd}{iv} lines. 
Additional \ion{Er}{iv} lines at 1355.931, 1362.934, 1379.405\,\AA\ are marginally detected at EWs close to 5\,m\AA. 

We used \ion{Tm}{iv} data from \citet{EnzongaYoca2017_TmIV}, based on experimental wavelengths from \citet{Meftah2007_TmIV}. The non-detection of \ion{Tm}{iv} 1183.631, 1407.304\,\AA\ yields an upper limit on the thulium abundance in \EC. 

\vspace{-10pt}\paragraph{Ytterbium.}

Several strong \ion{Yb}{iv} lines are detected in \EC, using atomic data from \citet{Wyart2001_YbIV}. Clear detections include transitions at 1316.040, 1326.361, 1350.259, 1355.791, 1407.050, 1407.998, and 1430.288\,\AA\ (Fig.\ \ref{fig:detail_EC_0}). 
Experimental wavelengths and energy levels for \ion{Yb}{v} were provided by \citet{Meftah2013_YbV}. 
Although oscillator strengths are also reported in that work, our HST spectra only cover the high-lying 4$f^{11}$6$s$ -- 4$f^{11}$6$p$ transition array, which is not detectable. 
There are no clearly detected ytterbium lines in \lsiv. The upper limit is mostly based on \ion{Yb}{iv} 1355.791, 1407.998\,\AA. 

\vspace{-10pt}\paragraph{Lutetium.} 

\ion{Lu}{iv} lines are detected at 1333.790 and 1429.082\,\AA\ in \EC, using atomic data from \citet{Motoumba2020_LuIV_HfV_TaVI}, which are based on the experimental line positions of \citet{Sugar1972_LuIV}. The latter line requires a higher abundance to be reproduced, possibly due to an uncertain continuum location. Owing to this inconsistency and the absence of additional lines, we assign a large uncertainty to the derived lutetium abundance in \EC. Oscillator strengths for \ion{Lu}{v} were provided by \citet{Maison2022Atoms_LuV}, and experimental wavelengths by \citet{Kaufman1978_LuV}; however, no \ion{Lu}{v} lines are detected in either star.

\subsection{Heavy trans-iron elements}

Given the detection of a high lanthanide abundance in \EC, in addition to its known extreme enrichment in lead, it is natural to search for other heavy trans-iron elements. This section examines elements from hafnium (Hf, $Z=72$) to bismuth (Bi, $Z=83$). \lsiv\ is discussed only where reliable measurements were possible; all elements were analysed in analogy to \EC. 

\vspace{-10pt}\paragraph{Hafnium.}

Experimental wavelengths and energy levels for \ion{Hf}{iv-v} were reported by \citet{Sugar1974_HfIV}. Many of these lines are prominent in \EC\ (Fig.\ \ref{fig:detail_EC_0}), at equivalent widths of up to 30\,m\AA. Because oscillator strengths were unavailable, we computed new radiative data for \ion{Hf}{iv}, specifically for the analysis of \EC; this work is described in Sect.\ \ref{sect:umons}. The strongest identified \ion{Hf}{iv} transitions occur at 1305.241, 1357.399, 1390.390, 1491.670, 1528.820, and 1717.210\,\AA. 
Atomic data for \ion{Hf}{v} were provided by \cite{Motoumba2020_LuIV_HfV_TaVI}. Although somewhat weaker than \ion{Hf}{iv}, prominent \ion{Hf}{v} lines are observed in \EC\ at 1232.030, 1396.664, 1407.169, 1408.381, 1413.511, 1433.434, 1437.267, 1437.734, and 1445.403\,\AA. 
The latter line is blended with \ion{Hg}{iv} 1445.347\,\AA\ and \ion{W}{v} 1445.413\,\AA. 

Two hafnium lines are detected in \lsiv, \ion{Hf}{iv} 1491.669, 1717.181\,\AA\ (see Fig.\ \ref{fig:detail_lsiv_4}), consistent with the blended \ion{Hf}{iv} 1357.339\,\AA\ line. All other hafnium lines are strongly blended or too weak to be detected. 

\begin{figure}
  \centering
  \includegraphics[width=0.99\columnwidth]{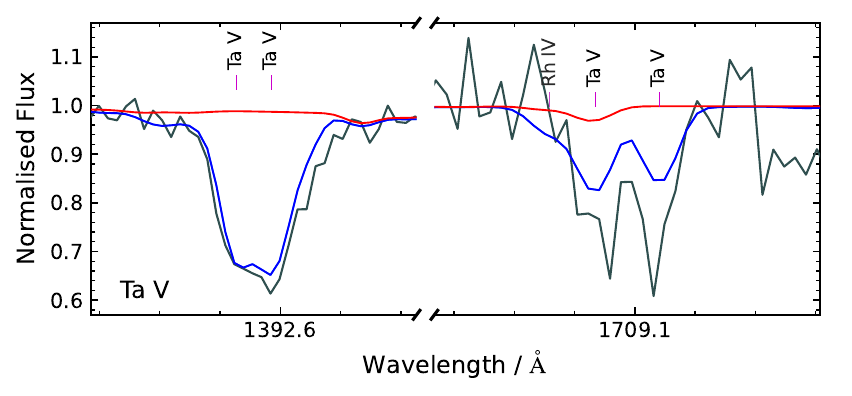}
  \vspace{-5pt}
  \caption{HFS-split \ion{Ta}{v} lines in \EC, similar to Fig.\ \ref{fig:detail_lsiv_1}. }
  \label{fig:detail_ec22_TaV}
\end{figure}

\vspace{-10pt}\paragraph{Tantalum.}

\citet{Kildiyarova1996_TaIV_WV_EUV} and \citet{Churilov1996_TaIV_WV_EUV} carried out detailed experimental studies of the 5$d^2$ -- 5$d$5$f$ and 5$d^2$ -- 5$d$\,$7p$ transition arrays of \ion{Ta}{iv}, respectively, providing accurate energy levels. While these transitions lie shortward of the HST wavelength range, the remaining low-lying arrays were analysed by \citet{Meijer1978_TaIV}. Their experimental wavelengths were used to search for \ion{Ta}{iv} features in \EC, but even the 5$d^2$ -- 5$d$6$p$ lines seem to be weak and are not clearly detected without oscillator strengths. 
Possible detections include \ion{Ta}{iv} 1211.94, 1223.73, 1238.12, 1311.35, 1607.70, and 1631.65\,\AA\ at strengths close to 5\,m\AA.  

\ion{Ta}{v} is the dominant ionisation stage in the photosphere of \EC. 
The wavelengths of \ion{Ta}{v} lines, including the strong 1393 and 1709\,\AA\ doublet from the ground 6s$^2$\,S$_{1/2}$ level, were measured by \citet{Meijer1973_TaV}, and oscillator strengths for these transitions were later calculated by \citet{Migdalek2014_LuIII_HfIV_TaV_WVI}. Hyperfine splitting of both lines was detected by \citet{Sugar1975_TaV_hfs} using sliding-spark spectroscopy, who also reported the corresponding split components. We adopted their data to account for this effect, which is clearly visible in the STIS spectrum of \EC\ (Fig.\ \ref{fig:detail_ec22_TaV}), confirming the identification of the \ion{Ta}{v} lines.
\ion{Ta}{v} 1392.527, 1392.585\,\AA\ are also detected in \lsiv, representing the only clearly identifiable tantalum features in that star. 
While oscillator strengths for \ion{Ta}{vi} were provided by \cite{Motoumba2020_LuIV_HfV_TaVI}, these lines are too weak to be observable in either star. 

\vspace{-10pt}\paragraph{Tungsten.}

Oscillator strengths for \ion{W}{iv-vi} were provided by \cite{EnzongaYoca2012_WIV, EnzongaYoca2012_WV, EnzongaYoca2012_WVI}, using experimental wavelengths. %
The dominant tungsten lines in \EC\ originate from \ion{W}{v}, in particular at 1262.707, 1264.066, 1286.456, 1305.601, 1311.462, 1367.341, 1373.439, 1445.413 (blend with \ion{Hf}{v}), 1591.804, 1604.126, and 1662.540\,\AA. 
\ion{W}{iv} 1515.932, 1519.813\,\AA\ and the \ion{W}{vi} 1168.151, 1467.958\,\AA\ doublet are also observed; the strongest lines are shown in Fig.\ \ref{fig:detail_EC_0}. 
To our knowledge, this is the first detection of \ion{W}{iv-vi} lines in any star. 
Tungsten lines are weaker in \lsiv, and the only clearly identified line is \ion{W}{v} 1305.611\,\AA. 
\ion{W}{vi} 1168.151\,\AA\ and \ion{W}{v} 1262.707\,\AA\ are close to the detection threshold, while all other tungsten lines are either too weak or blended with stronger lines. 

\ion{W}{v} 1305.601, 1311.462\,\AA\ are detected and clearly resolved in the STIS/E140H spectrum of \HD, at strengths close to 10\,m\AA. This means that tungsten can be detected at abundances close to 1000 times solar even at \teff\ $\approx$ 40\,000\,K if high-quality HST spectra are used. 

\vspace{-10pt}\paragraph{Rhenium.}

Extensive line lists using experimental wavelengths and theoretical oscillator strengths for \ion{Re}{iv-v} were provided by \cite{Azarov2018_ReIV,Azarov2018_ReV}, and are listed on NIST.
No rhenium lines are detected in \lsiv. 
The strongest predicted rhenium lines in \EC\ are \ion{Re}{v} 1220.879, 1244.996, and 1250.176\,\AA.
The 1220.879\,\AA\ line is blended with unidentified lines, while the 1244.996 and 1250.176\,\AA\ lines lie near the detection threshold. 
An upper limit is therefore adopted for \EC, which is likely close to the true abundance. 

\vspace{-10pt}\paragraph{Osmium.}

Experimental wavelengths and theoretical oscillator strengths for \ion{Os}{iv} and \ion{Os}{v} lines were provided by \cite{Ryabtsev1998_OsIV} and \cite{Azarov1997_OsV}, respectively. 
As shown in Fig.~\ref{fig:detail_EC_0}, the strongest osmium feature in the HST spectrum of \EC\ is \ion{Os}{v} 1193.388\,\AA. We also identify \ion{Os}{v} 1382.703, 1398.421, 1411.865, 1469.177, and 1492.470\,\AA, all arising from $5d^3\,6s$ -- $5d^3\,6p$ transitions and each with equivalent widths of roughly 8\,m\AA.
The \ion{Os}{iv} lines at 1266.236 and 1349.639\,\AA\ appear somewhat too strong at the abundances inferred from the \ion{Os}{v} lines, possibly indicating non-LTE effects in \ion{Os}{iv}. 
We detect no osmium lines in \lsiv. 

\vspace{-10pt}\paragraph{Iridium.}

It is very likely that both \ion{Ir}{iv} and \ion{Ir}{v} are present in the UV spectrum of \EC. 
We used line positions and oscillator strengths by \cite{Azarov2016_IrIV} to search for \ion{Ir}{iv}. 
\ion{Ir}{iv} 1502.57\,\AA\ matches a strong line in \EC, but the high abundance it implies is inconsistent with \ion{Ir}{iv} 1227.20, 1457.77\,\AA, which are close to the detection threshold. 

Energy levels and line positions for \ion{Ir}{v} were provided by \cite{Gayasov2000_IrV}, but oscillator strengths are not available. 
Several predicted \ion{Ir}{v} lines are close to unidentified lines, e.\,g.\ at 1226.397\,\AA. However, the lack of oscillator strengths and the limited accuracy in predicted line positions prevented clear identifications. 
Thus we cannot derive an iridium abundance. 

\begin{figure}
  \centering
  \includegraphics[width=0.99\columnwidth]{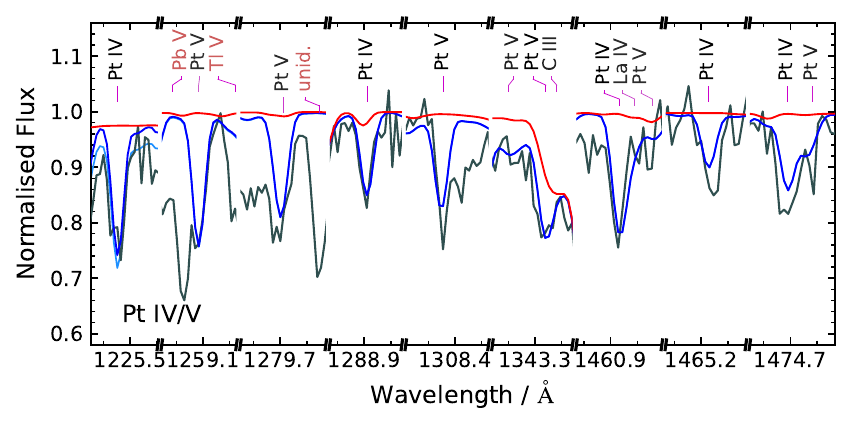}
  \vspace{-5pt}
  \caption{Strong platinum lines in \EC, like Fig.\ \ref{fig:detail_lsiv_1}. }
  \label{fig:detail_ec22_Pt}
\end{figure}
\begin{figure*}
\centering
\includegraphics[width=0.99\textwidth]{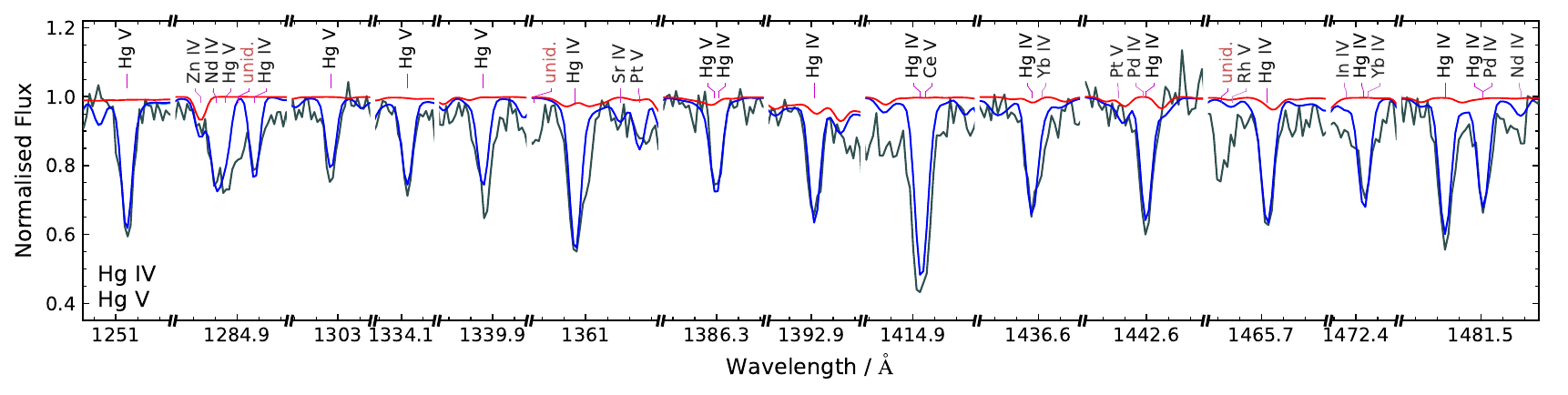}\\[-18.3pt]
\includegraphics[width=0.99\textwidth]{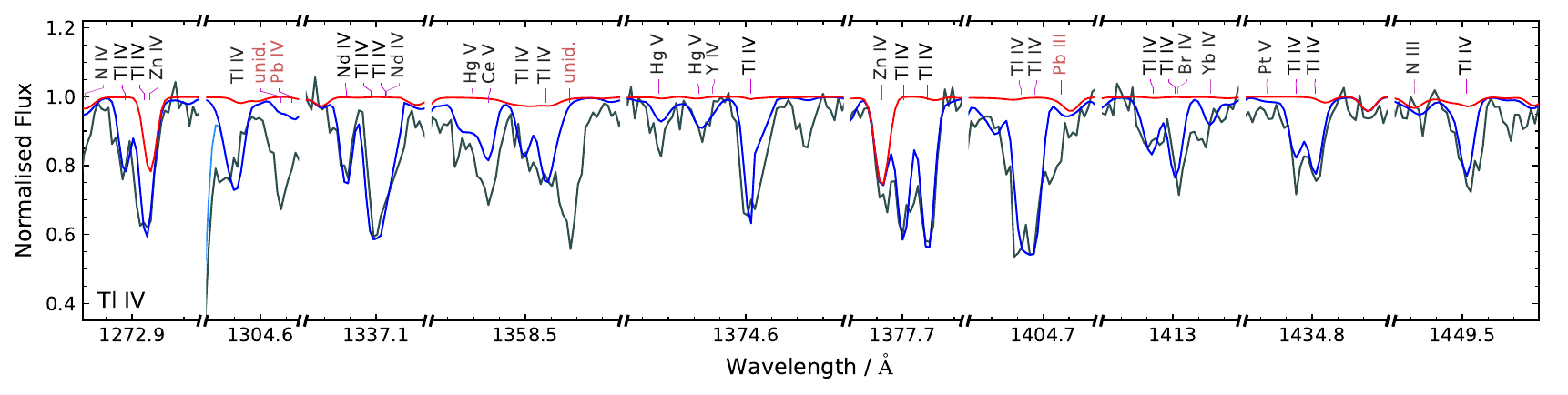}\\[-18.3pt]
\includegraphics[width=0.99\textwidth]{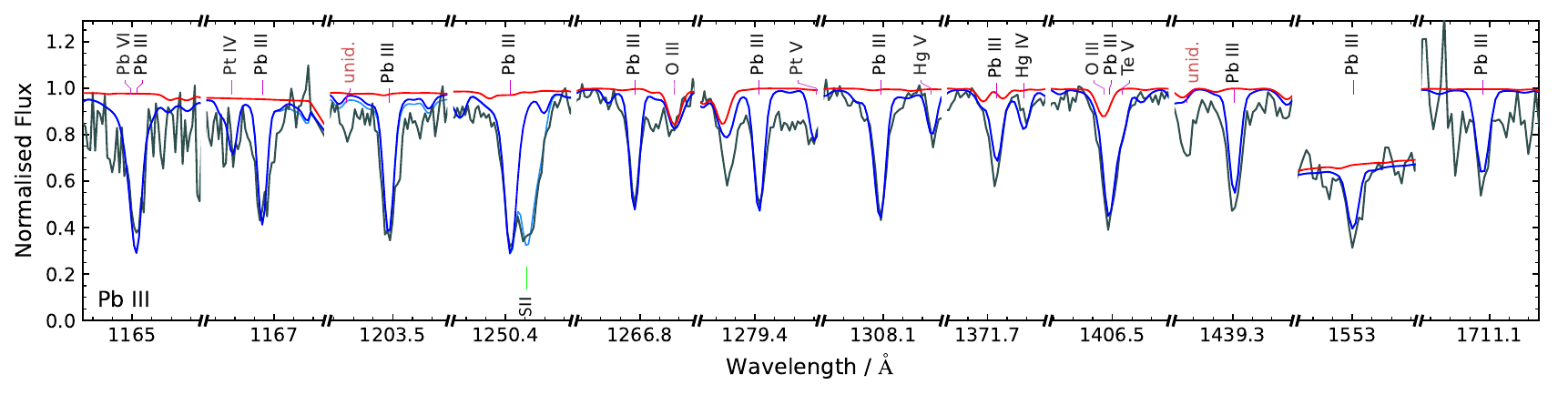}\\[-18.3pt]
\includegraphics[width=0.99\textwidth]{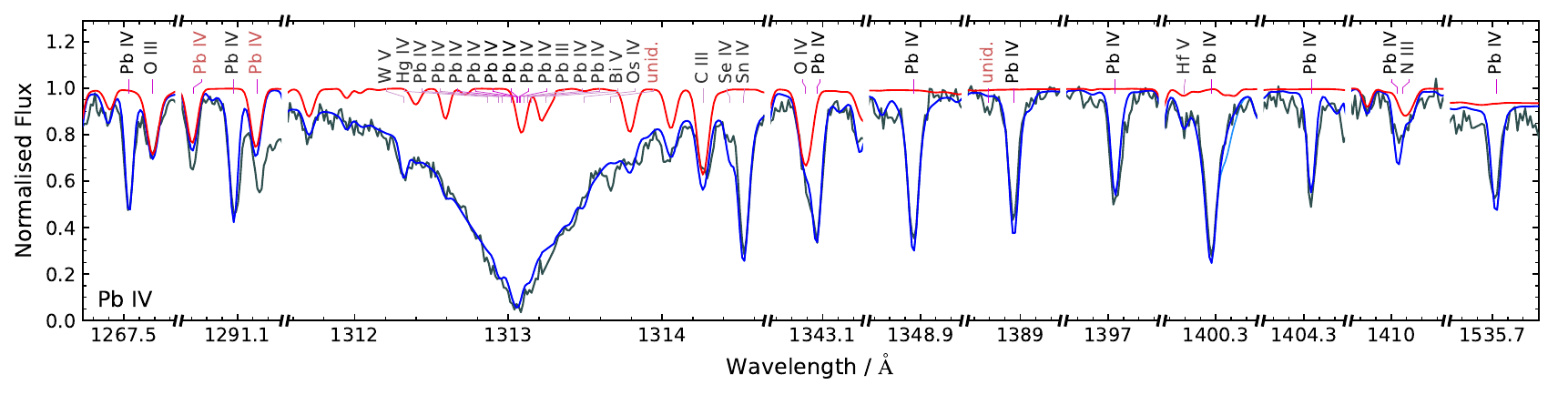}\\[-18.3pt]
\includegraphics[width=0.99\textwidth]{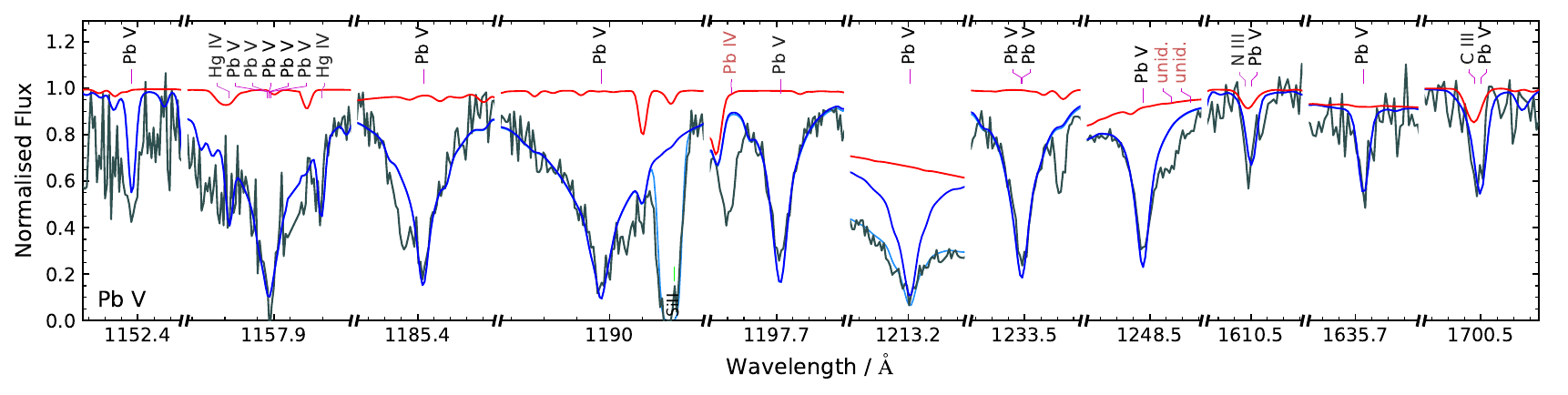}\\[-18.3pt]
\includegraphics[width=0.99\textwidth]{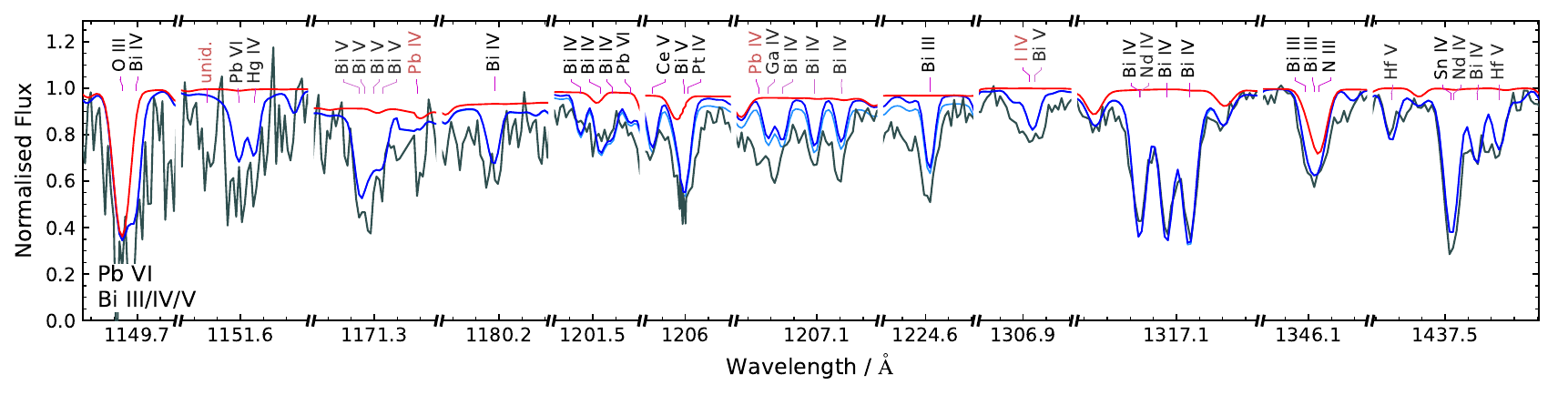}\\
\vspace{-7pt}
\caption{The strongest \ion{Hg}{iv-v}, \ion{Tl}{iv}, \ion{Pb}{iii-vi}, and \ion{Bi}{iii-v} lines, next to \ion{Sn}{iv}, in the STIS spectrum of \EC; like Fig.\ \ref{fig:detail_lsiv_1}.
}
\label{fig:detail_Pb}
\end{figure*}

\vspace{-10pt}\paragraph{Platinum.}

Using line positions and oscillator strengths from \citet{Azarov2016_PtIV}, several \ion{Pt}{iv} lines are clearly identified in \EC\ (Fig.\ \ref{fig:detail_ec22_Pt}), the strongest being 1225.453, 1288.911, and 1460.932\,\AA. \ion{Pt}{v} lines from \citet{Azarov2016_PtV} are also detected, including the prominent transitions at 1259.086, 1279.711, 1308.355\,\AA\ (blended with a similarly strong \ion{Hg}{v} line), and 1343.341\,\AA, with equivalent widths of about 10\,m\AA, as well as weaker lines at 1259.951, 1340.936, 1359.136, and 1361.247\,\AA.

In \lsiv, the least blended \ion{Pt}{iv} transitions occur at 1198.375, 1225.453, 1260.461, 1359.136, and 1465.227\,\AA. These lines are close to the detection threshold, but we prefer to state an upper limit. 

\vspace{-10pt}\paragraph{Gold.}

No gold lines are clearly detected in either star. 
Using experimental wavelengths for \ion{Au}{iv} at 1404.520 and 1450.948\,\AA\ from \citet{Zainab2023_AuIV}, we derive an upper limit for \EC, which may be close to its actual abundance, given that these transitions coincide with otherwise unidentified lines. 
We also searched for \ion{Au}{iii} using data from \cite{Zainab2019_AuIII}, but this ion is not abundant enough. 
Although \cite{Tauheed2021_PtIV_AuV_HgVI} do not predict any \ion{Au}{v} transitions within the STIS spectral range, \cite{Azarov2000_AuV} report experimental line positions that fall within this range. 
Given the weakness of gold lines in \EC, and the lack of oscillator strengths, we cannot identify any \ion{Au}{v} lines.

\vspace{-10pt}\paragraph{Mercury.} 

The UV spectrum of \ion{Hg}{iv} was analysed in detail by \cite{vanderValk1990_HgIV}, an analysis that was later updated by \cite{Rashid2021_HgIV}, who also provide oscillator strengths. 
Using their data, we detect more than 50 strong \ion{Hg}{iv} lines in \EC, the strongest of which are shown in Fig.\ \ref{fig:detail_Pb}, reaching equivalent widths of about 25\,m\AA. 
\ion{Hg}{iv} 1360.952, 1414.933, 1481.336\,\AA, are strong enough to be detected in He-sdO stars at abundances exceeding about 500$\times$ solar. 
In particular \ion{Hg}{iv} 1360.952\,\AA\ is typically isolated in relatively cool He-sdO stars, and not blended even at super-solar iron abundances. 
In addition, strong lines are predicted at 1082.778, 1109.970\,\AA; well outside the HST/STIS range, but covered by FUSE spectra of several He-sdO stars.

Line positions for \ion{Hg}{v} were measured by \citet{Wyart1993_HgV} using laboratory normal incidence spectroscopy, most importantly the 5d$^7$\,6s -- 5d$^7$\,6p transitions. 
Many of these transitions correspond to strong lines observed in the STIS spectrum of \EC. Theoretical oscillator strengths for the Os-like isoelectronic sequence, including \ion{Hg}{V}, were calculated by \citet{Taghadomi2022_HgV} using the multiconfiguration Dirac-Hartree-Fock method implemented in the GRASP2K package \citep{Jonsson2013_GRASP2K}. 
Because these ab initio calculations were not adjusted to observed energy levels, the resulting line positions are inaccurate. We therefore limited our analysis to the 5d$^7$\,6s -- 5d$^7$\,6p transitions, correcting their level energies using the experimental values from \citet{Wyart1993_HgV}.
The strongest \ion{Hg}{v} lines in \EC\ are shown in Fig.\ \ref{fig:detail_Pb}. 

Oscillator strengths,  experimental wavelengths and theoretical energies  for \ion{Hg}{vi} were reported by \citet{Tauheed2021_PtIV_AuV_HgVI}, but only for transitions in the extreme ultraviolet. No additional line positions have been published, preventing the identification of the expected $5d^66s-5d^66p$ transitions within the STIS wavelength range. 

There are no clearly isolated mercury lines in the spectrum of \lsiv. \ion{Hg}{iv} 1360.952\,\AA\ is dominant enough to estimate an abundance, consistent with \ion{Hg}{iv} 1157.627, 1158.191, and 1239.792\,\AA. 

\vspace{-10pt}\paragraph{Thallium.}

The strongest thallium features in \EC\ are \ion{Tl}{iv} transitions involving the lower $5d^9 6s$ levels. These lines exhibit pronounced hyperfine splitting, first measured experimentally by \cite{Arvidsson1930_TlIV_BiIV_HFS}, with a more detailed analysis of \ion{Tl}{iv} carried out later by \citet{Joshi1990_TlIV_hyperfine}. Additional line positions for the $6p$-$6d$ transitions, which show no hyperfine splitting, were reported by \citet{Wyart1992_TlIV_PbV_BiVI}.
Since no oscillator strengths are available in the literature, new values were computed for this work, as described in Sect.~\ref{sect:umons}. 
The two dominant thallium isotopes ($^{203}$Tl, $^{205}$Tl; nuclear spin $I = 1/2$) produce the observed hyperfine doublets if the $5d^9 6s$ levels have magnetic dipole constants of $A \simeq 70$\,GHz. 
The strongest detected lines are the hyperfine doublets at 1272.872 and 1272.946, 1337.081 and 1337.136, 1377.703 and 1377.797, and 1404.613 and 1404.668\,\AA\ (see Fig.\ \ref{fig:detail_Pb}), along with the single line at 1374.619\,\AA, in addition to several weaker transitions.

\cite{Raassen1991_TlV} performed spark spectroscopy for \ion{Tl}{v}, providing accurate positions for $5d^86s-5d^86p$ transitions, including many hyperfine doublets. More than 30 of these lines match otherwise unidentified lines in the HST spectrum of \EC, e.g.\ at 1258.148\,\AA, and some of the strongest are shown in Fig.\ \ref{fig:ec_missing_fosc}.
Unfortunately, no oscillator strengths are available for \ion{Tl}{v}. 
Energy levels for \ion{Tl}{vi} were taken from \citet{Raassen1994_TlVI} to compute the partition function. Several unidentified lines (1170.179, 1170.557, 1197.011\,\AA) are close to the their \ion{Tl}{vi} positions, but given the absence of oscillator strengths, these cannot be confidently identified.

The strongest \ion{Tl}{iii} lines originate from $5d^{10}6s$ -- $5d^{10}6p$ transitions, which \citet{Joshi1990_TlIII_HFS} showed to exhibit pronounced hyperfine splitting. We adopt their hyperfine constants and the oscillator strengths from \citet{deAndresGarcia2019_TlIII}. While \ion{Tl}{iii} 1558.78\,\AA\ is below the detection threshold in \EC, the \ion{Tl}{iii} 1266.19/1266.29\,\AA\ hyperfine doublet is detected, but is weak and blended with \ion{O}{iii} 1266.13\,\AA.
The feature is further predicted to be too strong relative to the more prominent \ion{Tl}{iv} lines. 

\ion{Tl}{iv} 1337.081,1337.136\,\AA\ are fairly strong and isolated in \lsiv.
\ion{Tl}{iv} 1404.613, 1404.668\,\AA\ are strong but blended with \ion{Ni}{iv} 1404.589\,\AA. 
While there are no other unblended lines, these two doublets are clearly detected and allow us to estimate the thallium abundance in \lsiv. 

\vspace{-10pt}\paragraph{Lead.}

As expected from its extreme enrichment, \EC\ shows very strong lead lines in the UV spectrum. Some of the most prominent features are displayed in Fig.\,\ref{fig:detail_Pb}, with a summary of all detected lines provided in Table~\ref{tab:idlines}. Our analysis uses the atomic data of \citet{Dougan2025_Pb}, calculated specifically for \EC\ and comprising photoionisation cross-sections and radiative data for \ion{Pb}{iii-vi}, allowing the construction of non-LTE model atoms. As discussed by \citet{Dougan2025_Pb}, non-LTE effects weaken predicted line strengths for the strongest UV transitions, leading to slightly higher inferred abundances. 
These effects are even stronger for optical \ion{Pb}{iii-iv} lines, which appear weaker in non-LTE and would imply higher lead abundances. However, these transitions arise from high-lying levels near the upper energy limits of the model atoms, and we therefore regard the low-lying UV transitions as more reliable.

For \ion{Pb}{iii}, transitions not included among the 20 levels analysed by \citet{Dougan2025_Pb} were supplemented with data from \citet{Alonso-Medina2009_PbIII}. These sources provide coverage for 16 observed \ion{Pb}{iii} lines in the UV spectrum of \EC. 
\citet{Lyall1965_PbIII} analysed the spark spectrum of \ion{Pb}{i-iv}, providing the most extensive experimental data for \ion{Pb}{iii} in the UV. Although they reported 0.01\,\AA\ precision, their wavelengths show systematic offsets of up to 0.04\,\AA\ relative to the STIS spectrum of \EC, from which we remeasured the \ion{Pb}{iii} line positions. Their measurements remain useful for identifying \ion{Pb}{iii} lines without published oscillator strengths, notably those at 1220.49, 1275.69, 1290.33, and 1327.14\,\AA. 

The \ion{Pb}{iv} structure of \citet{Dougan2025_Pb} includes the lowest 30 energy levels, leaving several upper levels of observed transitions unaccounted for. For 13 additional transitions from lower $5d^9\,6s6p$ levels to upper $8s\,^2$S, $7d\,^2$S, and $5g\,^2$G levels, we adopted data from \citet{Alonso-Medina2011_PbIV}. 
Additional \ion{Pb}{iv} lines measured by spark spectroscopy in \citet{Gutmann1969_Tl_Pb} are also strong in the STIS spectrum of \EC, but their oscillator strengths remain unavailable. 

The strongest observed \ion{Pb}{v} lines are $5d^9 6s$ -- $5d^9 6p$ transitions, which are covered by \citet{Dougan2025_Pb}. 
The line list was further extended to include $5d^9 6p$ -- $5d^9 6d$ transitions from \citet{Colon2014_PbV}; among these, only \ion{Pb}{v} 1240.07\,\AA\ is clearly detected, while the others are blended with stronger features. 
Three additional lines were identified using wavelengths reported by \citet{Gutmann1969_Tl_Pb}. 
The strongest \ion{Pb}{vi} lines in \EC\ are observed at 1151.59\,\AA\ and 1164.98\,\AA\ (blended with \ion{Pb}{iii}), next to an additional 12 weaker lines. 

An accurate treatment of Stark broadening is important for the strong  \ion{Pb}{iv-v} lines in the UV spectrum of \EC. 
This is most obvious for the strongly saturated \ion{Pb}{iv} 1313\,\AA, for which we adopted the electron-impact Stark widths of \citet{Hamdi2013_PbIV_Stark}, implemented via the temperature-dependent fitting formula provided in the STARK-B database \citep{STARKB}. 
Besides \ion{Pb}{iv} 1313\,\AA, Stark broadening is most significant for the \ion{Pb}{v} lines at 1157.9 and 1189.9\,\AA. Only the latter is included in the calculations of \cite{Alonso-Medina2012_PbV_Stark}. When these widths are adopted, the synthetic profile is stronger than observed; we therefore do not use their Stark widths for this transition and exclude the line from the abundance analysis. 

The \ion{Pb}{iv} 1313\,\AA\ resonance line is known to show HFS, as discussed by \cite{O'Toole2007}. We include HFS splitting using their implementation; a more detailed discussion of lead HFS is beyond the scope of this paper and will be addressed in the future. 
In addition, the various \ion{Pb}{iii-v} lines in \EC\ may be used to probe vertical abundance stratification, as demonstrated by \cite{Scott2024}. A detailed stratification analysis, is beyond the scope of this work; however, strong stratification is excluded by the good match between optical and UV lines, given that they form at different depths in the photosphere. 

\ion{Pb}{iv} 1313.1\,\AA\ and \ion{Pb}{v} 1157.9, 1185.4, 1189.9\,\AA\ are clearly detected in \lsiv, allowing a reliable abundance determination. 

\begin{figure*}
\centering
\includegraphics[width=0.92\textwidth]{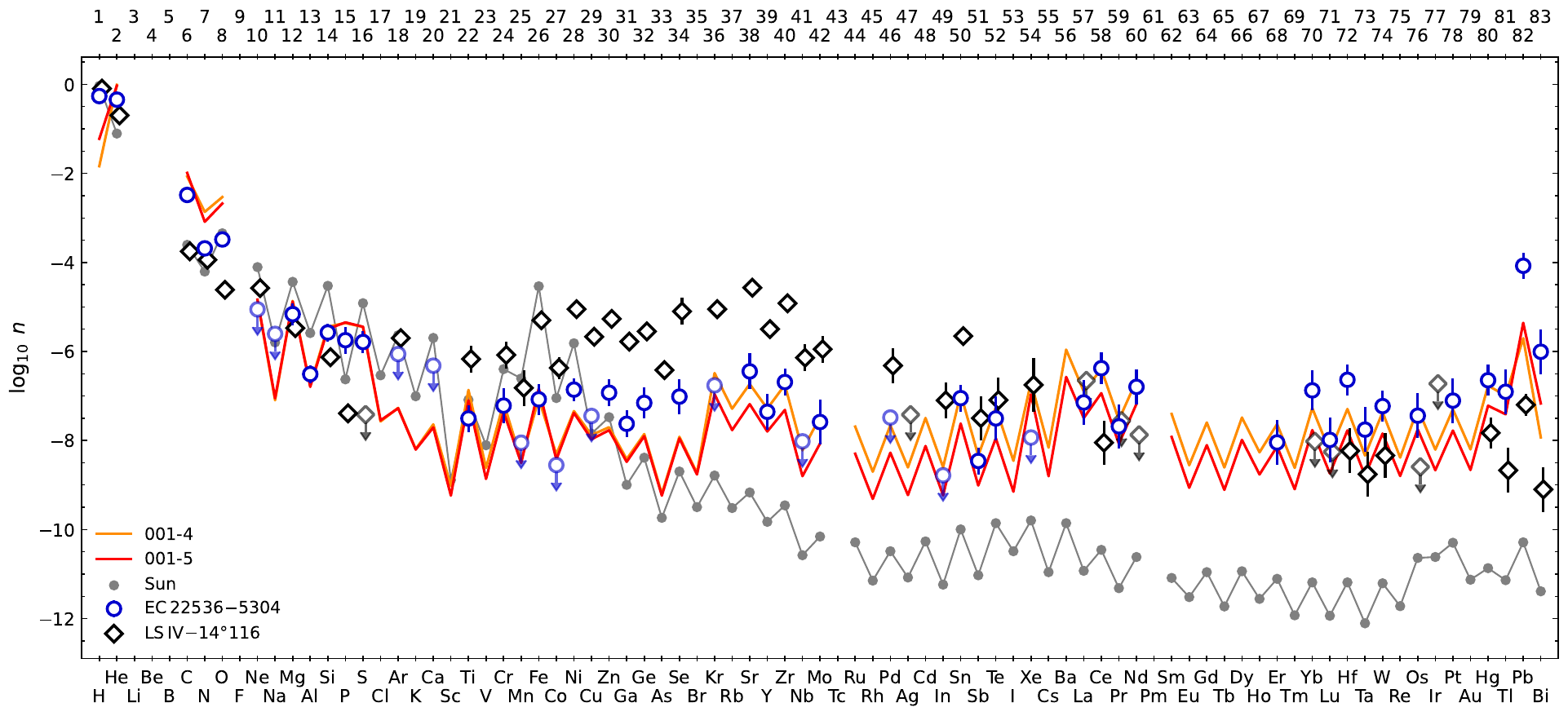}\\
\includegraphics[width=0.92\textwidth]{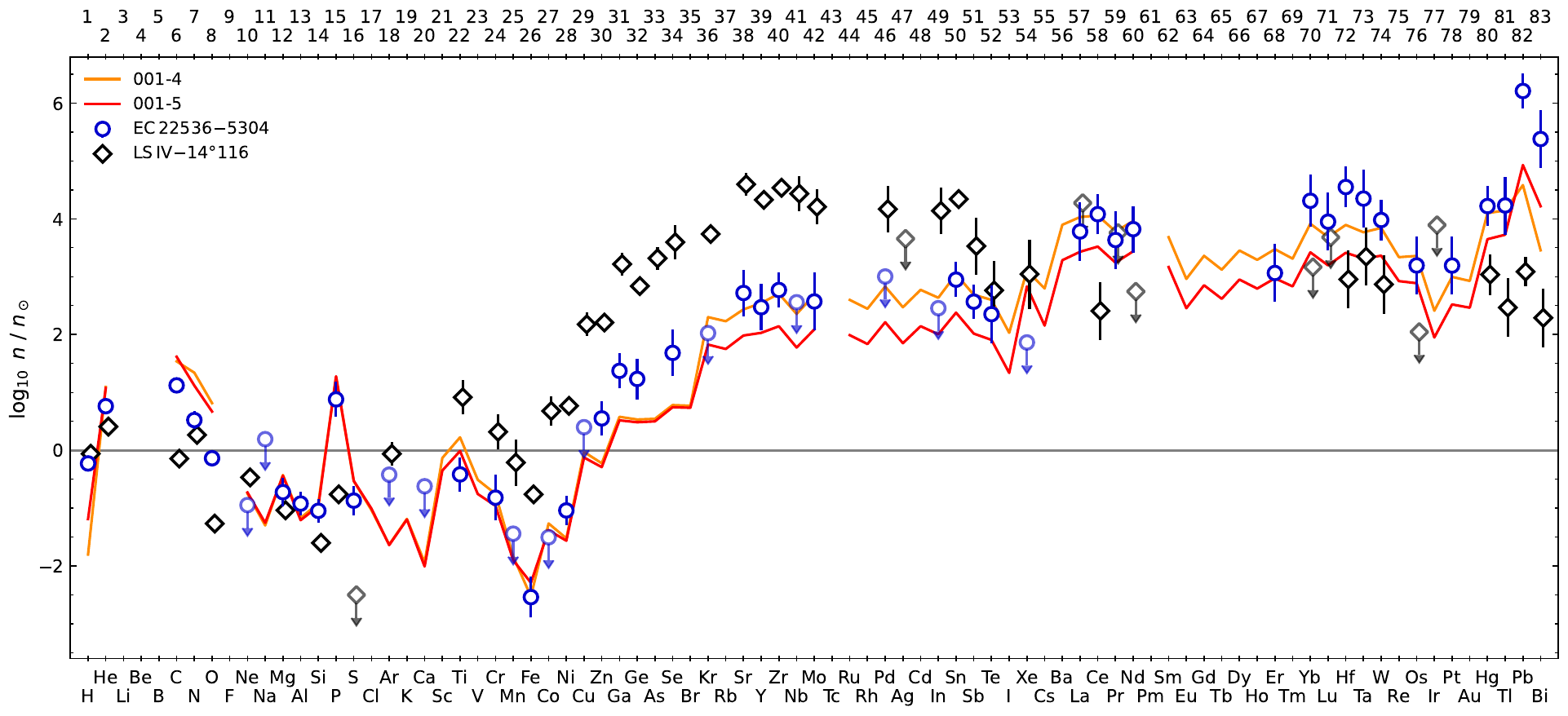}\\
\includegraphics[width=0.92\textwidth]{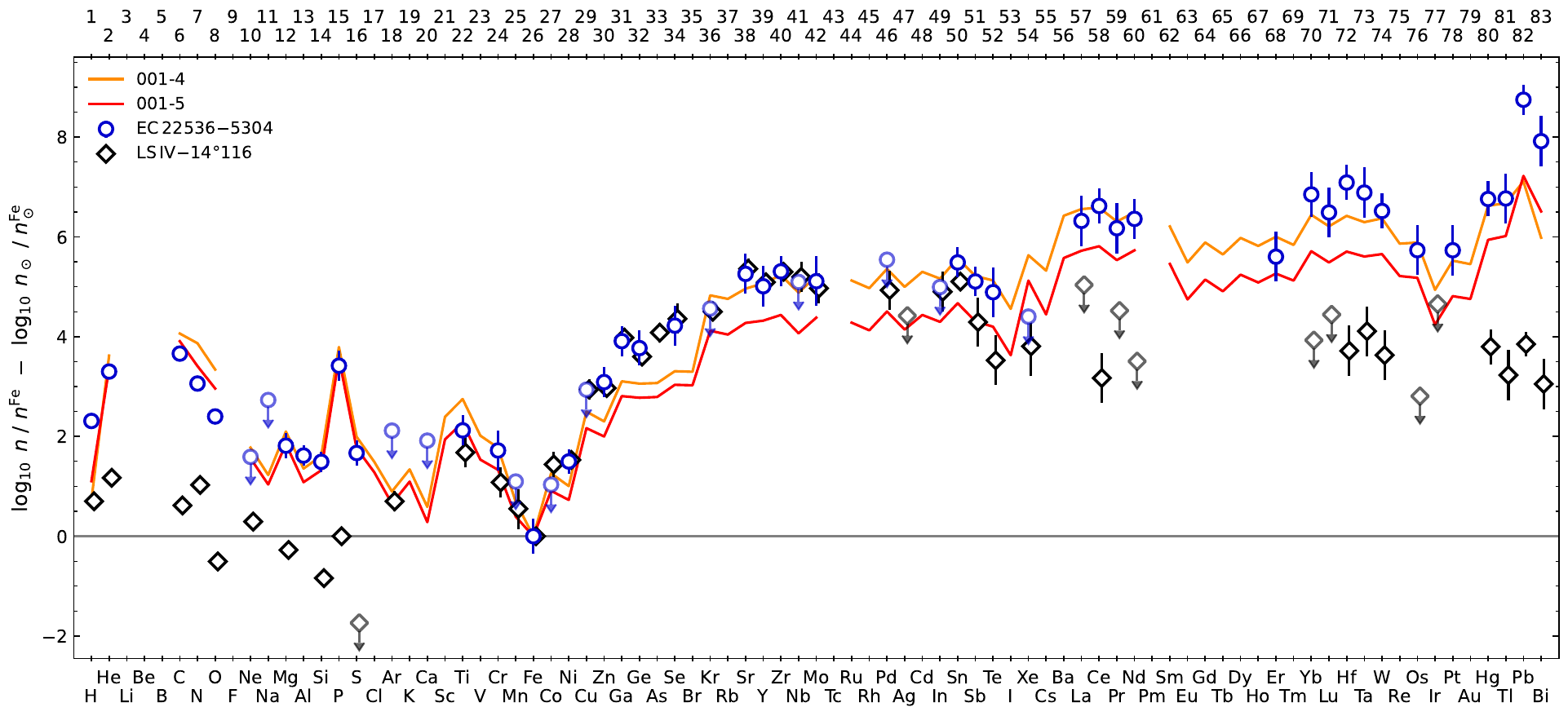}
\vspace{-3pt}
\caption{\textit{Top}: Surface abundances of \EC\ and \lsiv\ by number, compared to the solar pattern of \cite{asplund09}, supplemented with heavy-metal results from \cite{Grevesse2015}, \cite{Lodders2019}, and \cite{Asplund2021}. 
\textit{Middle}: Abundances relative to solar. 
\textit{Bottom}: Solar-relative abundances, scaled to the same Fe abundance. 
Also shown are two models from \cite{Battich2025} with initial metallicity [Fe/H]\,=\,$-1.15$ and envelope masses of 0.00055\,\msun\ (001-4) and 0.0006\,\msun\ (001-5).}
\label{fig:pattern_sun}
\end{figure*}

\vspace{-10pt}\paragraph{Bismuth.}

Several strong bismuth lines are observed in \EC, using atomic data from \cite{Arya2020_BiIII,Arya2022_BiIV,Arya2023_BiV} for \ion{Bi}{iii-v}. Because of its high nuclear spin ($I = 9/2$), bismuth exhibits pronounced hyperfine splitting \citep{Arvidsson1930_TlIV_BiIV_HFS}. 
The strongest feature is the well-resolved \ion{Bi}{iv} hyperfine triplet at 1316.939, 1317.067, and 1317.176\,\AA\ (Fig.\ \ref{fig:detail_Pb}). The \ion{Bi}{iv} 1149.7, 1180.18\,\AA, the 1201.5 and 1207.1\,\AA\ HFS triplets, and 1437.6\,\AA\ are also observed. 

Although \ion{Bi}{iii} is a minor ionisation stage at the high \teff\ of \EC, the 1224.62\,\AA\ line and the unresolved 1346.088/1346.124\,\AA\ hyperfine doublet are still detected. \ion{Bi}{iii} lines were also identified in the HgMn star HR\,7775 by \citet{Wahlgren2001_BiIII_CP_HFS}, who provide the relevant HFS constants. The intercombination resonance HFS multiplet \ion{Bi}{iii} 1423.4\,\AA\ is also present in \EC, albeit weak. 

The \ion{Bi}{v} 1171.233/1171.260/1171.296/1171.339\,\AA\ hyperfine complex is also detected in \EC\ but appears blended due to intrinsic line broadening; additional \ion{Bi}{v} are observed at 1202.96, 1206, and 1306.95\,\AA. 
The \ion{Bi}{v} 1139.55\,\AA\ resonance quadruplet is expected to be strong, but it lies outside the STIS range. It is likely present in other hot subdwarfs observed by the FUSE satellite. 
Overall, the predicted \ion{Bi}{v} lines are too weak relative to \ion{Bi}{iv}, possibly due to non-LTE effects.

The strongest bismuth feature in \lsiv\ is the \ion{Bi}{iv} 1317\,\AA\ triplet. While this triplet is partially blended with unidentified lines, it is strong enough for an abundance estimate. 

\section{Discussion}
\label{sec:discussion}

In total, we measure 40 surface metal abundances for \lsiv\ and 41 for \EC; these abundance patterns are by far the most extensive obtained for any hot subdwarf to date. The measured abundances are listed in Table~\ref{tab:abu} and compared with each other and with solar number fractions in Fig.~\ref{fig:pattern_sun}. 

\subsection{Surface composition of EC\,22536$-$5304}
\label{sect:discussion_ec}

\EC\ has an sdF-type companion with a metallicity of [Fe/H] = $-2$ and significant $\alpha$-enhancement \citep[+0.4\,dex;][]{Dorsch2021}.
In contrast, \EC\ itself is enriched relative to the solar number fractions in carbon (+1.1\,dex) and N (+0.5\,dex), and only mildly sub-solar in O ($-0.1$\,dex; see Table \ref{tab:abu}).  
The $\alpha$-elements Ne, Mg, Si, and S appear at approximately $-1$\,dex, consistently higher than the metallicity of the sdF companion. Phosphorus clearly deviates from this pattern: its abundance is about 0.9\,dex higher than solar. 
The P/Si ratio in \EC\ is strongly enhanced ($\log n(\mathrm{P})/n(\mathrm{Si})$ = $0.0$, compared with the solar value of $-2.1$). %
The iron-group elements display a distinct ``V-shaped'' pattern: the abundances decline from Ar and Ti (both $-0.4$\,dex) through Cr ($-0.8$\,dex) to a minimum at Fe ($-2.5$\,dex), before increasing again from Co ($-1.5$\,dex) and Ni ($-1.0$\,dex) up to Zn (+0.6\,dex). 
Heavy elements are moderately to extremely enhanced relative to solar, increasing from +1.4\,dex in Ga to +4.2\,dex in Tl. 
This corresponds to an approximately constant mass fraction (around $-5.7$\,dex) with a pronounced even-odd effect (upper panel of Fig.~\ref{fig:pattern_sun}).  
Lead is a notable exception, exhibiting an extremely high abundance of $-2.1$\,dex by mass, i.e.\ a +6.1\,dex enrichment relative to solar; bismuth shows a similar, though less extreme, enhancement, at $-4.1$\,dex by mass (+5.3\,dex enrichment).

A preliminary version of this surface abundance pattern was already discussed by \citet{Battich2025} in the context of their hot-flasher models, which focused on $i$-process nucleosynthesis. Their models reproduce the observed abundances remarkably well, including the CNO pattern, the phosphorous enrichment, the $\alpha$-element abundances at approximately $-1$ dex relative to solar, the characteristic ``V-shaped'' iron-group pattern, and the enhancements in heavy metals (cf.\ Fig.\ \ref{fig:pattern_sun}). This excellent match is achieved specifically for an initial metallicity of [Fe/H] = $-1.15$ (or $Z = 0.001$) for the sdOB component\footnote{\citet{Battich2025} did not consider $\alpha$-enhancement.}. 
Among their grid, models with hydrogen-envelope masses of $5.5 \times 10^{-4}$\,\msun\ (001-4) and $6.0 \times 10^{-4}$\,\msun\ (001-5) both provide good fits; as model 001-4 yields a marginally better agreement, we adopt it for the following comparisons.

Most of the observed elements are affected, to varying degrees, by nucleosynthesis during the formation of the sdOB. The abundances of H, He, C, N, and O are strongly modified by hydrogen and helium fusion. The increased phosphorus to silicon ratio is a clear signature of neutron capture\footnote{The connection between neutron capture and phosphorus enrichment has also been made for cool stars by \cite{Brauner2024_Prich}. Silicon depletion by weak selective winds \citep{Unglaub2008} is negligible here because both hot subdwarfs are compact at $\log g \approx 5.8$.}, given that P can be produced via $^{30}\mathrm{Si}(n,\gamma){}^{31}\mathrm{Si}\rightarrow{}^{31}\mathrm{P}+\beta^-$. 
The ``V-shaped'' abundance pattern among the iron-group elements also reflects neutron-capture processing, in which the lighter iron-group nuclei act as seeds for the production of heavier species. By contrast, the $\alpha$-elements Ne, Mg, Si, and S are only weakly affected by nuclear processing; their observed  abundances close to $-1$ dex are consistent with the initial metallicity required by \citet{Battich2025}. This agreement, however, raises the question why the sdF-companion has lower $\alpha$-element abundances closer to $-1.6$\,dex \citep[only Mg is directly measured in the sdF;][]{Dorsch2021}. 
For the trans-iron elements, the predicted abundances generally agree well with the observations. The main exceptions are Ga to Kr, which are under-predicted by about 1\,dex by \citet{Battich2025}, and Pb and Bi, which are under-predicted by about 1.6\,dex. Yb, Hf, and Ta are under-predicted by around 0.5\,dex.

We thus conclude that the surface composition of \EC\ is dominated by nucleosynthesis during its initial helium flash. Atomic diffusion is not only unnecessary to explain the observed abundances, but its effect in this star must also be small for the majority of elements. 
The remaining differences between the predictions of \citet{Battich2025} and the observed surface abundances of \EC\ may instead arise from limitations in the nucleosynthesis models. In particular, the treatment of convection in one-dimensional, mixing-length-based models, may overestimate the entropy barrier between the H- and He-burning convection zones \citep{Herwig2014, Battich2025}. The under-predicted abundances of Yb to Ta, Pb and Bi, may point to a longer neutron exposure. However, differences in abundances particularly for $Z>40$ may also arise from uncertainties in the neutron-capture reaction rates \citep{Martinet2024}.

\subsection{Surface composition of LS\,IV\-$-$14$\degr$116}

For $Z<25$ and $Z>52$, the surface composition of \lsiv\ is broadly similar to \EC. 
Helium is enriched to a lesser degree (\logy\ = $-0.6$ versus $-0.1$ in \EC), and the C and O abundances are also lower, whereas N and Mg are nearly identical in the two stars. 
Ne is more abundant in \lsiv, while Si is slightly less abundant. 
As in \EC, the P/Si ratio in \lsiv\ is strongly enhanced ($\log n(\mathrm{P})/n(\mathrm{Si})$ = $-0.4$, compared with the solar value of $-2.1$). 
The Ar abundance is close to solar in \lsiv\ and therefore significantly higher than in \EC. 
The iron-group elements show a similar  ``V-shaped'' pattern in both stars, but in \lsiv\ the pattern is shifted upward by about $+2$\,dex relative to \EC. Despite this, \lsiv\ remains Fe-poor with respect to the Sun ($-0.8$ dex). 

These differences in the light-metal abundance pattern likely reflect differences in both initial metallicity and the formation histories of the two stars. 
\EC\ is known to have formed through Roche lobe overflow onto its sdF-type companion \citep{Dorsch2021}, which corresponds to the hot-flasher scenario modelled by \cite{Battich2025}. 
In contrast, \lsiv\ is single \citep{Dorsch2020} and has been proposed to originate from the merger of a He-core white dwarf with a less massive hybrid HeCO-WD \citep{MillerBertolami2022}\footnote{This scenario is motivated largely by the rich pulsation spectrum of \lsiv, which seems to require C/O enrichment in its envelope \citep{Saio2019}.}. 
Although detailed nucleosynthesis predictions for such WD mergers are not available, the surface composition of \lsiv\ suggests material processed by H- and He-burning, with slightly less He-processed material than in \EC.  
Its strong phosphorus enrichment is plausibly produced by the same neutron-capture mechanism inferred for \EC. 
The higher iron-group abundances may indicate a somewhat higher initial metallicity than that of \EC, although \lsiv\ was likely still initially metal-poor, consistent with its halo-like kinematics \citep{Randall2015}. 

\lsiv\ is substantially richer in heavy metals up to tellurium compared to \EC, reaching maximum enhancements of 4 to 4.6\,dex relative to solar in Sr to Sn (Fig.\ \ref{fig:pattern_sun}, middle panel). 
However, a more meaningful comparison is obtained by using Fe-normalised, solar-relative abundances, [$X$/Fe] = $\log n / n_\mathrm{Fe} - \log n_\odot / n_\mathrm{Fe,\odot}$ (Fig.\ \ref{fig:pattern_sun}, bottom panel), which removes the offset caused by the different residual Fe abundances in the two stars. 
On this scale, the agreement between \lsiv\ and \EC\ in the Fe to Mo range is striking. 
Beyond Sb, \lsiv\ shows Fe-normalised abundances relative to solar of about 3 to 4\,dex, whereas \EC\ reaches 6 to 7\,dex up to Tl, 8\,dex for Bi and 8.7\,dex for Pb.

These similarities suggest that the neutron-capture process operated in a broadly comparable way in both stars between Fe and Mo; the differences in absolute abundances may reflect variations in the duration of neutron exposure or in the efficiency with which processed material was mixed to the surface.
In contrast, the production of the heaviest elements appears less efficient in \lsiv\ than in \EC. %
This difference may arise from variations in the initial metallicity, as well as in neutron densities and timing of mixing. In particular, a shorter nucleosynthesis timescale in \lsiv\ could limit the build-up of the heaviest species. Such variations are naturally expected given the different formation channels of \EC\ and \lsiv.

We therefore propose that neutron capture plays a dominant role in the heavy-element enrichment of \lsiv, operating in an analogous fashion to \EC\ but under conditions altered by the merger process. This highlights the need for nucleosynthesis calculations for both double He-WD mergers and He-WD + hybrid He/CO-WD mergers. Finally, we note that diffusion processes may still influence the surface composition of \lsiv\ without contradicting the $i$-process interpretation; their significance can only be assessed once detailed nucleosynthesis models for such mergers become available.

\subsection{Diffusion}

Atomic diffusion, encompassing gravitational settling and radiative levitation, operates efficiently in hot stars with radiative envelopes and is widely invoked to explain the surface abundance anomalies observed in He-poor hot subdwarfs and blue horizontal-branch stars. In He-poor sdB stars, diffusion leads to strong helium depletion\footnote{Diffusion-only models fail to reproduce the observed helium abundances in sdB stars, implying the presence of additional atmospheric mixing processes, such as radiatively driven stellar winds \citep[e.g.][]{Fontaine1997} or an as-yet unidentified form of turbulence \citep{Hu2011,Michaud2011}. } \citep{Michaud2011} and pronounced heavy-element overabundances, in some cases reaching 3\,dex above solar \citep[][]{Chayer2006, O'Toole2006b, Blanchette2008}. Similar chemical peculiarities are observed in HgMn and Ap/Bp stars, where diffusion is generally accepted as the dominant mechanism shaping the atmospheric composition \citep{Michaud2015}.
Heavy-metal iHe-sdOB stars also show some resemblance to a small group of metal-rich DO and DAO white dwarfs, such as \RE, HD\,149499B, HZ\,21 \citep{Chayer2005,Rauch2012_Ge}, and BD$-$22$\degr$3467 \citep{Loebling2020}. These objects are thought to be post-AGB stars that experienced a late thermal pulse, or descendants of mergers involving a He-core white dwarf. Their surface compositions may therefore reflect $s$-process enrichment during the AGB phase, subsequently modified by radiative levitation. Interestingly, very late thermal pulses are also candidates for $i$-process nucleosynthesis \citep{Herwig2011}.

Given their high effective temperatures and largely radiative atmospheres, one might expect atomic diffusion to operate in iHe-sdOB stars similarly to hot DO and DAO white dwarfs. However, the heavy-metal enriched DO and DAO stars are significantly hotter (50\,000 to 80\,000 K) than the iHe-sdOBs discussed here (35\,000 to 40\,000 K), as well as less compact (\logg\ $\approx$ 5.8, rather than $\approx$ 7.5 in DOs). This implies not only weaker gravitational settling, but also different dominant ions in the atmospheres of heavy metal subdwarfs, whose atomic structure yields different opacities (lines and ionisation), thereby affecting radiative acceleration and diffusion efficiency. 
For \EC, the close agreement between the observed abundances and the $i$-process predictions discussed in Sect.\ \ref{sect:discussion_ec} argues against diffusion playing a dominant role in shaping the observed abundance pattern. 

\begin{figure}
\centering
\includegraphics[width=0.99\columnwidth]{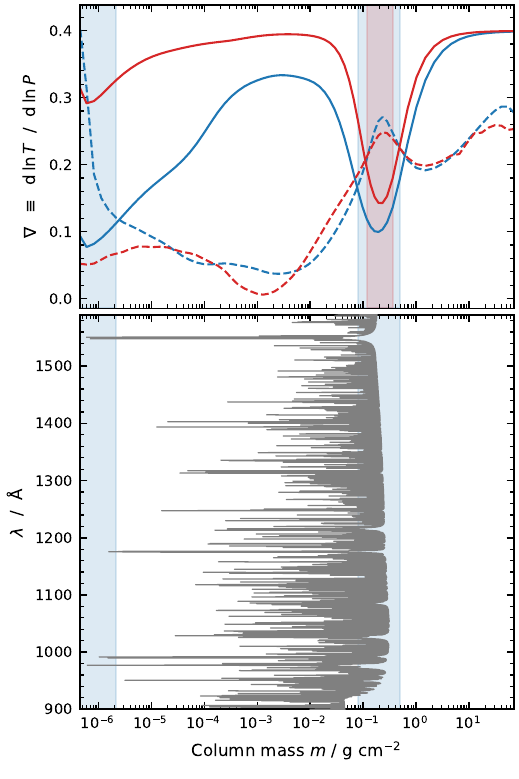}
\vspace{-5pt}
\caption{Convection test for \lsiv\ (red) and \EC\ (blue). \textit{Top}: Adiabatic gradients (solid), computed following \citet{Groth1985}, compared with the \tlusty\ model gradients (dashed); shaded regions indicate layers unstable according to the Schwarzschild criterion. \textit{Bottom}: UV line formation depths in \EC; the strongest lines arise from \ion{C}{iii-iv}, while most others are due to heavy metals.}
\label{fig:convection}
\end{figure}

A plausible explanation for the decreased importance of diffusion is provided by the elevated helium abundances of iHe-sdOB stars and their intermediate temperature. The ionisation of \ion{He}{ii} produces an opacity peak that can drive shallow convection \citep{Groth1985}. Recent work shows that this instability extends into the iHe-sdOB regime, but not into He-poor sdO/B stars (see appendix~A.1 of \citealt{Dorsch2024_PhD}). Consistent with this, our \tlusty\ models of both \lsiv\ and \EC\ exhibit convective instability in the continuum-forming layers according to the Schwarzschild criterion (Fig.\ \ref{fig:convection}), enclosing a total mass of only $\sim\!10^{-12}$\,\msun. Although this convection does not significantly affect the radiative energy transport, even weak mixing may in principle inhibit atomic diffusion.
Hydrodynamical simulations of analogous shallow convection zones in DA  WDs suggest that overshoot can extend the mixed region by up to 2.5 orders of magnitude beyond the Schwarzschild boundary \citep{Freytag1996, Cunningham2019} in terms of mixed mass. No quantitative diffusion-convection models exist for iHe-sdOB stars. 

The mixed mass we find falls well short of the $\sim\!10^{-8}$\,\msun\ that the simplified models of \citet{Unglaub2010} require to preserve the initial surface composition of a \teff\ = $40\,000$\,K He-sdO, and of the $\sim\!10^{-7.5}$\,\msun\ of outer mixing invoked by \citet{Michaud2011} to reproduce the abundance pattern of He-poor sdB stars. 
In the latter case, He-driven convection is absent altogether, implying that additional mixing mechanisms are required. Line-driven wind calculations for hot subdwarfs similar to our targets predict mass-loss rates of $\dot{M} \sim 10^{-14} \textnormal{ to } 10^{-13}$\,\msun\,yr$^{-1}$  \citep{Krticka2016}, which can compete with gravitational settling in the outermost layers \citep{Unglaub2001,Unglaub2008}\footnote{Neither star shows signatures of a metallic wind, such as the C or Si depletions seen in some He-poor sdOBs \citep{O'Toole2006b}.}.  However, neither the shallow convection seen in our He-rich models nor a weak wind at these predicted rates match the mixing depths required by \citet{Unglaub2010} and \citet{Michaud2011}. The physical origin of the mixing, if it indeed exists, thus remains uncertain for hot subdwarf stars, and for iHe-sdOBs in particular.

\subsection{Stellar neutron capture in context} 

Since the discovery of radioactive technetium in AGB stars \citep{Merrill1952}, neutron-capture nucleosynthesis has been firmly established in these objects \citep{Burbidge1957}, primarily via the slow neutron-capture $s$-process at low neutron densities \citep{Busso1999}. 
Clear $s$-process signatures are observed in several classes of chemically peculiar stars, including barium stars and related binaries \citep{Jorissen2019}, S stars \citep{Shetye2020}, carbon-enhanced metal-poor stars with $s$-process enrichment (CEMP-$s$; \citealt{Aoki2007}), post-AGB stars \citep{DeSmedt2016}, and planetary nebulae \citep{Sharpee2007}.
By contrast, metal-poor stars with strong europium enhancements and actinide-rich stars exhibit abundance patterns consistent with the high-density $r$-process nucleosynthesis \citep{Roederer2010, Holmbeck2018}. 

A growing number of heavy-element-enriched stars show abundance patterns that cannot be explained by either a pure $s$- or $r$-process, nor by simple mixtures of the two \citep[e.g.][]{Hampel2016,Karinkuzhi2021}. These stars typically display enhanced first- and second-peak neutron-capture elements together with relatively high Eu abundances, implying neutron capture at intermediate neutron densities. Such patterns are naturally produced by the $i$-process (see \citealt{Wiedeking2025_irev} for a  review). 

The $i$-process occurs at neutron densities of about $10^{13}$ to $10^{16}$\,cm$^{-3}$. These densities are only achievable in specific environments involving the sudden ingestion of protons into C-rich He-burning convective regions. Neutrons are released by the reaction chain $^{12}\mathrm{C}(p,\gamma)^{13}\mathrm{N}(\beta^+)^{13}\mathrm{C}(\alpha,n)^{16}\mathrm{O}$. The $^{12}\mathrm{C}(p,\gamma)^{13}\mathrm{N}$ reaction occurs at the outer regions of the convective zone at about 150\,MK. $^{13}\mathrm{N}$ is carried by convection to the hotter interior of the convective region where He burning is taking place. This allows the formation of $^{13}\mathrm{C}$ via $\beta^+$ decay in an already very hot environment (300\,MK) where the reaction $^{13}\mathrm{C}(\alpha,n)^{16}\mathrm{O}$ takes place.

As reviewed by \cite{Wiedeking2025_irev}, such conditions may arise during the AGB phase of low-mass, low-metallicity stars, in very late thermal pulses of post-AGB stars, and in rapidly accreting white dwarfs. In hot subdwarfs, the $i$-process can occur during the initial helium flash, when helium ignition drives convection that ingests protons and establishes a secondary, hydrogen-burning convective zone. The resulting $^{13}\mathrm{C}$ is mixed into the helium-burning layer, where neutron production ensues. The best-fitting models of \citet[][cf.\ Fig.\ \ref{fig:pattern_sun}]{Battich2025} predict peak neutron densities of $1.7\times10^{12}$ (model 001-5) and $3.0\times10^{13}\,\mathrm{cm^{-3}}$ (model 001-4), at the low end of $i$-process densities. 
In these models, the two convective zones remain connected for about 500\,yr\footnote{Given the lower abundances of very heavy elements in \lsiv, the $i$-process in this star may have operated for a shorter duration, but with more efficient transport of processed material to the surface.} before an entropy barrier formed by the hydrogen-burning shell \citep{Iben1976} separates them, halting further $^{13}\mathrm{C}$ supply and surface enrichment. 
Similar nucleosynthesis during the helium flash may also occur in very low-metallicity stars \citep{Campbell2010}. 
The specific conditions required for mixing i-processed material to the surface, e.\,g.\ a low initial metallicity \citep[][]{Battich2025}, may in part explain the relative rarity of heavy-metal sdOBs. 

Against this theoretical background, \EC\ and \lsiv\ currently provide some of the most direct evidence for $i$-process nucleosynthesis. This conclusion is supported by two main observations. First, their heavy-element enrichments exceed those of cool stars by several orders of magnitude, consistent with relatively undiluted in situ production. While even strongly enriched post-AGB stars typically show iron-scaled enhancements of $[X/\mathrm{Fe}]\lesssim2.5$\,dex \citep[e.g.][]{DeSmedt2016}, \lsiv\ reaches values >$5$\,dex and \EC\ exceeds $8$\,dex. 
Secondly, \EC\ is securely identified as a post-RGB star, implying that its heavy elements must have been synthesised during the helium-flash phase itself, rather than being inherited from prior AGB evolution.

Although it cannot be excluded that \EC\ and \lsiv\ formed from material that was mildly enriched in neutron-capture elements, any such pre-enrichment would be negligible compared to their present abundance levels. In contrast to most other heavy-element-enhanced stars, whose compositions are likely influenced at least in part by $s$-process nucleosynthesis, the extreme and evolutionarily constrained abundance patterns of \EC\ and \lsiv\ strongly point to helium-flash-driven $i$-process nucleosynthesis as their dominant origin.

\section{Summary and conclusions}
\label{sec:conclusions}

We have presented a detailed abundance analysis of the heavy-metal intermediate-helium hot subdwarfs \lsiv\ and \EC\ based on high-resolution HST/STIS spectroscopy. These stars exhibit some of the most extreme surface chemical compositions known among all stars. Our analysis provides the strongest observational evidence to date that their heavy-element ($Z>30$) enrichment is predominantly the result of in situ nucleosynthesis via the intermediate neutron-capture ($i$-) process, linked directly to their formation histories.

We performed a comprehensive spectral analysis, deriving photospheric abundances for 40 metals in \lsiv\ and 39 metals in \EC, making this the most detailed abundance study of any hot subdwarf star to date. %
At effective temperatures close to 40\,000\,K and high absolute abundances, both stars show many strong trans-iron transitions arising from ionisation stages  \textsc{iii--v}. 
Because standard atomic line lists are incomplete for these ions, we assembled a large atomic database by collecting experimental wavelengths and theoretical oscillator strengths from a wide range of individual atomic-physics studies. 

Where no suitable data were available, we computed new atomic data, including oscillator strengths for \ion{As}{iii}, \ion{Se}{iii}, \ion{Hf}{iv}, and \ion{Tl}{iv} (Sect.\ \ref{sect:umons}; Deprince et al., in prep.), as well as photoionisation cross sections for \ion{Pb}{iii-vi} \citep[see][]{Dougan2025_Pb}. Using this expanded dataset, we identified numerous heavy-metal transitions for the first time in a stellar spectrum. In addition, several ions were found to exhibit clear signatures of hyperfine splitting, including \ion{In}{iii}, \ion{Sb}{v}, \ion{Ta}{v}, \ion{Tl}{iv-v}, \ion{Pb}{iv}, and \ion{Bi}{iii-v}, highlighting the importance of detailed atomic structure effects when modelling chemically extreme stars. 

Both \lsiv\ and \EC\ are iron-poor compared to solar values: $-0.8$ dex in \lsiv\ and $-2.5$\,dex in \EC. 
Despite this, both stars are extraordinarily enriched in all detectable trans-iron elements up to bismuth. 
Given the consistent enrichment across more than 20 heavy elements in each star, it is likely that all stable elements in this mass range are similarly enhanced. 
The two stars nevertheless show distinct abundance patterns. In \lsiv, abundances rise to $\sim$4.3\,dex above solar up to the lanthanides, declining to $\sim$3\,dex at higher atomic numbers. In contrast, \EC\ exhibits a steady increase with atomic number, from $\sim$1\,dex in Ga to peak enrichments of 6.2 and 5.4\,dex in Pb and Bi, respectively. Although \lsiv\ is known as a ``zirconium star'', it does show $\sim$3\,dex enrichment even in the heaviest elements, thus lower than \EC. These systematic differences indicate that neutron capture operated under different physical conditions in the two stars.

The abundance pattern of \EC\ is remarkably well reproduced by the late hot-flasher models of \citet{Battich2025}. In this scenario, a star is stripped to a thin hydrogen envelope near the RGB tip before undergoing a delayed helium flash, during which hydrogen-rich material is ingested into the helium-burning region, enabling $i$-process nucleosynthesis. 
This scenario is consistent with its nature as binary system with an sdF-type companion on an orbital period of 457 days \citep{Dorsch2021}. 
Most predicted abundances agree with the observations within 0.5\,dex, with larger deviations (1.5\,dex) confined to Pb and Bi. This close agreement strongly supports self-enrichment through $i$-process nucleosynthesis as the origin of the heavy-metal pattern in \EC. The survival of a clear signature of nucleosynthesis, despite the possible action of diffusion, demonstrates that surface abundances in heavy-metal subdwarfs preserve a measurable imprint of their formation channel. 

The abundance pattern of \lsiv\ closely resembles that of \EC\ up to approximately Mo, but shows systematically lower abundances for heavier elements. We conclude that \lsiv\ also experienced $i$-process nucleosynthesis, most likely during its formation as a merger product \citep{MillerBertolami2022}, where hydrogen ingestion into a helium-burning layer can occur under conditions distinct from the late hot-flasher scenario. This demonstrates that the $i$-process can operate in multiple post-RGB evolutionary pathways\footnote{Note that not all lead-rich heavy-metal hot subdwarfs are found in binary systems like \EC, meaning that mergers may be able to produce similar surface compositions. }. 

Several avenues for future work follow naturally from this study. From an atomic physics perspective, oscillator strengths remain unavailable for several detected ions, including \ion{Rb}{iv}, \ion{Cd}{iv}, \ion{I}{iv}, \ion{Ta}{iv}, and \ion{Tl}{v}, while data are incomplete for \ion{As}{iv}, \ion{Se}{iv}, and \ion{Sn}{v}. Experimental wavelengths are also lacking for a number of important ions, such as \ion{Rb}{v}, \ion{Ru}{iv-v}, \ion{Rh}{iv-v}, and partially for \ion{Se}{iii} and \ion{Mo}{iii}, while \ion{Sm}{iv}, \ion{Eu}{iv}, \ion{Dy}{iv}, and \ion{Ho}{iv} lack both oscillator strengths and wavelength measurements (see Table \ref{tab:ion_detection} for a summary). 
Photoionisation cross sections for stages \textsc{iii-vi} are presently only available for lead; new data for important species such as Ga, Ge, Zr, and Sn would be valuable to improve the accuracy of future abundance measurements by including non-LTE effects. 
In addition, unresolved hyperfine splitting and isotopic shifts may still influence measured line strengths; improved atomic data, including nuclear magnetic moments, magnetic dipole constants, and isotopic shifts, would benefit future modelling. A future study based on the same HST dataset used here will determine the isotopic composition of lead in \EC\ and \HD, providing a direct observational test of $i$-process nucleosynthesis predictions.

Further constraints on the origin of the heavy-metal enrichment in \EC\ may be obtained by analysing the surface composition of its sdF-type companion, which likely preserves the system's initial abundance pattern. To date, only its overall metallicity and $\alpha$-enhancement have been measured \citep{Dorsch2021} and a more detailed abundance analysis, including heavy elements such as Ba, would be valuable.

More broadly, our results suggest that heavy-element self-enrichment via the $i$-process may be more widespread among He-rich hot subdwarfs than currently recognised, but often remains hidden owing to the lack of ultraviolet spectroscopy. Extremely He-rich sdO stars, thought to form predominantly through mergers similar to \lsiv\ \citep{Zhang2012}, may undergo comparable nucleosynthetic processing that is undetectable in optical spectra. Heavy-metal hot subdwarfs therefore provide unique empirical laboratories for studying $i$-process nucleosynthesis under well-constrained evolutionary conditions. As some He-rich subdwarfs are sufficiently massive to evolve into helium giants and possibly toward the AGB, they may also represent an unrecognised channel for returning $i$-process material to the interstellar medium. Cooler extreme helium stars should likewise be examined for signatures of trans-iron element enrichment. 
Finally, nucleosynthesis models including $i$-process neutron capture should be applied to both double He-WD mergers and He-WD + hybrid He/CO-WD mergers. 

\section*{Data Availability}

The updated \synspec\ 51 version and the line list used here are available at \url{https://github.com/mattidorsch/synspec51_fork}. 
Plots of the full HST-STIS/E140M spectra of \lsiv\ and \EC\ with their best-fit models, as well as the Pb model atoms are available at \url{https://zenodo.org/records/20073988}.

\begin{acknowledgements}
We thank Claudio Mendoza for his preliminary \ion{Ge}{iii} data. 
MD was supported by the Deutsches Zentrum für Luft- und Raumfahrt (DLR) through grant 50-OR-2304. JD is supported by the FWO and F.R.S.-FNRS under the Excellence of Science (EOS) programme (numbers O.0004.22 and O022818F). PQ is F.R.S.-FNRS Research Director. 
The Armagh Observatory and Planetarium is funded by direct grant from the Northern Ireland Department for Communities.
DJD thanks the Science and Technology Facilities Council (STFC) of the UK Research and Innovation (UKRI) body for their support through his studentship (Project Code: ST/Y509504/1).
LJAS gratefully acknowledges UKRI for support in the form of a Frontier Research grant under the UK government’s ERC Horizon Europe funding guarantee (SYMPHONY; PI Bowman; grant number: EP/Y031059/1).
This research is based on observations made with the NASA/ESA Hubble Space Telescope obtained from the Space Telescope Science Institute, which is operated by the Association of Universities for Research in Astronomy, Inc., under NASA contract NAS 5-26555. These observations are associated with program(s) 17072. 
The TOSS service (\url{http://dc.g-vo.org/TOSS}) used for this paper was constructed as part of the activities of the German Astrophysical Virtual Observatory. 
This research has made use of NASA's Astrophysics Data System. 

\end{acknowledgements}

\bibliographystyle{aa}
\bibliography{hst_heavy}

\begin{appendix}

\section{Additional material}

\subsection{Atomic structure models}
\label{sect:atomic}

\begin{table}[!h]
\caption{Atomic structure models for \ion{As}{iii}, \ion{Se}{iii}, \ion{Hf}{iv}, and \ion{Tl}{iv}. Braces indicate multiple configurations generated by coupling the enclosed orbitals with those outside the braces.}
\label{tab:atomic_conf}
\centering
\begin{tabular}{lp{0.8\columnwidth}}
\toprule
\toprule
Ion & \multicolumn{1}{c}{Details}\\
\midrule
\ion{As}{iii} & \textit{Model:} 4s$^2$\{4p, 4d, 5s, 5p, 5d, 5f, 6s, 6p, 6d, 6f\}, 4s4p\{4d, 4f, 5s, 5p, 5d, 5f, 6s, 6p, 6d, 6f\}, 4s4d\{4f, 5s\}, 4s\{4p$^2$, 4d$^2$, 4f$^2$\}, 4p\{4d$^2$, 4f$^2$\}, 4p$^2$4d, 4p$^3$\\
              & \textit{CPOL:} \ion{As}{vi} core, $\alpha_d = 0.54$ $a_0^3$ \citep{Fraga1976}, $r_c = 0.66$ $a_0$\\
              & \textit{Refs:} Experimental energy levels from \citet{Lang1928}; model similar to the one used by \citet{Rauch2016b} for \ion{Kr}{vi}\\
\midrule
\ion{Se}{iii} & \textit{Model:} 4s$^2$4p$^2$, 4s$^2$4p\{4d, 4f, 5s, 5p, 5d, 5f, 5g, 6s, 6p, 6d, 6f, 7s, 7d, 8s\}, 4s4p$^2$\{4d, 4f, 5s, 5p, 5d, 5f, 6s, 6p, 6d, 6f\}, 4s\{4p4d4f, 4p4d5s, 4p4d$^2$, 4p4f$^2$, 4p$^3$\}, 4p$^2$\{4d$^2$, 4f$^2$\}, 4p$^3$\{4d, 4f, 5s\}, 4p$^4$\\
              & \textit{CPOL:} \ion{Se}{vii} core, $\alpha_d = 0.36$ $a_0^3$ \citep{Johnson1983}, $r_c = 0.62$ $a_0$ \\
              & \textit{Refs:} Experimental energy levels from \citet{Tauheed2012}; model similar to the one used by \citet{Rauch2016b} for \ion{Kr}{v}\\
\midrule
\ion{Hf}{iv} & \textit{Model:} 4f$^{14}$\{5d, 6s, 6p, 6d, 6f, 6g, 7s, 7p, 7d, 7f, 7g, 8s, 8p, 8d, 8f, 8g, 9s, 9p, 9d, 9f, 8g, 10s, 10p, 10d, 10f, 10g\}, 4f$^{13}$\{5d6s, 5d6p, 6s$^2$, 6s6p\}\\
             & \textit{CPOL:} \ion{Hf}{vi} core, $\alpha_d = 3.48$ $a_0^3$ \citep[extrapolated from][]{Fraga1976}, $r_c = 0.63$ $a_0$\\
             & \textit{Refs:} Experimental energy levels from \citet{Klinkenberg1961} and \citet{Sugar1974_HfIV}; model similar to the one used by \citet{Quinet2004} for \ion{Lu}{iii}\\
\midrule
\ion{Tl}{iv} & \textit{Model:} 5d$^{10}$, 5d$^9$\{5f, 6s, 6p, 6d, 6f, 7s, 7p, 7d, 8s\}, 5d$^8$\{6s$^2$, 6s5f, 6s6p, 6s6d, 6s6f, 6s7s, 6s7p, 6p$^2$, 7s$^2$, 7p$^2$\}, 5d$^7$\{6s$^2$5f, 6s$^2$6p, 6s$^2$6d, 6s$^2$6f, 6s6p$^2$\}\\
             & \textit{CPOL:} \ion{Tl}{vii} core, $\alpha_d = 3.25$ $a_0^3$ \citep[extrapolated from][]{Fraga1976}, $r_c = 1.25$ $a_0$\\
             & \textit{Refs:} Experimental energy levels from \citet{Joshi1990} and \citet{Wyart1992_TlIV_PbV_BiVI}; model similar to the one used by \citet{Fivet2006} for \ion{Au}{ii}\\
\bottomrule
\end{tabular}
\end{table}

\begin{table}[!h]
\begin{minipage}{1\columnwidth}
\setstretch{1.1}
\caption{Non-LTE model atoms used in the \tlusty\ atmospheres for \lsiv\ and \EC, listing the number of explicit levels (L) and superlevels (SL). Each element also includes a higher ionisation stage as a one-level atom.}
\vspace{-1pt}
\label{tab:modelatoms}
\centering
\begin{threeparttable}
\begin{center}

\begin{minipage}{0.32\columnwidth}
\centering
\begin{tabular}{lcc}
\toprule
Ion & L & SL \\
\midrule
H\,\textsc{i}    & 17 & $-$\\
He\,\textsc{i}   & 24 & $-$\\
He\,\textsc{ii}  & 20 & $-$\\
C\,\textsc{ii}   & 34 & \phantom{1}5\\
C\,\textsc{iii}  & 34 & 12\\
C\,\textsc{iv}   & 35 & \phantom{1}2\\
N\,\textsc{ii}   & 32 & 10\\
N\,\textsc{iii}  & 40 & \phantom{1}8\\
N\,\textsc{iv}   & 34 & 14\\
N\,\textsc{v}    & 21 & \phantom{1}4\\
O\,\textsc{ii}   & 36 & 12\\
O\,\textsc{iii}  & 28 & 13\\
O\,\textsc{iv}   & 31 & \phantom{1}8\\
O\,\textsc{v}    & 35 & \phantom{1}5\\
O\,\textsc{vi}   & 15 & \phantom{1}5\\
\bottomrule
\end{tabular}
\end{minipage}
\begin{minipage}{0.32\columnwidth}
\centering
\begin{tabular}{lcc}
\toprule
Ion & L & SL \\
\midrule
Ne\,\textsc{ii}  & 23 & \phantom{1}9\\
Ne\,\textsc{iii} & 22 & 12\\
Ne\,\textsc{iv}  & 10 & \phantom{1}2\\
Mg\,\textsc{ii}  & 21 & \phantom{1}4\\
Mg\,\textsc{iii} & 37 & \phantom{1}3\\
Mg\,\textsc{iv}  & 29 & \phantom{1}5\\
Al\,\textsc{ii}  & 20 & \phantom{1}9\\
Al\,\textsc{iii} & 19 & \phantom{1}4\\
Si\,\textsc{ii}  & 23 & 10\\
Si\,\textsc{iii} & 31 & 15\\
Si\,\textsc{iv}  & 19 & \phantom{1}4\\
P\,\textsc{iv}   & 14 & $-$\\
P\,\textsc{v}    & 13 & \phantom{1}4\\
S\,\textsc{ii}   & 23 & 10\\
S\,\textsc{iii}  & 29 & 12\\
\bottomrule
\end{tabular}
\end{minipage}
\begin{minipage}{0.32\columnwidth}
\centering
\begin{tabular}{lcc}
\toprule
Ion & L & SL \\
\midrule
S\,\textsc{iv}   & 33 & \phantom{1}5\\
S\,\textsc{v}    & 20 & \phantom{1}5\\
S\,\textsc{vi}   & 13 & \phantom{1}3\\
Fe\,\textsc{ii}  & 35 & $-$\\
Fe\,\textsc{iii} & 50 & $-$\\
Fe\,\textsc{iv}  & 43 & $-$\\
Fe\,\textsc{v}   & 42 & $-$\\
Ni\,\textsc{iii} & 36 & $-$\\
Ni\,\textsc{iv}  & 38 & $-$\\
Ni\,\textsc{v}   & 48 & $-$\\
Pb\,\textsc{iii} & 20 & $-$\\
Pb\,\textsc{iv}  & 30 & $-$\\
Pb\,\textsc{v}   & 20 & $-$\\
Pb\,\textsc{vi}  & 40 & $-$\\
 &  &  \\
\bottomrule
\end{tabular}
\end{minipage}

\end{center}
\end{threeparttable}
\end{minipage}
\end{table}

\subsection{Hyperfine structure}
\label{sect:HFS}

The UV spectra of \lsiv\ and \EC\ feature strong resonance lines originating from low-lying states of heavy ions. Ground states, particularly $s$-configurations, exhibit large magnetic dipole hyperfine constants ($A$) due to substantial electron density at the nucleus. 
For isotopes with non-zero nuclear spin $I$, this can result in significant hyperfine splitting (HFS) of spectral lines involving these ground or low-lying $s$-states. 
In earlier UV studies of hot subdwarfs, HFS was considered only for the \ion{Pb}{iv}\,1313\,\AA\ resonance line, in attempts to constrain the $^{207}$Pb/$^{208}$Pb isotopic ratio \citep{O'Toole2007}; this is feasible because $^{207}$Pb exhibits a substantial HFS-induced triplet split. Beyond such wavelength shifts, HFS also increases line equivalent widths making it important to account for in abundance analyses. 

To model this splitting, we redistribute the original fine-structure line strength among the hyperfine components using Wigner 6j symbols \citep{Wigner1931_gruppentheorie}, as implemented in \textsc{SymPy} \citep{SymPy_2017}. 
In cases where experimentally measured wavelengths of the split components are not available, the hyperfine $A$ constants of the lower and upper levels were adjusted to reproduce the observed line splits. As first described by \citet{Fermi1930_HFS}, the hyperfine energy\footnote{The electric quadrupole term ($B$ constant) can be neglected because it is typically two orders of magnitude smaller than the magnetic dipole term and vanishes entirely for states with $J=1/2$.} due to the magnetic dipole interaction is $E_\mathrm{hfs} = K\cdot(A/2)$, where  $K = F(F+1) - I(I+1) - J(J+1)$ is the coupling strength between the nuclear spin $I$ and electronic angular momentum $J$, and $F$ is the total atomic angular momentum quantum number, ranging from $I + J$ to $|I - J|$. 
This implies that significant HFS splits require at least three conditions to be met: $J$ > $0$, $I$ > 0, and a significant hyperfine constant $A$, which is particularly large for low-lying $s$-orbital states. 
Even-$Z$, even-neutron isotopes have $I$ = 0 due to nucleon pairing and thus exhibit no HFS. 

\subsection{Unidentified lines}
\label{sect:unid}

Although many lines in \EC\ are attributable to heavy-metal transitions, nearly 200 remain unidentified. Because the \lsiv\ spectrum is more crowded and likely contaminated by unclassified iron-group lines, we restrict our analysis to unidentified features detected in \EC. Given its low iron abundance, these are strong candidates for heavy-metal transitions, including unclassified lines of known ions or transitions from yet unidentified species. Some broader features may instead arise from very high-lying \ion{C}{iii} transitions, which are prominent in \EC.
A list of unidentified lines with equivalent widths greater than 5\,m\AA\ is provided in Table \ref{tab:EC:unid}, measured from Gaussian fits to the normalised spectrum. Lines are classified as sharp (``s'') or broad (``b'').

\begin{table*}
    \centering
    \caption{Abundances of \lsiv\ and \EC\ from HST/STIS-E140M and optical data, stated as mass fraction, number fraction, and number fraction relative to solar \citep{asplund09,Asplund2021,Grevesse2015,Lodders2019}. Uncertainties are rough reliability estimates. 
    Upper limits are given at the best-fit abundance; their uncertainty indicates clearly too strong lines. 
    }
    \vspace{-3pt}
    \label{tab:abu}
\begin{tabular}{lrrrrrr}
\toprule
\toprule
 \hspace{-2pt}  & \multicolumn{2}{c}{Mass fraction ($\log X$)}& \multicolumn{2}{c}{Number fraction ($\log n/\sum_i n_i$)} & \multicolumn{2}{c}{Abundance ($\log n  / n _\mathrm{\odot}$)}\\
\cmidrule(lr){2-3}\cmidrule(lr){4-5}\cmidrule(lr){6-7}
Element \hspace{-10pt} & \small{\lsiv} & \small{\EC}  & \small{\lsiv} & \small{\EC} & \small{\lsiv} & \small{\EC} \\
\midrule
H & $-0.31\pm0.04$ & $-0.65\pm0.03$ & $-0.10\pm0.02$ & $-0.26\pm0.02$ & $-0.06\pm0.02$ & $-0.23\pm0.02$ \\
He & $-0.30\pm0.04$ & $-0.13\pm0.01$ & $-0.70\pm0.08$ & $-0.34\pm0.03$ & $0.41\pm0.08$ & $0.76\pm0.03$ \\
C & $-2.88\pm0.12$ & $-1.79\pm0.10$ & $-3.75\pm0.12$ & $-2.48\pm0.10$ & $-0.14\pm0.12$ & $1.12\pm0.11$ \\
N & $-3.00\pm0.07$ & $-2.92\pm0.15$ & $-3.94\pm0.06$ & $-3.68\pm0.15$ & $0.27\pm0.08$ & $0.52\pm0.16$ \\
O & $-3.62\pm0.11$ & $-2.66\pm0.10$ & $-4.61\pm0.10$ & $-3.48\pm0.10$ & $-1.27\pm0.11$ & $-0.14\pm0.11$ \\
Ne & $-3.48\pm0.07$ & <$-3.74^{+0.40}_{}$ & $-4.57\pm0.06$ & <$-4.66^{+0.40}_{}$ & $-0.47\pm0.11$ & <$-0.56^{+0.40}_{}$ \\
Na &  & <$-4.17^{+0.30}_{}$ &  & <$-5.15^{+0.30}_{}$ &  & <$0.65^{+0.30}_{}$ \\
Mg & $-4.30\pm0.05$ & $-4.16\pm0.25$ & $-5.47\pm0.03$ & $-5.16\pm0.25$ & $-1.04\pm0.05$ & $-0.72\pm0.26$ \\
Al &  & $-5.46\pm0.20$ &  & $-6.51\pm0.20$ &  & $-0.92\pm0.20$ \\
Si & $-4.89\pm0.08$ & $-4.51\pm0.20$ & $-6.13\pm0.07$ & $-5.57\pm0.20$ & $-1.60\pm0.08$ & $-1.05\pm0.20$ \\
P & $-6.10\pm0.15$ & $-4.64\pm0.30$ & $-7.39\pm0.15$ & $-5.74\pm0.30$ & $-0.76\pm0.15$ & $0.88\pm0.30$ \\
S & <$-5.67^{+0.30}_{}$ & $-4.66\pm0.25$ & <$-6.96^{+0.30}_{}$ & $-5.78\pm0.25$ & <$-2.05^{+0.30}_{}$ & $-0.87\pm0.25$ \\
Ar & $-4.30\pm0.21$ & <$-4.59^{+0.60}_{}$ & $-5.70\pm0.20$ & <$-5.80^{+0.60}_{}$ & $-0.06\pm0.24$ & <$-0.17^{+0.60}_{}$ \\
Ca &  & <$-4.71^{+0.40}_{}$ &  & <$-5.93^{+0.40}_{}$ &  & <$-0.23^{+0.40}_{}$ \\
Ti & $-4.70\pm0.30$ & $-6.21\pm0.30$ & $-6.17\pm0.30$ & $-7.50\pm0.30$ & $0.92\pm0.30$ & $-0.42\pm0.30$ \\
Cr & $-4.57\pm0.30$ & $-5.88\pm0.40$ & $-6.08\pm0.30$ & $-7.21\pm0.40$ & $0.32\pm0.30$ & $-0.82\pm0.41$ \\
Mn & $-5.29\pm0.41$ & <$-6.31^{+0.40}_{}$ & $-6.82\pm0.40$ & <$-7.66^{+0.40}_{}$ & $-0.21\pm0.41$ & <$-1.06^{+0.40}_{}$ \\
Fe & $-3.76\pm0.15$ & $-5.71\pm0.35$ & $-5.30\pm0.15$ & $-7.07\pm0.35$ & $-0.76\pm0.15$ & $-2.54\pm0.35$ \\
Co & $-4.81\pm0.26$ & <$-6.78^{+0.40}_{}$ & $-6.37\pm0.25$ & <$-8.16^{+0.40}_{}$ & $0.68\pm0.26$ & <$-1.12^{+0.40}_{}$ \\
Ni & $-3.49\pm0.14$ & $-5.47\pm0.25$ & $-5.05\pm0.13$ & $-6.86\pm0.25$ & $0.77\pm0.14$ & $-1.04\pm0.26$ \\
Cu & $-4.07\pm0.21$ & <$-5.65^{+0.40}_{}$ & $-5.67\pm0.20$ & <$-7.06^{+0.40}_{}$ & $2.18\pm0.21$ & <$0.78^{+0.40}_{}$ \\
Zn & $-3.66\pm0.09$ & $-5.49\pm0.30$ & $-5.27\pm0.08$ & $-6.92\pm0.30$ & $2.21\pm0.09$ & $0.55\pm0.30$ \\
Ga & $-4.14\pm0.21$ & $-6.17\pm0.30$ & $-5.78\pm0.20$ & $-7.62\pm0.30$ & $3.22\pm0.22$ & $1.37\pm0.32$ \\
Ge & $-3.89\pm0.11$ & $-5.68\pm0.35$ & $-5.55\pm0.10$ & $-7.15\pm0.35$ & $2.84\pm0.14$ & $1.23\pm0.38$ \\
As & $-4.75\pm0.21$ &  & $-6.42\pm0.20$ &  & $3.32\pm0.21$ &  \\
Se & $-3.41\pm0.30$ & $-5.50\pm0.40$ & $-5.10\pm0.30$ & $-7.01\pm0.40$ & $3.60\pm0.30$ & $1.68\pm0.40$ \\
Kr & $-3.33\pm0.11$ & <$-4.83^{+0.40}_{}$ & $-5.05\pm0.11$ & <$-6.37^{+0.40}_{}$ & $3.74\pm0.12$ & <$2.41^{+0.40}_{}$ \\
Sr & $-2.83\pm0.21$ & $-4.89\pm0.40$ & $-4.57\pm0.20$ & $-6.44\pm0.40$ & $4.60\pm0.21$ & $2.72\pm0.41$ \\
Y & $-3.76\pm0.04$ & $-5.79\pm0.40$ & $-5.50\pm0.01$ & $-7.35\pm0.40$ & $4.33\pm0.05$ & $2.47\pm0.41$ \\
Zr & $-3.17\pm0.10$ & $-5.11\pm0.30$ & $-4.92\pm0.09$ & $-6.68\pm0.30$ & $4.54\pm0.10$ & $2.77\pm0.30$ \\
Nb & $-4.38\pm0.30$ & <$-6.05^{+0.40}_{}$ & $-6.14\pm0.30$ & <$-7.63^{+0.40}_{}$ & $4.44\pm0.30$ & <$2.95^{+0.40}_{}$ \\
Mo & $-4.17\pm0.30$ & $-5.99\pm0.50$ & $-5.95\pm0.30$ & $-7.58\pm0.50$ & $4.21\pm0.31$ & $2.57\pm0.52$ \\
Pd & $-4.50\pm0.41$ & <$-5.52^{+0.50}_{}$ & $-6.32\pm0.40$ & <$-7.16^{+0.50}_{}$ & $4.17\pm0.41$ & <$3.32^{+0.50}_{}$ \\
Ag & <$-5.27^{+0.50}_{}$ &  & <$-7.10^{+0.50}_{}$ &  & <$3.98^{+0.50}_{}$ &  \\
In & $-5.25\pm0.41$ & <$-6.79^{+0.50}_{}$ & $-7.10\pm0.40$ & <$-8.46^{+0.50}_{}$ & $4.14\pm0.47$ & <$2.77^{+0.51}_{}$ \\
Sn & $-3.79\pm0.08$ & $-5.35\pm0.30$ & $-5.65\pm0.07$ & $-7.04\pm0.30$ & $4.34\pm0.12$ & $2.95\pm0.33$ \\
Sb & $-5.62\pm0.51$ & $-6.75\pm0.30$ & $-7.50\pm0.50$ & $-8.45\pm0.30$ & $3.53\pm0.51$ & $2.57\pm0.30$ \\
Te & $-5.19\pm0.51$ & $-5.78\pm0.50$ & $-7.09\pm0.50$ & $-7.50\pm0.50$ & $2.77\pm0.50$ & $2.35\pm0.50$ \\
Xe & $-4.84\pm0.60$ & <$-5.88^{+0.50}_{}$ & $-6.75\pm0.60$ & <$-7.61^{+0.50}_{}$ & $3.05\pm0.61$ & <$2.18^{+0.50}_{}$ \\
La & <$-4.26^{+0.30}_{}$ & $-5.39\pm0.50$ & <$-6.20^{+0.30}_{}$ & $-7.14\pm0.50$ & <$4.73^{+0.30}_{}$ & $3.78\pm0.51$ \\
Ce & $-6.11\pm0.51$ & $-4.61\pm0.35$ & $-8.05\pm0.50$ & $-6.37\pm0.35$ & $2.41\pm0.51$ & $4.08\pm0.35$ \\
Pr & <$-5.16^{+0.30}_{}$ & $-5.92\pm0.50$ & <$-7.10^{+0.30}_{}$ & $-7.68\pm0.50$ & <$4.22^{+0.30}_{}$ & $3.63\pm0.51$ \\
Nd & <$-5.67^{+0.60}_{}$ & $-5.02\pm0.40$ & <$-7.62^{+0.60}_{}$ & $-6.79\pm0.40$ & <$3.00^{+0.60}_{}$ & $3.82\pm0.41$ \\
Er &  & $-6.21\pm0.50$ &  & $-8.04\pm0.50$ &  & $3.06\pm0.51$ \\
Yb & <$-5.67^{+0.50}_{}$ & $-5.02\pm0.45$ & <$-7.70^{+0.50}_{}$ & $-6.87\pm0.45$ & <$3.49^{+0.50}_{}$ & $4.31\pm0.47$ \\
Lu & <$-5.76^{+0.30}_{}$ & $-6.13\pm0.50$ & <$-7.80^{+0.30}_{}$ & $-7.98\pm0.50$ & <$4.14^{+0.31}_{}$ & $3.95\pm0.52$ \\
Hf & $-6.18\pm0.51$ & $-4.77\pm0.35$ & $-8.23\pm0.50$ & $-6.63\pm0.35$ & $2.96\pm0.51$ & $4.55\pm0.35$ \\
Ta & $-6.71\pm0.51$ & $-5.88\pm0.50$ & $-8.76\pm0.50$ & $-7.75\pm0.50$ & $3.35\pm0.67$ & $4.35\pm0.67$ \\
W & $-6.28\pm0.51$ & $-5.34\pm0.35$ & $-8.34\pm0.50$ & $-7.22\pm0.35$ & $2.87\pm0.53$ & $3.98\pm0.38$ \\
Os & <$-6.20^{+0.50}_{}$ & $-5.55\pm0.50$ & <$-8.27^{+0.50}_{}$ & $-7.44\pm0.50$ & <$2.37^{+0.50}_{}$ & $3.19\pm0.51$ \\
Ir & <$-4.32^{+0.50}_{}$ &  & <$-6.40^{+0.50}_{}$ &  & <$4.22^{+0.50}_{}$ &  \\
Pt &  & $-5.20\pm0.50$ &  & $-7.10\pm0.50$ &  & $3.19\pm0.58$ \\
Hg & $-5.73\pm0.35$ & $-4.73\pm0.35$ & $-7.83\pm0.35$ & $-6.64\pm0.35$ & $3.04\pm0.42$ & $4.22\pm0.42$ \\
Tl & $-6.57\pm0.51$ & $-4.98\pm0.50$ & $-8.67\pm0.50$ & $-6.90\pm0.50$ & $2.47\pm0.58$ & $4.23\pm0.58$ \\
Pb & $-5.09\pm0.26$ & $-2.14\pm0.30$ & $-7.20\pm0.25$ & $-4.07\pm0.30$ & $3.09\pm0.28$ & $6.21\pm0.33$ \\
Bi & $-6.99\pm0.51$ & $-4.07\pm0.50$ & $-9.10\pm0.50$ & $-6.00\pm0.50$ & $2.29\pm0.51$ & $5.38\pm0.51$ \\
\bottomrule
\end{tabular}
\end{table*}

\begin{table}[!h]
\caption{Availability of atomic data and detection by ionisation stage in \EC\ and \lsiv.}
\label{tab:ion_detection}
\centering
\begin{tabular}{lllll}
\toprule
\toprule
  & \textsc{iii} & \textsc{iv} & \textsc{v} & \textsc{vi} \\
\midrule
Ga & $\detboth$ & $\detboth$ & $\nodet$  & \\
Ge & $\detLS$ & $\detboth$ & $\detLS$ & \\
As & $\detLS$ & $\detLS^\mathrm{f}$ & $\nodet$ & \\
Se & $\detLS^\mathrm{w}$ & $\detLS^\mathrm{f}$ & $\detboth$ & \\
Br & $\nodet$ & $\detLS$ & $\nodet$ & \\
Kr & $\detLS$ & $\detLS$ & $\detLS$ & \\
Rb &  $\nodet^\mathrm{w}$ & $\detLS^\mathrm{f}$ & $\nodet^\mathrm{w}$ & \\
Sr & $\detLS$ & $\detboth$ & $\detLS$ & \\
Y & $\detLS$ & $\detboth$ & $\nodet$ & \\
Zr & $\detLS$ & $\detboth$ & $\detLS$ & \\
Nb &  $\nodet$ & $\detLS$ & $\detLS$ & \\
Mo & $\nodet^\mathrm{w}$ & $\detLS$ & $\detboth$ & \\
Ru & & $\nodet^\mathrm{w}$ & $\nodet^\mathrm{w}$ & \\
Rh & & $\nodet^\mathrm{w}$ & $\nodet^\mathrm{w}$ & \\
Pd & & $\detLS$ & $\nodet$ & \\
Ag & & $\detLS$ & $\nodet^\mathrm{f}$ & \\
Cd & & $\detLS^\mathrm{f}$ & $\nodet^\mathrm{f}$ & \\
In & $\detLS$ & $\detLS$ & $\nodet$ & \\
Sn & $\detboth$ & $\detboth$ & $\detLS^\mathrm{f}$ & \\
Sb & $\nodet$ & $\detLS$ & $\detboth$ & \\
Te & & $\detboth^\mathrm{f}$ & $\detboth$ &  $\nodet$ \\
I & & $\nodet^\mathrm{f}$ & $\detLS$ & \\
Xe & & $\detLS$ & $\detLS$ & \\
Cs & & $\nodet^\mathrm{f}$ & $\nodet$ & \\
Ba & & $\nodet$ & $\nodet$ & \\
La & & $\detEC$ & $\nodet$ & \\
\bottomrule
\end{tabular}
\hspace{1pt}
\begin{tabular}{lllll}
\toprule
\toprule
  & \textsc{iii} & \textsc{iv} & \textsc{v} & \textsc{vi} \\
\midrule
Ce & & $\detboth$ & $\detEC$ & \\
Pr & & $\detEC$ & $\nodet$ & \\
Nd & & $\detEC$ & $\nodet$ & \\
Sm & & $\nodet^\mathrm{wf}$ & & \\
Eu & & $\nodet^\mathrm{wf}$ & & \\
Gd & & $\nodet^\mathrm{f}$ & & \\
Tb & & $\nodet^\mathrm{f}$ & & \\
Dy & & $\nodet^\mathrm{wf}$ & $\nodet^\mathrm{wf}$  & \\
Ho & & $\nodet^\mathrm{wf}$ & $\nodet^\mathrm{wf}$ & \\
Er & & $\detEC$ & & \\
Tm & & $\nodet$ & & \\
Yb & & $\detEC$ & $\nodet$ & \\
Lu & & $\detEC$ & $\nodet$ & \\
Hf & & $\detboth$ & $\detEC$ & \\
Ta & & $\detEC^\mathrm{f}$ & $\detboth$ & $\nodet$ \\
W & & $\detEC$ & $\detboth$ & $\detEC$ \\
Re & & $\nodet$ & $\nodet$ & \\
Os & & $\detEC$ & $\detEC$ & \\
Ir & & $\nodet$ & $\nodet^\mathrm{f}$ & \\
Pt & & $\detEC$ & $\detEC$ & $\detEC$ \\
Au & $\nodet$ & $\nodet$ & $\nodet^\mathrm{f}$ & \\
Hg & & $\detboth$ & $\detEC$ & $\nodet^\mathrm{wf}$ \\
Tl & $\nodet$ & $\detboth$ & $\detEC^\mathrm{f}$ & $\nodet^\mathrm{f}$ \\
Pb & $\detEC$ & $\detboth$ & $\detEC$ &  $\detEC$ \\
Bi & $\detEC$ & $\detboth$ & $\detEC$ & \\
Th & $\nodet$ & $\nodet$ & $\nodet^\mathrm{wf}$ & \\
\bottomrule
\end{tabular}
\tablefoot{
Detection:  $\detboth$ = in both stars; $\detLS$ = \lsiv\ only; $\detEC$ = \EC\ only; $\nodet$ = not detected. 
Atomic data requirements: $^\mathrm{w}$ = needs wavelengths; $^\mathrm{f}$ = needs oscillator strengths; $^\mathrm{wf}$ = needs both. 
The radioactive Tc and Pm are not listed, but are missing accurate wavelengths. }
\end{table}

\begin{table}
\caption{
\ion{Se}{iv} lines identified in \lsiv\ using energy levels from \cite{Kelly1987} with identifications. 
Observed wavelengths $\lambda_\mathrm{obs}$ were measured using the STIS/E140M spectrum of \lsiv. 
}
\label{tab:seiv}
\centering
\setstretch{1.1}
\resizebox{\columnwidth}{!}{
\begin{tabular}{clrrr}
\toprule
\toprule
$\lambda_\mathrm{obs}$  / \AA & $\lambda_\mathrm{Ritz}$ / \AA & Lower level & Upper level & $\log gf^\dagger$\\
\midrule
1157.297 & 1157.340 & \termconfig[]{4s}{4p^2}{2}{D}{5/2} & \termconfig[\circ]{4s^2}{5p}{2}{P}{3/2} & $-0.372$\\
1166.795 & 1166.834 & \termconfig[]{4s}{4p^2}{2}{D}{3/2} & \termconfig[\circ]{4s^2}{5p}{2}{P}{1/2} & $-0.690$\\
1259.540 & 1259.519 & \termconfig[\circ]{4s^2}{4p}{2}{P}{1/2} & \termconfig[]{4s}{4p^2}{4}{P}{1/2} \\ %
1262.436 & 1262.416 & \termconfig[\circ]{4s^2}{4p}{2}{P}{3/2} & \termconfig[]{4s}{4p^2}{4}{P}{5/2} \\ %
1307.238 & 1307.241 & \termconfig[]{4s^2}{4d}{2}{D}{3/2} & \termconfig[\circ]{4s^2}{4f}{2}{F}{5/2} & 0.488\\
 & 1314.440$^\ast$ & \termconfig[]{4s^2}{4d}{2}{D}{5/2} & \termconfig[\circ]{4s^2}{4f}{2}{F}{7/2} & 0.546\\ %
1318.131 & 1318.220 & \termconfig[]{4s}{4p^2}{4}{P}{5/2} & \termconfig[\circ]{4p^3}{}{4}{S}{3/2}  \\ %
\bottomrule
\end{tabular}
}
\tablefoot{
$^\dagger$ from lifetime measurements by \cite{Bahr1982_Se}.
$^\ast$ blended with \ion{Sn}{iv} 1314.530\,\AA, so not clearly detected. 
}
\end{table}

\begin{table}
\caption{
Strongest detected lead lines in \lsiv\ and \EC. Asterisk-marked ions indicate lines for which oscillator strengths are unavailable in the literature.
}
\label{tab:idlines}
\centering
\resizebox{1\columnwidth}{!}{
\begin{tabular}{l>{\raggedright\arraybackslash}p{0.985\columnwidth}}
\toprule
\toprule
Ion & \multicolumn{1}{c}{Wavelength / \AA} \\
\midrule
\ion{Pb}{iii} & 1165.025, 1166.940, 1203.479, 1250.425, 1266.773, 1274.542, 1279.420, 1308.088, 1371.744, 1406.485, 1439.308, 1553.004, 1587.834, 1597.823, 1610.199, 1711.065 \\
\ion{Pb}{iii}$^\ast$ & 1220.479, 1275.691, 1290.333, 1309.333, 1321.045, 1327.139, 1355.038, 1385.996, 1404.769, 1581.782, 1593.242, 1668.806, 1699.582 \\
\ion{Pb}{iv} & 1156.269, 1267.533, 1291.076, 1312.936, 1312.958, 1313.018, 1313.052, 1313.057, 1313.059, 1313.069, 1313.076, 1313.126, 1313.148, 1343.062, 1348.854, 1388.958, 1397.045, 1400.269, 1404.346, 1410.041, 1510.772, 1535.723\\
\ion{Pb}{iv}$^\ast$ & 1151.300, 1154.878, 1158.706, 1165.431, 1171.488, 1189.202, 1191.214, 1194.194, 1196.519, 1196.277, 1197.423, 1198.712, 1204.046, 1206.846, 1211.738, 1251.936, 1265.136, 1283.121, 1290.816, 1291.227, 1269.801, 1304.720, 1315.239, 1420.398, 1484.041, 1622.000\\
\ion{Pb}{v} & 1152.375, 1157.871, 1183.733, 1185.441, 1189.955, 1197.712, 1213.195, 1233.492, 1240.069, 1248.465, 1301.343, 1610.506, 1635.734, 1700.500 \\
\ion{Pb}{v}$^\ast$ & 1193.034, 1240.138, 1247.038, 1258.986, 1286.872, 1509.092 \\
\ion{Pb}{vi} & 1151.593, 1162.345, 1164.991, 1184.264, 1201.661, 1222.718, 1252.515, 1277.358, 1280.427, 1331.056, 1332.437, 1335.034, 1402.964, 1468.152 \\
\bottomrule
\end{tabular}
}
\end{table}

\begin{table}
\begin{minipage}{1\textwidth}
\caption{
Positions, equivalent widths, FWHM, and line types for unidentified lines in the STIS spectrum of \EC.
}
\label{tab:EC:unid}
\vspace{-14pt}
\setstretch{1.05}
\begin{center}
\begin{minipage}{0.32\textwidth}
\begin{tabular}{crrr}
\toprule
\toprule
$\lambda$ / \AA\ & EW / m\AA\ & $\Delta\lambda$ / m\AA &  T  \\
\midrule
1147.980 & $34 \pm 6$ & $76 \pm 13$ &  \\
1148.255 & $36 \pm 9$ & $74 \pm 19$ & b \\
1148.324 & $30 \pm 8$ & $51 \pm 12$ & b \\
1150.709 & $34 \pm 8$ & $57 \pm 14$ &  \\
1151.454 & $24 \pm 9$ & $66 \pm 19$ &  \\
1154.085 & $30 \pm 9$ & $86 \pm 6$ &  \\
1155.472 & $40 \pm 11$ & $119 \pm 35$ & b \\
1155.964 & $38 \pm 12$ & $96 \pm 27$ & b \\
1156.657 & $27 \pm 7$ & $64 \pm 18$ & b \\
1159.377 & $37 \pm 11$ & $130 \pm 40$ & b \\
1159.526 & $34 \pm 8$ & $88 \pm 22$ & b \\
1159.917 & $56 \pm 10$ & $138 \pm 25$ &  \\
1164.325 & $32 \pm 6$ & $97 \pm 20$ &  \\
1168.013 & $45 \pm 5$ & $119 \pm 14$ & s \\
1170.533 & $22 \pm 4$ & $52 \pm 9$ & b\\
1171.753 & $35 \pm 5$ & $68 \pm 11$ & \\
1173.239 & $16 \pm 4$ & $76 \pm 18$ & \\
1178.556 & $22 \pm 3$ & $65 \pm 11$ &  \\
1182.570 & $16 \pm 4$ & $53 \pm 14$ & s \\
1183.395 & $12 \pm 3$ & $43 \pm 6$ & s \\
1184.165 & $28 \pm 8$ & $101 \pm 29$ & b \\
1184.261 & $29 \pm 7$ & $72 \pm 15$ & b \\
1185.307 & $35 \pm 5$ & $113 \pm 15$ & b \\
1186.522 & $24 \pm 4$ & $66 \pm 11$ &  \\
1188.302 & $21 \pm 2$ & $57 \pm 6$ & s \\
1192.065 & $28 \pm 4$ & $65 \pm 12$ & s \\
1193.736 & $10 \pm 2$ & $46 \pm 11$ &  \\
1198.167 & $23 \pm 4$ & $74 \pm 12$ & b \\
1198.841 & $13 \pm 3$ & $46 \pm 13$ & b \\
1199.843 & $25 \pm 3$ & $59 \pm 7$ & s \\
1201.289 & $13 \pm 3$ & $53 \pm 14$ &  \\
1203.263 & $7 \pm 2$ & $51 \pm 18$ &  \\
1204.438 & $20 \pm 2$ & $58 \pm 7$ & s \\
1205.736 & $28 \pm 3$ & $99 \pm 10$ & b \\
1209.167 & $17 \pm 3$ & $81 \pm 14$ &  \\
1211.068 & $6 \pm 1$ & $44 \pm 8$ &  \\
1211.192 & $11 \pm 2$ & $63 \pm 9$ &  \\
1220.876 & $12 \pm 3$ & $73 \pm 19$ & b \\
1221.125 & $19 \pm 2$ & $67 \pm 9$ & s \\
1222.099 & $12 \pm 2$ & $70 \pm 13$ &  \\
1223.390 & $17 \pm 2$ & $88 \pm 13$ &  \\
1225.130 & $16 \pm 2$ & $60 \pm 9$ & \\
1225.299 & $13 \pm 3$ & $73 \pm 16$ & b \\
1226.397 & $16 \pm 2$ & $74 \pm 10$ &  \\
1228.850 & $18 \pm 2$ & $66 \pm 7$ &  \\
1229.131 & $17 \pm 2$ & $60 \pm 8$ &  \\
1229.613 & $13 \pm 3$ & $85 \pm 19$ &  \\
1233.146 & $13 \pm 2$ & $61 \pm 11$ & b \\
1233.717 & $16 \pm 2$ & $57 \pm 8$ & b \\
1234.841 & $32 \pm 3$ & $85 \pm 7$ &  \\
1237.154 & $16 \pm 2$ & $75 \pm 8$ &  \\
1242.384 & $18 \pm 1$ & $54 \pm 4$ &  \\
1243.117 & $17 \pm 2$ & $75 \pm 10$ &  \\
1243.772 & $22 \pm 3$ & $82 \pm 13$ & b \\
1243.873 & $10 \pm 3$ & $55 \pm 15$ & b \\
1245.272 & $15 \pm 2$ & $47 \pm 5$ & b \\
1246.175 & $18 \pm 2$ & $74 \pm 7$ &  \\
\bottomrule
\end{tabular}
\end{minipage}
\begin{minipage}{0.32\textwidth}
\begin{tabular}{crrr}
\toprule
\toprule
$\lambda$ / \AA\ & EW / m\AA\ & $\Delta\lambda$ / m\AA &  T  \\
\midrule
1248.630 & $19 \pm 2$ & $127 \pm 16$ &  b \\
1248.750 & $5 \pm 2$ & $60 \pm 18$ & b  \\
1254.031 & $17 \pm 4$ & $99 \pm 23$ & b \\
1254.352 & $9 \pm 2$ & $57 \pm 11$ & b \\
1254.443 & $8 \pm 2$ & $70 \pm 20$ & b \\
1255.434 & $20 \pm 2$ & $67 \pm 7$ & s \\
1256.122 & $8 \pm 2$ & $54 \pm 11$ &  \\
1257.176 & $22 \pm 2$ & $64 \pm 5$ & s \\
1256.903 & $11 \pm 4$ & $84 \pm 26$ &  \\
1256.984 & $8 \pm 3$ & $58 \pm 16$ &  \\
1260.800 & $23 \pm 3$ & $81 \pm 9$ &  \\
1261.001 & $46 \pm 4$ & $139 \pm 10$ &  \\
1262.041 & $8 \pm 2$ & $71 \pm 18$ &  \\
1262.635 & $16 \pm 2$ & $76 \pm 11$ &  \\
1264.235 & $19 \pm 2$ & $71 \pm 6$ &  \\
1264.511 & $56 \pm 2$ & $102 \pm 4$ &  \\
1269.518 & $8 \pm 2$ & $46 \pm 11$ &  \\
1269.915 & $20 \pm 2$ & $68 \pm 7$ & b \\
1270.747 & $13 \pm 1$ & $70 \pm 8$ &  \\
1271.585 & $18 \pm 2$ & $61 \pm 6$ &  \\
1272.416 & $9 \pm 2$ & $53 \pm 12$ &  \\
1272.679 & $16 \pm 2$ & $61 \pm 9$ &  \\
1275.499 & $20 \pm 2$ & $115 \pm 14$ & b \\
1276.875 & $9 \pm 3$ & $63 \pm 19$ &  \\
1278.564 & $8 \pm 2$ & $47 \pm 10$ &  \\
1279.846 & $13 \pm 2$ & $52 \pm 6$ &  \\
1281.455 & $11 \pm 1$ & $69 \pm 7$ &  \\
1282.932 & $13 \pm 2$ & $50 \pm 8$ &  \\
1283.594 & $15 \pm 2$ & $50 \pm 6$ &  \\
1284.904 & $8 \pm 4$ & $49 \pm 21$ & b \\
1288.327 & $26 \pm 2$ & $64 \pm 4$ &  \\
1290.006 & $15 \pm 3$ & $62 \pm 12$ &  \\
1292.146 & $20 \pm 3$ & $78 \pm 11$ &  \\
1294.328 & $12 \pm 2$ & $54 \pm 9$ &  b \\
1294.902 & $41 \pm 3$ & $99 \pm 6$ &  b \\
1295.335 & $13 \pm 2$ & $62 \pm 11$ &  \\
1297.193 & $16 \pm 3$ & $82 \pm 11$ &  b \\
1301.436 & $22 \pm 3$ & $89 \pm 11$ &  b \\
1301.679 & $28 \pm 5$ & $79 \pm 13$ &  \\
1302.462 & $82 \pm 2$ & $131 \pm 4$ &  \\
1303.198 & $12 \pm 2$ & $65 \pm 8$ &  \\
1303.744 & $20 \pm 2$ & $87 \pm 9$ &  \\
1303.981 & $14 \pm 2$ & $64 \pm 8$ &  \\
1305.966 & $11 \pm 2$ & $66 \pm 13$ &  b \\
1306.442 & $8 \pm 3$ & $47 \pm 14$ &  \\
1313.663 & $11 \pm 2$ & $88 \pm 15$ &  \\
1319.454 & $13 \pm 3$ & $80 \pm 16$ &  \\
1320.223 & $5 \pm 1$ & $52 \pm 13$ &  \\
1323.227 & $14 \pm 2$ & $67 \pm 9$ &  \\
1323.536 & $19 \pm 3$ & $90 \pm 13$ &  \\
1324.771 & $14 \pm 2$ & $73 \pm 13$ &  \\
1328.171 & $22 \pm 3$ & $78 \pm 10$ & s \\
1330.775 & $10 \pm 2$ & $77 \pm 11$ &  \\
1331.032 & $13 \pm 2$ & $84 \pm 14$ &  \\
1332.786 & $6 \pm 2$ & $60 \pm 19$ &  \\
1332.909 & $5 \pm 2$ & $45 \pm 17$ &  \\
1336.362 & $9 \pm 1$ & $44 \pm 6$ &  \\
\bottomrule
\end{tabular}
\end{minipage}
\begin{minipage}{0.32\textwidth}
\begin{tabular}{crrr}
\toprule
\toprule
$\lambda$ / \AA\ & EW / m\AA\ & $\Delta\lambda$ / m\AA &  T  \\
\midrule
1340.318 & $16 \pm 2$ & $60 \pm 7$ & b \\
1345.614 & $30 \pm 2$ & $103 \pm 8$ &  \\
1354.249 & $8 \pm 1$ & $52 \pm 6$ &  \\
1358.673 & $32 \pm 2$ & $83 \pm 6$ &  \\
1359.261 & $25 \pm 2$ & $99 \pm 9$ &  \\
1360.672 & $6 \pm 3$ & $84 \pm 38$ & b \\
1360.763 & $7 \pm 2$ & $60 \pm 19$ & b \\
1371.250 & $22 \pm 2$ & $69 \pm 5$ & b \\
1379.887 & $56 \pm 2$ & $113 \pm 5$ &  \\
1380.394 & $17 \pm 2$ & $84 \pm 11$ & b \\
1380.778 & $11 \pm 1$ & $58 \pm 8$ &  \\
1385.810 & $13 \pm 2$ & $77 \pm 9$ & b \\
1388.790 & $5 \pm 1$ & $60 \pm 16$ &  \\
1393.153 & $16 \pm 2$ & $57 \pm 9$ &  \\
1397.596 & $12 \pm 19$ & $53 \pm 26$ &  \\
1401.773 & $7 \pm 2$ & $51 \pm 12$ &  \\
1402.018 & $28 \pm 4$ & $78 \pm 8$ & s \\
1407.607 & $25 \pm 3$ & $73 \pm 7$ & s \\
1408.553 & $12 \pm 2$ & $69 \pm 12$ &  \\
1429.689 & $11 \pm 2$ & $69 \pm 14$ &  \\
1429.916 & $9 \pm 1$ & $44 \pm 5$ &  \\
1439.065 & $16 \pm 2$ & $73 \pm 8$ & b \\
1440.043 & $29 \pm 10$ & $113 \pm 22$ & b \\
1441.058 & $16 \pm 3$ & $60 \pm 8$ & s \\
1445.631 & $12 \pm 2$ & $87 \pm 18$ &  \\
1450.281 & $12 \pm 2$ & $56 \pm 8$ & s \\
1450.747 & $9 \pm 2$ & $75 \pm 16$ &  \\
1451.880 & $10 \pm 2$ & $50 \pm 10$ & s \\
1465.516 & $15 \pm 2$ & $73 \pm 11$ & b \\
1468.391 & $8 \pm 2$ & $44 \pm 11$ &  \\
1472.071 & $25 \pm 5$ & $130 \pm 24$ & b \\
1473.864 & $13 \pm 2$ & $84 \pm 15$ &  \\
1474.146 & $21 \pm 3$ & $133 \pm 19$ &  \\
1476.101 & $31 \pm 5$ & $138 \pm 23$ &  \\
1484.742 & $14 \pm 3$ & $57 \pm 12$ & s \\
1487.327 & $17 \pm 3$ & $81 \pm 17$ &  \\
1493.618 & $11 \pm 2$ & $63 \pm 15$ & b \\
1493.752 & $46 \pm 3$ & $111 \pm 7$ & b \\
1495.174 & $66 \pm 5$ & $173 \pm 13$ & b \\
1501.305 & $27 \pm 15$ & $98 \pm 33$ & b \\
1501.421 & $63 \pm 16$ & $151 \pm 33$ & b \\
1502.559 & $34 \pm 4$ & $120 \pm 15$ & b \\
1502.062 & $30 \pm 8$ & $134 \pm 41$ & b \\
1502.245 & $15 \pm 7$ & $115 \pm 61$ & b \\
1503.531 & $15 \pm 3$ & $60 \pm 11$ & b \\
1509.357 & $11 \pm 7$ & $72 \pm 35$ & b \\
1509.438 & $18 \pm 7$ & $84 \pm 29$ & b \\
1514.978 & $18 \pm 3$ & $73 \pm 12$ &  \\
1527.868 & $27 \pm 16$ & $85 \pm 29$ &  \\
1540.568 & $30 \pm 3$ & $144 \pm 15$ & b \\
1541.872 & $20 \pm 2$ & $92 \pm 10$ & b \\
1557.029 & $51 \pm 7$ & $294 \pm 37$ & b \\
1559.447 & $19 \pm 3$ & $70 \pm 13$ & s \\
1569.962 & $29 \pm 10$ & $143 \pm 47$ & b \\
1570.122 & $17 \pm 10$ & $132 \pm 68$ & b \\
1570.347 & $28 \pm 6$ & $165 \pm 33$ & b \\
1627.008 & $12 \pm 3$ & $40 \pm 10$ & s \\
\bottomrule
\end{tabular}
\end{minipage}
\end{center}
\end{minipage}
\end{table}

\begin{onecolumn}
\FloatBarrier

\begin{figure}[!h]
\centering
\includegraphics[width=0.99\textwidth]{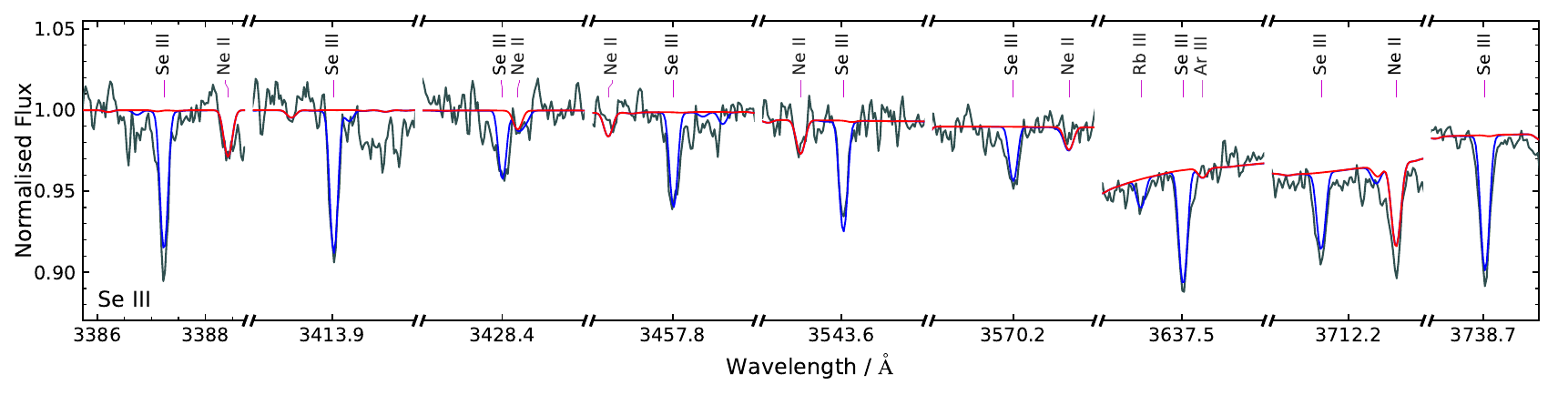}\\[-17pt]
\includegraphics[width=0.99\textwidth]{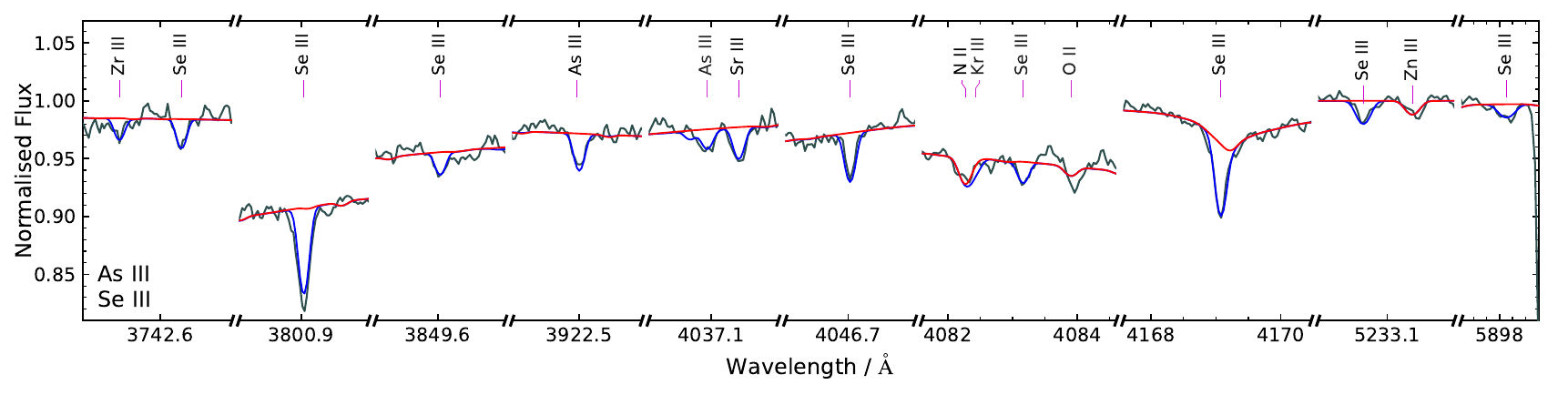}\\%
\vspace{-10pt}
\caption{The strongest \ion{As}{iii} and \ion{Se}{iii} lines in the UVES spectrum of \lsiv.
}
\label{fig:seiii_uves}
\end{figure}

\begin{figure}[!h]
\centering
\includegraphics[width=0.99\textwidth]{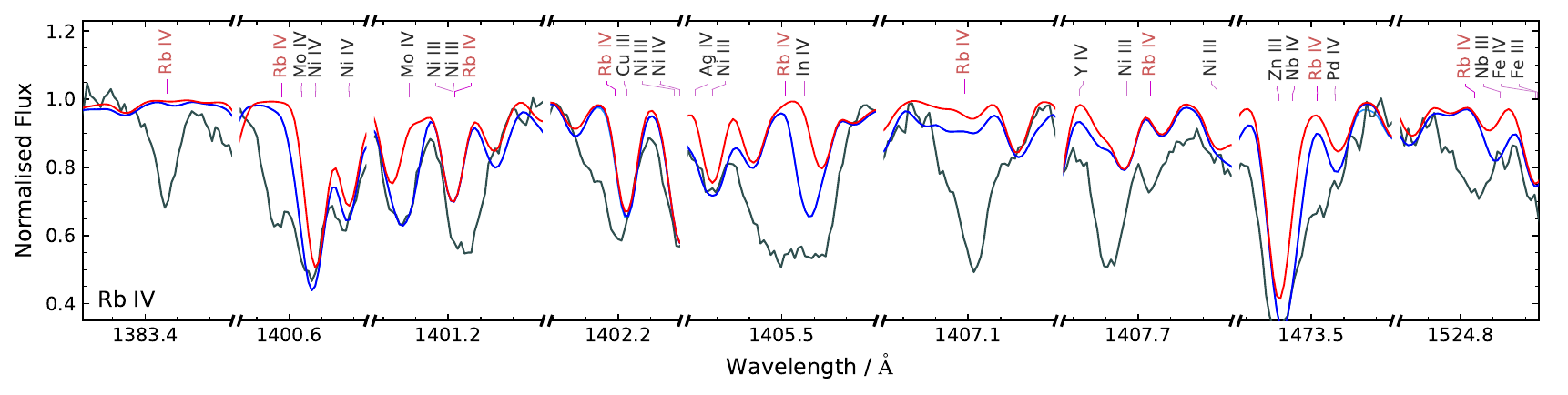}\\[-18.3pt]
\includegraphics[width=0.99\textwidth]{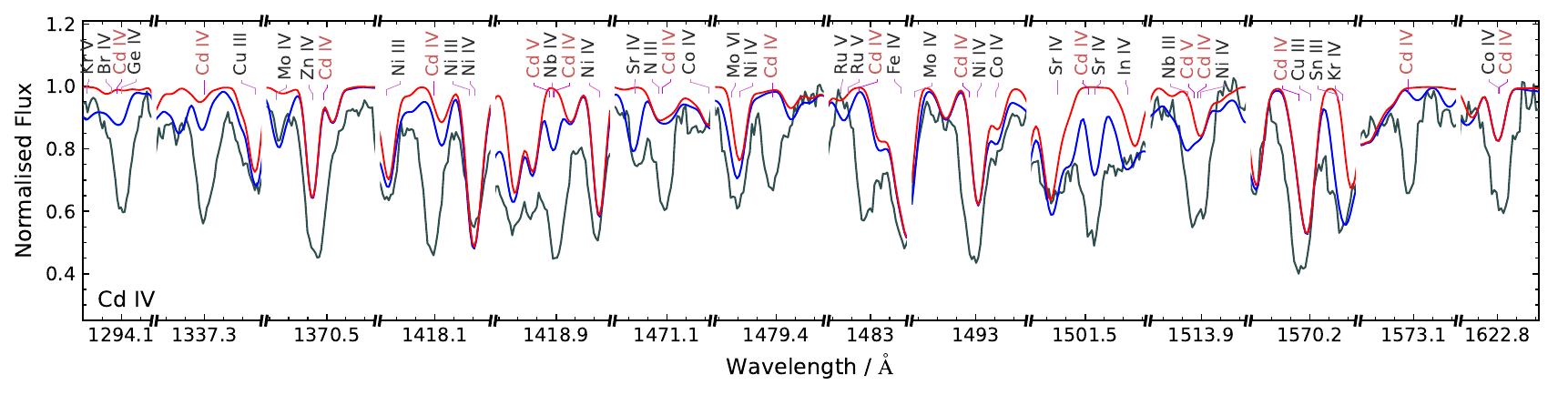}\\[-18.3pt]
\includegraphics[width=0.99\textwidth]{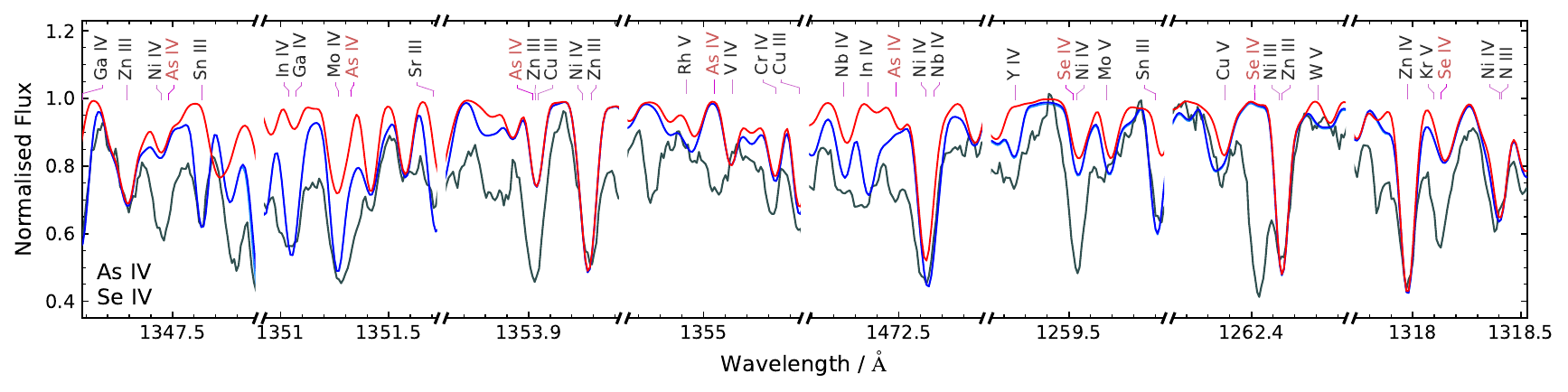}\\
\vspace{-10pt}
\caption{Possible identifications of \ion{As}{iv}, \ion{Se}{iv}, and \ion{Rb}{iv} lines in the UV spectrum of \lsiv, lacking oscillator strengths. 
}
\label{fig:asiv_seiv_rbiv}
\end{figure}

\begin{figure}[!h]
\centering
\includegraphics[width=0.99\textwidth]{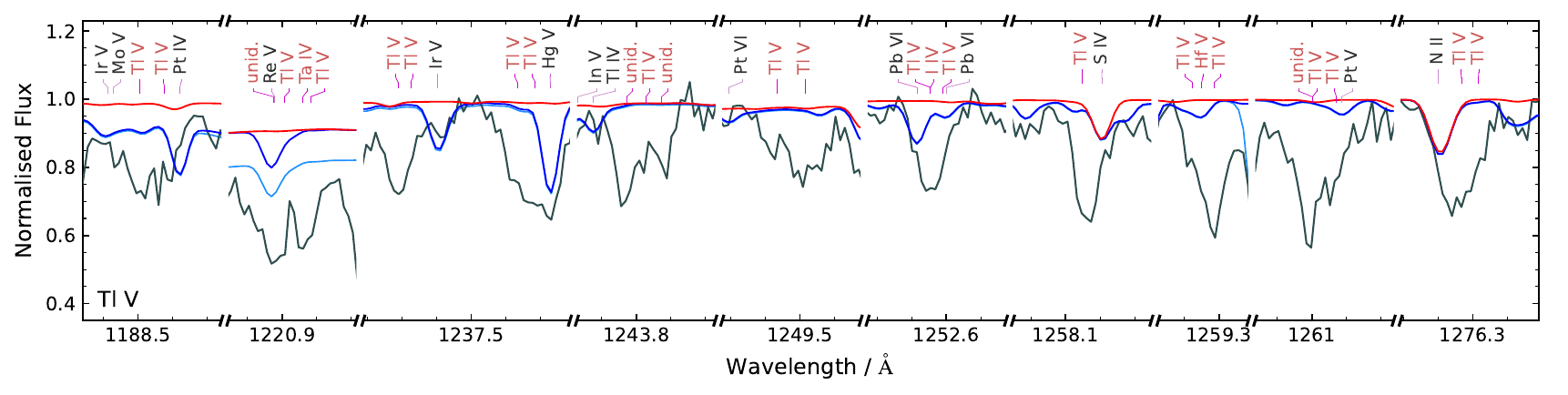}\\%
\vspace{-10pt}
\caption{\ion{Tl}{v} line identifications in \EC, all of which lack oscillator strengths. }
\label{fig:ec_missing_fosc}
\end{figure}

\begin{figure}[!h]
\centering
\includegraphics[width=0.495\textwidth]{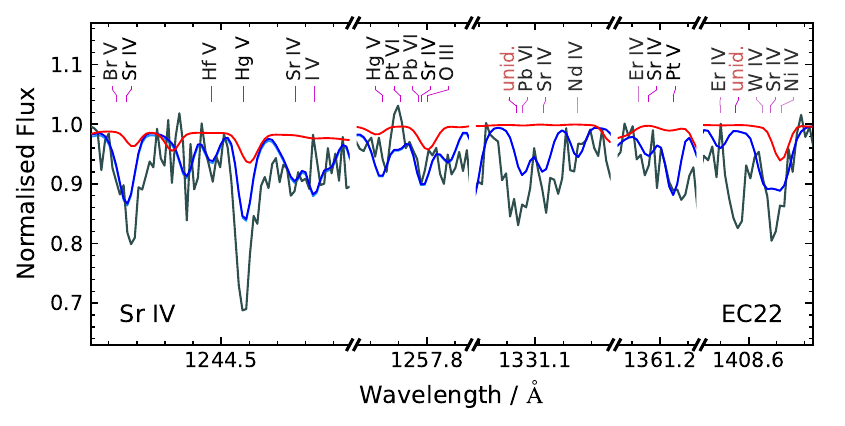}
\includegraphics[width=0.495\textwidth]{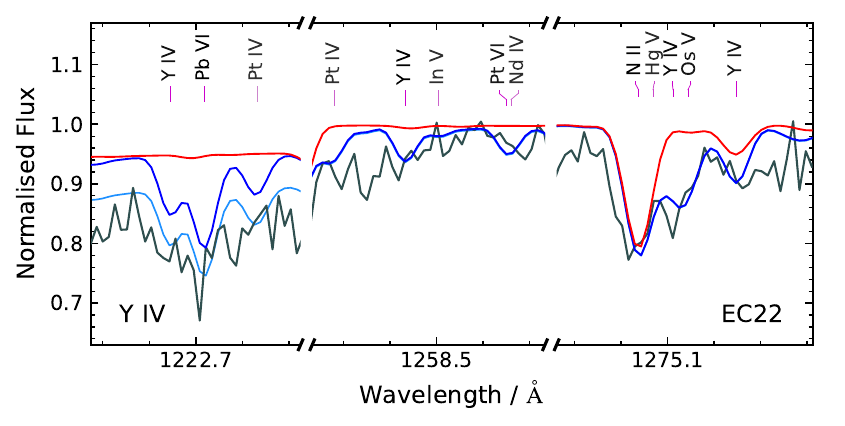}
\\[-18.3pt]
\includegraphics[width=0.99\textwidth]{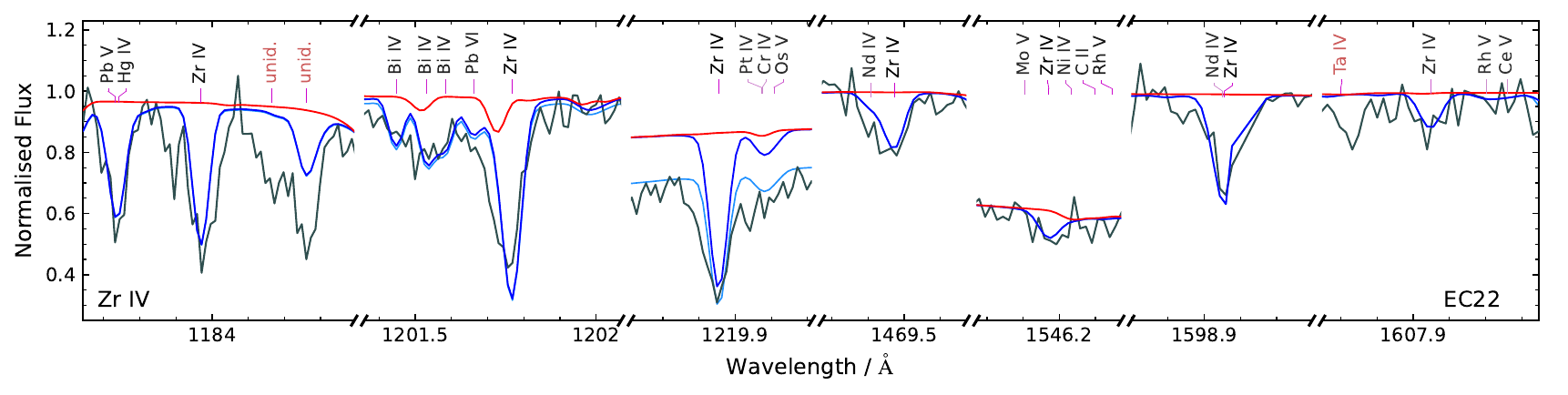}\\[-18.3pt]
\includegraphics[width=0.99\textwidth]{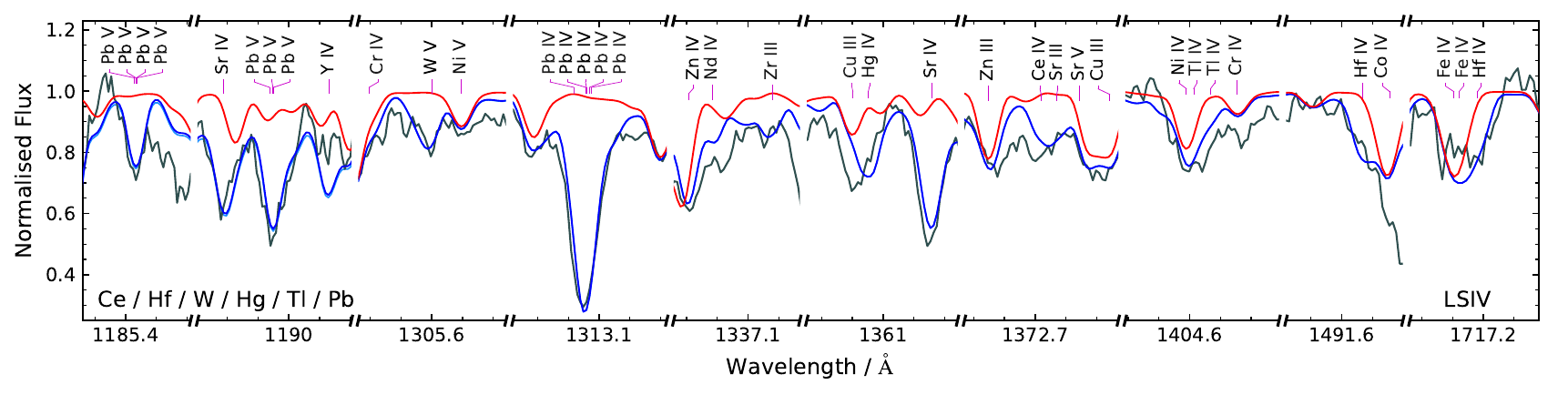}\\%
\caption{The strongest Sr, Y, and Zr lines in \EC, and Ce, Hf, W, Hg, Tl, and Pb in \lsiv. 
}
\label{fig:detail_lsiv_4}
\end{figure}

\FloatBarrier

\end{onecolumn}

\end{appendix}

\end{document}